\documentclass[fleqn,usenatbib]{mnras}

\usepackage{newtxtext,newtxmath}

\usepackage[T1]{fontenc}
\usepackage{ae,aecompl} 

\usepackage{graphicx}	
\usepackage{amsmath}	
\usepackage{amssymb}	
\usepackage{natbib}
\usepackage{footnote}
\usepackage[usenames]{color}
\usepackage{tabularx}
\usepackage{hhline}
\usepackage{array}
\usepackage{url}
\usepackage{multirow}
\usepackage{booktabs}
\usepackage{xspace}
\usepackage{xcolor}
\usepackage{pdflscape}
\usepackage{longtable}
\usepackage{tablefootnote}
\usepackage{lineno}   
\usepackage{eso-pic} 
\usepackage{subfigure}
\usepackage{captcont}
\usepackage{lineno}
\usepackage{longtable}

\newcommand{\cref}[1]{{\textbf{#1}}}

\def\HOLI{H0LiCOW\xspace}
\def\STRIDES{STRIDES\xspace}

\def\HEofor{HE\,0435$-$1223\,}
\def\WFItwenty{WFI\,2033$-$4723\,}

\def\PGeleven{PG\,1115$+$080\,}

\newcommand{\DESone}{DES\,J0408$-$5354\,}
\newcommand{\zsone}{2.375}
\newcommand{\zdone}{0.5967}

\newcommand{\DEStwo}{WGD\,2038$-$4008\,}
\newcommand{\zstwo}{0.777}
\newcommand{\zdtwo}{0.2283}

\newcommand{\vel}{$\sigma_{\rm {obs}}$\, }
\def\kms{km\,s$^{-1}$}
\def\kmsMpc{\rm km\,s$^{-1}$\,Mpc$^{-1}$}
\newcommand{\angdist}{$D_{\rm \theta}(z)$\,}

\newcommand\T{\rule{0pt}{2.6ex}}       
\newcommand\B{\rule[-1.2ex]{0pt}{0pt}} 

\def\ucla{Department of Physics and Astronomy, PAB, 430 Portola Plaza, Box 951547, Los Angeles, CA 90095-1547, USA}

\def\kipac{Kavli Institute for Particle Astrophysics and Cosmology, Stanford University, 452 Lomita Mall, Stanford, CA 94305, USA}

\def\ucd{Department of Physics, University of California Davis, 1 Shields Avenue, Davis, CA 95616, USA}

\def\epfl{Laboratoire d'Astrophysique, Ecole Polytechnique F\'ed\'erale de Lausanne (EPFL), Observatoire de Sauverny, CH-1290 Versoix, Switzerland}

\def\ports{Institute of Cosmology \& Gravitation, University of Portsmouth, Portsmouth, PO1 3FX, UK}

\newcommand{\tabdesgroups}
{
\begin{table*}
	\caption{Group properties in the field of view of \DESone and \DEStwo. The columns show the group ID, group redshift, number of spectroscopically identified galaxies in that group, the group rest-frame velocity dispersion (rounded to the nearest 10\,\kms), the group centroid (in RA and Dec), projected distance of the centroid to the lens ($\Delta\theta$), and median flexion shift $\log(\Delta_3 x (\rm{arcsec}))$ (see \S \ref{sec:flexion} for methodology). All reported values are median quantities of 1000 bootstrapped samples, with uncertainties given by the 16th and 84th confidence intervals of the distribution of the bootstrapped quantity. Velocity dispersion estimates are rounded to nearest 10 \kms. See \S \ref{subsec:flex_groups} for further discussion.}
	\label{tab:groups}
	\centering
	\begin{minipage}{\linewidth}
		\centering
	\begin{tabular}{llrccccccc}
		\hline
ID & $\bar{z}_{\rm group}$ & N & $\sigma_{\rm{rest}}$ & $\sigma_{\rm{int}}$ & $R_{200}$ & RA$_{\rm ctr}$, DEC$_{\rm ctr}$ & error(RA$_{\rm ctr}$, DEC$_{\rm ctr}$) &  $\Delta\theta$ & $\log_{10}(\Delta_3 x)$  \\
& & & (\kms) & (\kms) & (Mpc) & (deg) &  (arcmin) &  (arcsec) & ($\log_{10}(\rm{arcsec})$) \T \B \\
\hline
\multicolumn{10}{c}{\DESone} \T \B \\ 
1 & 0.272 & 11 & $110\pm20$ & $<70$ & $0.30\substack{+0.05\\-0.06}$ & 62.096099, -53.903136 & 0.67, 0.46 & $32\substack{+22\\-16}$ & $<-5.88$  \T \B \\ 
2 & 0.307 & 6 & $290\substack{+60\\-90}$ & $300\substack{+160\\-80}$ & $0.81\substack{+0.15\\-0.24}$ & 62.065082, -53.919339 & 0.71, 0.31 & $94\substack{+10\\-10}$ & $-4.79\substack{+0.74\\-0.63}$  \T \B \\ 
3 & 0.428 & 7 & $110\pm30$ & $<70$ & $0.31\substack{+0.07\\-0.09}$ & 62.130389, -53.936509 & 0.29, 0.29 & $156\substack{+10\\-8}$ & $<-8.04$  \T \B \\ 
4 & 0.570 & 5 & $140\substack{+30\\-90}$ & $<160$ & $0.41\substack{+0.09\\-0.25}$ & 62.042510, -53.918109 & 1.00, 0.52 & $122\substack{+28\\-21}$ & $<-6.53$  \T \B \\ 
5 $\star$ & 0.599 & 17 & $240\substack{+30\\-40}$ & $220\substack{+60\\-50}$ & $0.68\substack{+0.08\\-0.11}$ & 62.079749, -53.901149 & 0.54, 0.18 & $26\substack{+16\\-12}$ & $-3.86\substack{+0.97\\-0.72}$  \T \B \\ 
6 & 0.751 & 3 & $80\pm40$ & $<100$ & $0.21\substack{+0.11\\-0.11}$ & 62.147320, -53.872270 & 0.05, 0.09 & $156\substack{+3\\-2}$ & $<-8.06$  \T \B \\ 
7 & 0.768 & 5 & $120\substack{+30\\-50}$ & $<120$ & $0.34\substack{+0.10\\-0.15}$ & 62.084144, -53.898689 & 0.28, 0.24 & $17\substack{+12\\-7}$ & $<-4.83$  \T \B \\ 
8 & 0.799 & 6 & $340\substack{+60\\-80}$ & $370\substack{+170\\-100}$ & $0.96\substack{+0.18\\-0.22}$ & 62.143634, -53.894864 & 0.64, 0.71 & $120\substack{+22\\-22}$ & $-5.46\substack{+0.72\\-0.59}$  \T \B \\ 
9 & 0.918 & 9 & $300\substack{+40\\-50}$ & $310\substack{+120\\-70}$ & $0.86\substack{+0.10\\-0.15}$ & 62.051793, -53.892228 & 0.49, 0.45 & $90\substack{+16\\-17}$ & $-5.70\substack{+0.58\\-0.52}$  \T \B \\ 
10 & 1.252 & 7 & $240\substack{+30\\-50}$ & $240\substack{+110\\-70}$ & $0.67\substack{+0.09\\-0.15}$ & 62.124690, -53.913336 & 0.69, 0.12 & $87\substack{+22\\-21}$ & $-6.92\substack{+0.80\\-0.71}$  \T \B \\ 
\hline
\multicolumn{10}{c}{\DEStwo} \T \B \\
1 $\star$ & 0.229 & 7 & $200\substack{+70\\-60}$ & $200\substack{+90\\-70}$ & $0.53\substack{+0.18\\-0.16}$ & 309.528309, -40.126006 & 0.41, 0.28 & $62\substack{+23\\-24}$ & $-5.30\substack{+0.90\\-0.83}$  \T \B \\ 
2 & 0.342 & 8 & $380\substack{+70\\-90}$ & $400\substack{+150\\-100}$ & $1.05\substack{+0.19\\-0.26}$ & 309.543495, -40.142963 & 0.54, 0.28 & $91\substack{+26\\-27}$ & $-5.39\substack{+0.73\\-0.59}$  \T \B \\ 
\hline
\end{tabular}
\begin{flushleft}
{\small 
	{\bf Note:} $\star$ Group contains the lens galaxy.
}
\end{flushleft}
\end{minipage}
\end{table*}
}

\newcommand{\tabdesgalflex}{
\begin{table*}
	\caption{Properties of the 10 galaxies with the largest flexion shifts, sorted in order of decreasing 68th percentile upper limits, in the field of \DESone and \DEStwo. The columns display, in order, the DES Y3 Object ID (and ID used in future papers), coordinates (RA, DEC in degrees; ICRS), redshift $z$, distances to the lensing galaxy and flexion shifts, calculated using the scaling relations by \citet{Zahid2016} and \citet{Auger2010} respectively. Galaxies marked with $*$ are not in the spectroscopic survey and only have photometric redshifts; we report the \texttt{DNF\_ZMEAN\_SOF} redshift value and \texttt{DNF\_ZSIGMA\_SOF} uncertainties. Spectroscopic redshift uncertainties are about 100~\kms, or 0.00033 in redshift. Stellar masses and corresponding uncertainties were calculated using the \texttt{Le PHARE} galaxy template fitting code and DES Y3 photometry (see \S \ref{sec:photozmstar}). Galaxies marked with $\dagger$ have spurious \texttt{MOF} magnitudes. For these galaxies, we use \texttt{MAG\_AUTO\_CORRECTED} photometry to calculate stellar masses instead. Flexion shifts and uncertainties are calculated following the method described in \S \ref{sec:flexion}-\ref{subsec:flex_galax}. For a complete list of galaxies, see Table \ref{tab:flexion_galaxy_all}}
	\label{tab:flexion_galaxy}
\begin{tabular}{lcccccccc}
\hline
ID & RA & DEC & $z$ & $i$-band Magnitude& $\log_{10}(M_{*})$  & $\Delta\theta$  & $\log_{10}(\Delta_{3}x_{\rm Zahid})$ & $\log_{10}(\Delta_{3}x_{\rm Auger})$  \\
& (deg) & (deg) & & & ($log_{10}(M_{\odot})$) & (arcsec) & ($\log_{10}(\rm{arcsec})$) & ($\log_{10}(\rm{arcsec})$) \\
\hline
\multicolumn{9}{c}{\DESone} \T \B \\ 
488068102 (G3) & 62.090965 & -53.901634 & 0.76866 & 20.096 & $11.42\substack{+0.10\\-0.12}$ & 6.4 & $-2.21\substack{+0.25\\-0.29}$ & $-2.20\substack{+0.21\\-0.22}$ \T \B \\ 
488065185 (G6) & 62.095575 & -53.898291 & 0.59441 & 21.867 & $10.29\substack{+0.10\\-0.17}$ & 12.4 & $-3.95\substack{+0.42\\-0.58}$ & $-3.42\substack{+0.21\\-0.24}$ \T \B \\ 
488066144 (G4) & 62.090243 & -53.903609 & 0.77069 & 22.015 & $10.66\substack{+0.13\\-0.17}$ & 13.4 & $-4.07\substack{+0.38\\-0.51}$ & $-3.72\substack{+0.22\\-0.24}$ \T \B \\ 
488066768 (G5) & 62.092923 & -53.900091 & 1.03197 & 23.682 & $9.68\substack{+0.24\\-0.25}$ & 5.4 & $-4.93\substack{+0.63\\-0.93}$ & $-3.87\substack{+0.26\\-0.27}$ \T \B \\ 
488066462 & 62.083912 & -53.903969 & 0.60048 & 22.304 & $9.88\substack{+0.12\\-0.19}$ & 20.2 & $-5.24\substack{+0.52\\-0.79}$ & $-4.37\substack{+0.22\\-0.25}$ \T \B \\ 
MUSE6 / 488065214 $\dagger$ & 62.077898 & -53.903878 & 0.59797 & 22.303 & $10.37\substack{+0.11\\-0.16}$ & 30.2 & $-5.02\substack{+0.41\\-0.56}$ & $-4.53\substack{+0.21\\-0.23}$ \T \B \\
488070148 & 62.084648 & -53.885204 & 0.27247 & 18.309 & $10.99\substack{+0.04\\-0.04}$ & 54.3 & $-4.81\substack{+0.31\\-0.37}$ & $-4.61\substack{+0.20\\-0.20}$ \T \B \\
488066886 & 62.090541 & -53.904725 & 0.77568 & 22.925 & $9.91\substack{+0.16\\-0.30}$ & 17.4 & $-5.46\substack{+0.53\\-0.90}$ & $-4.61\substack{+0.23\\-0.30}$ \T \B \\
488070807 & 62.072770 & -53.910296 & 0.59811 & 19.975 & $11.27\substack{+0.05\\-0.06}$ & 53.0 & $-4.69\substack{+0.25\\-0.30}$ & $-4.61\substack{+0.20\\-0.21}$ \T \B \\
MUSE8 / 488067782 & 62.103108 & -53.897861 & 0.59832 & 22.100 & $10.11\substack{+0.13\\-0.20}$ & 27.9 & $-5.29\substack{+0.48\\-0.71}$ & $-4.62\substack{+0.22\\-0.25}$ \T \B \\
\hline
\multicolumn{9}{c}{\DEStwo} \T \B \\ 
169192350 $*$ & 309.514393 & -40.137815 & $0.349 \pm 0.033$  & 21.999 & $9.53\substack{+0.15\\-0.29}$ & 8.8 & $-5.71\substack{+0.59\\-1.03}$ & $-4.52\substack{+0.22\\-0.30}$  \T \B \\ 
169193255 $*$ & 309.529391 & -40.151522 & $0.240 \pm 0.138$  & 18.806 & $10.76\substack{+0.06\\-0.10}$ & 72.0 & $-5.94\substack{+0.35\\-0.45}$ & $-5.63\substack{+0.20\\-0.21}$ \T \B \\ 
169190952 $*$ & 309.503723 & -40.120144 & $0.277 \pm 0.021$  & 19.052 & $10.78\substack{+0.05\\-0.09}$ &  64.3 & $-6.02\substack{+0.34\\-0.44}$ & $-5.72\substack{+0.20\\-0.21}$ \T \B \\ 
169192931 $*$ & 309.508988 & -40.144869 & $0.421 \pm 0.044$  & 20.446 & $10.62\substack{+0.05\\-0.07}$ & 29.0 & $-6.17\substack{+0.37\\-0.47}$ & $-5.80\substack{+0.20\\-0.21}$ \T \B \\ 
169191098 $*$ & 309.487406 & -40.124006 & $0.238 \pm 0.016$  & 18.360 & $10.69\substack{+0.04\\-0.03}$ & 80.9 & $-6.16\substack{+0.36\\-0.45}$ & $-5.82\substack{+0.20\\-0.20}$ \T \B \\ 
169191973 $*$ & 309.536238 & -40.134562 & $0.331 \pm 0.011$  & 18.444 & $11.15\substack{+0.03\\-0.03}$ & 69.0 & $-6.05\substack{+0.27\\-0.33}$ & $-5.92\substack{+0.20\\-0.20}$ \T \B \\ 
169193929 $*$ & 309.520149 & -40.160086 & $0.250 \pm 0.005$  & 19.137 & $10.63\substack{+0.05\\-0.09}$ & 86.5 & $-6.41\substack{+0.36\\-0.47}$ & $-6.04\substack{+0.20\\-0.21}$ \T \B \\ 
169190452 & 309.539131 & -40.115839 & 0.22917 & 18.213 & $10.76\substack{+0.04\\-0.04}$ & 107.9 & $-6.39\substack{+0.35\\-0.43}$ & $-6.08\substack{+0.20\\-0.20}$ \T \B \\ 
169192596 & 309.492815 & -40.140131 & 0.23003 & 21.389 & $9.28\substack{+0.11\\-0.18}$ & 52.3 & $-7.61\substack{+0.60\\-0.96}$ & $-6.21\substack{+0.21\\-0.24}$ \T \B \\ 
169191228 & 309.535586 & -40.122881 & 0.22900 & 20.025 & $10.13\substack{+0.07\\-0.11}$ & 83.9 & $-6.86\substack{+0.45\\-0.62}$ & $-6.21\substack{+0.21\\-0.22}$ \T \B \\ 
\hline
\end{tabular}
\begin{flushleft}
	{\small 
		{\bf Note:} $*$ Galaxy is not in the spectroscopy survey; stellar masses and flexion shift estimates are from photometric redshifts. \\ $\dagger$ Galaxy has spurious \texttt{MOF} photometry. For these galaxies, we use \texttt{MAG\_AUTO\_CORRECTED} photometry to calculate stellar masses instead. \\
	}
\end{flushleft}
\end{table*}
}

\newcommand{\tabdesgalflexall}{
\begin{longtable}{lcccccccc}
	\caption{Properties of all 199 galaxies in spectroscopic survey of the field of \DESone and 54 galaxies in the spectroscopic survey of the field of \DEStwo, arranged in order of decreasing flexion shift on the lens galaxy. The columns display, in order, the galaxy ID, coordinates (RA, DEC in degrees; ICRS), redshift $z$, DES \texttt{MOF\_CM\_MAG\_CORRECTED} $i$-band magnitudes (whenever possible), distances to the lensing galaxy and flexion shifts. Spectroscopic redshift uncertainties are about 100~\kms, or 0.00033 in redshift. Galaxies with bad \texttt{MOF} magnitudes are indicated with $\dagger$; for these galaxies, we use \texttt{MAG\_AUTO\_CORRECTED} photometry and the corresponding stellar mass estimates from that photometric data. For the galaxy IDs, DES Y3 galaxies have 9-digit IDs, DES Y1 galaxies have 10-digit IDs, and MUSE galaxies were labeled with the prefix "MUSE" and sorted by ascending redshifts. Stellar masses and corresponding uncertainties were calculated using the \texttt{Le PHARE} galaxy template fitting code and DES Y3 photometry (see \S \ref{sec:photozmstar}). Flexion shift and uncertainties are calculated following the method described in \S \ref{sec:flexion}-\ref{subsec:flex_galax}.} 
	\label{tab:flexion_galaxy_all} \\
\hline
ID & RA & DEC & $z$ & $i$-band Mag & $\log_{10}(M_{*})$  & $\Delta\theta$  & $\log_{10}(\Delta_{3}x_{\rm Zahid})$ & $\log_{10}(\Delta_{3}x_{\rm Auger})$  \\
& (deg) & (deg) & & & ($log_{10}(M_{\odot})$) & (arcsec) & ($\log_{10}(\rm{arcsec})$) & ($\log_{10}(\rm{arcsec})$) \\
\hline
\multicolumn{9}{c}{\DESone} \T \B \\ 
488068102 & 62.090965 & -53.901634 & 0.76866 & 20.096 & $11.42\substack{+0.10\\-0.12}$ & 6.4 & $-2.21\substack{+0.25\\-0.29}$ & $-2.20\substack{+0.21\\-0.22}$ \T \B \\
488065185 & 62.095575 & -53.898291 & 0.59441 & 21.867 & $10.29\substack{+0.10\\-0.17}$ & 12.4 & $-3.95\substack{+0.42\\-0.58}$ & $-3.42\substack{+0.21\\-0.24}$ \T \B \\
488066144 & 62.090243 & -53.903609 & 0.77069 & 22.015 & $10.66\substack{+0.13\\-0.17}$ & 13.4 & $-4.07\substack{+0.38\\-0.51}$ & $-3.72\substack{+0.22\\-0.24}$ \T \B \\
488066768 & 62.092923 & -53.900091 & 1.03197 & 23.682 & $9.68\substack{+0.24\\-0.25}$ & 5.4 & $-4.93\substack{+0.63\\-0.93}$ & $-3.87\substack{+0.26\\-0.27}$ \T \B \\
488066462 & 62.083912 & -53.903969 & 0.60048 & 22.304 & $9.88\substack{+0.12\\-0.19}$ & 20.2 & $-5.24\substack{+0.52\\-0.79}$ & $-4.37\substack{+0.22\\-0.25}$ \T \B \\
MUSE6/488065214$\dagger$ & 62.077898 & -53.903878 & 0.59797 & 22.303 & $10.37\substack{+0.11\\-0.16}$ & 30.2 & $-5.02\substack{+0.41\\-0.56}$ & $-4.53\substack{+0.21\\-0.23}$ \T \B \\
488070148 & 62.084648 & -53.885204 & 0.27247 & 18.309 & $10.99\substack{+0.04\\-0.04}$ & 54.3 & $-4.81\substack{+0.31\\-0.37}$ & $-4.61\substack{+0.20\\-0.20}$ \T \B \\
488066886 & 62.090541 & -53.904725 & 0.77568 & 22.925 & $9.91\substack{+0.16\\-0.30}$ & 17.4 & $-5.46\substack{+0.53\\-0.90}$ & $-4.61\substack{+0.23\\-0.30}$ \T \B \\
488070807 & 62.072770 & -53.910296 & 0.59811 & 19.975 & $11.27\substack{+0.05\\-0.06}$ & 53.0 & $-4.69\substack{+0.25\\-0.30}$ & $-4.61\substack{+0.20\\-0.21}$ \T \B \\
MUSE8/488067782 & 62.103108 & -53.897861 & 0.59832 & 22.100 & $10.11\substack{+0.13\\-0.20}$ & 27.9 & $-5.29\substack{+0.48\\-0.71}$ & $-4.62\substack{+0.22\\-0.25}$ \T \B \\
488071104 & 62.077670 & -53.913521 & 0.29469 & 18.555 & $11.05\substack{+0.04\\-0.04}$ & 56.0 & $-4.80\substack{+0.30\\-0.36}$ & $-4.62\substack{+0.20\\-0.20}$ \T \B \\
488070774 & 62.075134 & -53.912907 & 0.29368 & 18.519 & $11.01\substack{+0.04\\-0.04}$ & 57.0 & $-4.87\substack{+0.30\\-0.37}$ & $-4.67\substack{+0.20\\-0.20}$ \T \B \\
MUSE15/488066060$\dagger$ & 62.080992 & -53.904565 & 0.76785 & 22.762 & $10.44\substack{+0.12\\-0.18}$ & 26.1 & $-5.19\substack{+0.40\\-0.55}$ & $-4.74\substack{+0.22\\-0.24}$ \T \B \\
488066584 & 62.071692 & -53.906245 & 0.59727 & 21.426 & $10.69\substack{+0.04\\-0.03}$ & 45.8 & $-5.18\substack{+0.35\\-0.45}$ & $-4.84\substack{+0.20\\-0.20}$ \T \B \\
MUSE12/488067196 & 62.093482 & -53.905711 & 0.64363 & 23.311 & $9.18\substack{+0.22\\-0.28}$ & 21.9 & $-6.60\substack{+0.65\\-1.05}$ & $-5.10\substack{+0.25\\-0.29}$ \T \B \\
MUSE10/488070958 & 62.085862 & -53.910180 & 0.60029 & 22.304 & $10.01\substack{+0.14\\-0.22}$ & 38.3 & $-5.86\substack{+0.50\\-0.77}$ & $-5.10\substack{+0.22\\-0.26}$ \T \B \\
489523481 & 62.078380 & -53.881917 & 0.59451 & 19.918 & $11.04\substack{+0.07\\-0.18}$ & 69.6 & $-5.31\substack{+0.30\\-0.43}$ & $-5.13\substack{+0.21\\-0.24}$ \T \B \\
MUSE9/488069251$\dagger$ & 62.085632 & -53.891330 & 0.59906 & 23.333 & $9.49\substack{+0.26\\-0.35}$ & 32.4 & $-6.48\substack{+0.67\\-1.12}$ & $-5.26\substack{+0.27\\-0.33}$ \T \B \\
488070002 & 62.070997 & -53.882405 & 0.37961 & 20.614 & $10.39\substack{+0.05\\-0.06}$ & 75.2 & $-6.02\substack{+0.39\\-0.51}$ & $-5.54\substack{+0.20\\-0.20}$ \T \B \\
488070966 & 62.067875 & -53.910488 & 0.59977 & 21.729 & $10.11\substack{+0.15\\-0.19}$ & 61.2 & $-6.32\substack{+0.49\\-0.70}$ & $-5.65\substack{+0.23\\-0.25}$ \T \B \\
488067363 & 62.134647 & -53.901754 & 0.59479 & 21.047 & $10.86\substack{+0.05\\-0.05}$ & 94.1 & $-5.92\substack{+0.33\\-0.41}$ & $-5.66\substack{+0.20\\-0.20}$ \T \B \\
MUSE14/488070770 & 62.099785 & -53.908977 & 0.76669 & 23.533 & $9.85\substack{+0.21\\-0.28}$ & 38.3 & $-6.56\substack{+0.58\\-0.89}$ & $-5.66\substack{+0.24\\-0.29}$ \T \B \\
488070443 & 62.101056 & -53.907011 & 1.17764 & 21.911 & $11.01\substack{+0.34\\-0.32}$ & 34.2 & $-5.88\substack{+0.48\\-0.55}$ & $-5.68\substack{+0.31\\-0.31}$ \T \B \\
488071406 & 62.054830 & -53.913785 & 0.44241 & 20.276 & $10.58\substack{+0.06\\-0.07}$ & 90.5 & $-6.08\substack{+0.37\\-0.48}$ & $-5.69\substack{+0.20\\-0.21}$ \T \B \\
405340030 & 62.118966 & -53.880474 & 0.57772 & 20.663 & $10.71\substack{+0.07\\-0.18}$ & 92.5 & $-6.06\substack{+0.36\\-0.50}$ & $-5.73\substack{+0.21\\-0.24}$ \T \B \\
488071428 & 62.068708 & -53.913477 & 0.59843 & 21.581 & $10.15\substack{+0.12\\-0.17}$ & 67.2 & $-6.36\substack{+0.46\\-0.66}$ & $-5.73\substack{+0.22\\-0.24}$ \T \B \\
488072005 & 62.068079 & -53.917283 & 0.54122 & 21.624 & $10.32\substack{+0.10\\-0.18}$ & 78.5 & $-6.27\substack{+0.41\\-0.58}$ & $-5.76\substack{+0.21\\-0.24}$ \T \B \\
MUSE19/488068048 & 62.088268 & -53.890642 & 0.91504 & 23.065 & $9.92\substack{+0.30\\-0.29}$ & 33.6 & $-6.64\substack{+0.64\\-0.87}$ & $-5.80\substack{+0.29\\-0.29}$ \T \B \\
488067814 & 62.109798 & -53.896296 & 0.78734 & 22.001 & $9.90\substack{+0.20\\-0.24}$ & 43.1 & $-6.68\substack{+0.56\\-0.83}$ & $-5.83\substack{+0.24\\-0.27}$ \T \B \\
MUSE22/488065663 & 62.090044 & -53.891845 & 1.09148 & 23.419 & $10.22\substack{+0.28\\-0.27}$ & 29.0 & $-6.40\substack{+0.59\\-0.75}$ & $-5.83\substack{+0.28\\-0.29}$ \T \B \\
488072697 & 62.061177 & -53.921532 & 0.27257 & 19.767 & $10.33\substack{+0.04\\-0.05}$ & 99.6 & $-6.37\substack{+0.40\\-0.52}$ & $-5.87\substack{+0.20\\-0.20}$ \T \B \\
MUSE20/488065095 & 62.091419 & -53.890585 & 0.94698 & 23.108 & $9.91\substack{+0.34\\-0.26}$ & 33.6 & $-6.74\substack{+0.69\\-0.85}$ & $-5.89\substack{+0.31\\-0.28}$ \T \B \\
488070825 & 62.106067 & -53.909453 & 0.77562 & 21.908 & $9.93\substack{+0.21\\-0.26}$ & 47.8 & $-6.74\substack{+0.56\\-0.84}$ & $-5.91\substack{+0.24\\-0.28}$ \T \B \\
488066235 & 62.023367 & -53.910705 & 0.57025 & 19.314 & $11.28\substack{+0.06\\-0.17}$ & 147.4 & $-6.00\substack{+0.26\\-0.36}$ & $-5.92\substack{+0.20\\-0.24}$ \T \B \\
489523492 & 62.067524 & -53.881235 & 0.78001 & 20.704 & $10.92\substack{+0.12\\-0.18}$ & 82.9 & $-6.17\substack{+0.34\\-0.46}$ & $-5.93\substack{+0.22\\-0.24}$ \T \B \\
489521356 & 62.083229 & -53.863995 & 0.57765 & 20.123 & $11.02\substack{+0.07\\-0.17}$ & 130.1 & $-6.14\substack{+0.31\\-0.42}$ & $-5.95\substack{+0.21\\-0.24}$ \T \B \\
488070625 & 62.057786 & -53.908625 & 0.93778 & 21.744 & $11.27\substack{+0.05\\-0.07}$ & 76.0 & $-6.04\substack{+0.26\\-0.31}$ & $-5.96\substack{+0.20\\-0.21}$ \T \B \\
488073290 & 62.106161 & -53.923866 & 0.57035 & 20.933 & $10.36\substack{+0.07\\-0.12}$ & 92.5 & $-6.47\substack{+0.40\\-0.54}$ & $-5.98\substack{+0.21\\-0.22}$ \T \B \\
488074578 & 62.144676 & -53.934498 & 0.27176 & 18.315 & $11.11\substack{+0.05\\-0.12}$ & 169.6 & $-6.16\substack{+0.29\\-0.37}$ & $-6.01\substack{+0.20\\-0.22}$ \T \B \\
488070449 & 62.025309 & -53.910519 & 0.42954 & 19.664 & $10.95\substack{+0.05\\-0.07}$ & 143.3 & $-6.24\substack{+0.32\\-0.39}$ & $-6.01\substack{+0.20\\-0.21}$ \T \B \\
488073735 & 62.139479 & -53.928501 & 0.42858 & 19.429 & $10.95\substack{+0.05\\-0.07}$ & 146.4 & $-6.27\substack{+0.32\\-0.39}$ & $-6.04\substack{+0.20\\-0.21}$ \T \B \\
488072433 & 62.093088 & -53.919040 & 0.30909 & 21.573 & $9.45\substack{+0.09\\-0.11}$ & 69.2 & $-7.31\substack{+0.58\\-0.88}$ & $-6.06\substack{+0.21\\-0.22}$ \T \B \\
489522058 & 62.078867 & -53.872863 & 0.09083 & 18.298 & $9.90\substack{+0.04\\-0.04}$ & 100.3 & $-6.92\substack{+0.49\\-0.68}$ & $-6.07\substack{+0.20\\-0.20}$ \T \B \\
488066362 & 62.052505 & -53.883271 & 0.94641 & 21.835 & $11.50\substack{+0.03\\-0.03}$ & 100.2 & $-6.14\substack{+0.21\\-0.24}$ & $-6.17\substack{+0.20\\-0.20}$ \T \B \\
488072414 & 62.148320 & -53.919275 & 0.42744 & 20.060 & $10.70\substack{+0.10\\-0.07}$ & 141.2 & $-6.50\substack{+0.37\\-0.45}$ & $-6.17\substack{+0.21\\-0.21}$ \T \B \\
488072532 & 62.024006 & -53.921066 & 0.44478 & 19.809 & $10.87\substack{+0.06\\-0.08}$ & 160.1 & $-6.49\substack{+0.33\\-0.42}$ & $-6.23\substack{+0.20\\-0.21}$ \T \B \\
488067185 & 62.020675 & -53.907526 & 0.30566 & 19.692 & $10.58\substack{+0.04\\-0.05}$ & 150.5 & $-6.64\substack{+0.37\\-0.47}$ & $-6.25\substack{+0.20\\-0.20}$ \T \B \\
488076699 & 62.096336 & -53.944173 & 0.30792 & 19.259 & $10.67\substack{+0.05\\-0.07}$ & 159.9 & $-6.62\substack{+0.36\\-0.46}$ & $-6.27\substack{+0.20\\-0.21}$ \T \B \\
MUSE21/488071544 & 62.087179 & -53.913732 & 0.98821 & 22.458 & $10.21\substack{+0.23\\-0.32}$ & 50.3 & $-6.88\substack{+0.53\\-0.81}$ & $-6.30\substack{+0.25\\-0.31}$ \T \B \\
488067820 & 62.129495 & -53.889946 & 0.25191 & 21.049 & $9.53\substack{+0.09\\-0.11}$ & 90.3 & $-7.49\substack{+0.57\\-0.86}$ & $-6.30\substack{+0.21\\-0.22}$ \T \B \\
MUSE3/488069277 & 62.103394 & -53.889095 & 0.40779 & 23.661 & $8.52\substack{+0.27\\-0.32}$ & 47.6 & $-8.40\substack{+0.43\\-0.53}$ & $-6.31\substack{+0.27\\-0.31}$ \T \B \\
488068050 & 62.133568 & -53.897135 & 0.46159 & 20.744 & $9.75\substack{+0.09\\-0.10}$ & 92.1 & $-7.31\substack{+0.53\\-0.77}$ & $-6.32\substack{+0.21\\-0.22}$ \T \B \\
MUSE26/488070824 & 62.095831 & -53.909607 & 1.25062 & 22.383 & $10.49\substack{+0.19\\-0.22}$ & 36.8 & $-6.76\substack{+0.43\\-0.57}$ & $-6.33\substack{+0.24\\-0.26}$ \T \B \\
488071089 & 62.049904 & -53.911313 & 0.30581 & 21.203 & $9.64\substack{+0.10\\-0.12}$ & 95.3 & $-7.43\substack{+0.56\\-0.83}$ & $-6.34\substack{+0.21\\-0.22}$ \T \B \\
488065930 & 62.137580 & -53.888784 & 0.57825 & 21.113 & $10.08\substack{+0.11\\-0.12}$ & 107.7 & $-7.08\substack{+0.47\\-0.65}$ & $-6.38\substack{+0.21\\-0.22}$ \T \B \\
488069700 & 62.083413 & -53.882657 & 0.27216 & 22.456 & $8.80\substack{+0.16\\-0.21}$ & 63.8 & $-8.23\substack{+0.50\\-0.72}$ & $-6.39\substack{+0.23\\-0.25}$ \T \B \\
489521068 & 62.092960 & -53.861224 & 0.27153 & 20.306 & $10.21\substack{+0.07\\-0.12}$ & 139.3 & $-6.98\substack{+0.43\\-0.60}$ & $-6.40\substack{+0.20\\-0.22}$ \T \B \\
489521448 & 62.103449 & -53.864717 & 0.43985 & 20.616 & $10.23\substack{+0.10\\-0.15}$ & 129.6 & $-6.97\substack{+0.44\\-0.62}$ & $-6.41\substack{+0.21\\-0.23}$ \T \B \\
488072561 & 62.053602 & -53.920059 & 0.62960 & 21.216 & $10.16\substack{+0.09\\-0.11}$ & 106.6 & $-7.04\substack{+0.45\\-0.61}$ & $-6.41\substack{+0.21\\-0.22}$ \T \B \\
488074608 & 62.140970 & -53.931443 & 0.43054 & 20.683 & $10.53\substack{+0.05\\-0.05}$ & 156.2 & $-6.84\substack{+0.38\\-0.48}$ & $-6.43\substack{+0.20\\-0.20}$ \T \B \\
488070133 & 62.169886 & -53.887360 & 0.59607 & 20.158 & $10.84\substack{+0.14\\-0.12}$ & 174.5 & $-6.74\substack{+0.36\\-0.44}$ & $-6.47\substack{+0.22\\-0.22}$ \T \B \\
488073762 & 62.131777 & -53.926465 & 0.42904 & 21.665 & $10.11\substack{+0.08\\-0.14}$ & 129.8 & $-7.15\substack{+0.45\\-0.65}$ & $-6.49\substack{+0.21\\-0.23}$ \T \B \\
488073021 & 62.080583 & -53.922670 & 0.30741 & 21.184 & $9.21\substack{+0.08\\-0.09}$ & 84.6 & $-7.96\substack{+0.59\\-0.89}$ & $-6.49\substack{+0.21\\-0.21}$ \T \B \\
489522199 & 62.081990 & -53.870601 & 0.39730 & 21.866 & $9.72\substack{+0.14\\-0.24}$ & 106.9 & $-7.51\substack{+0.56\\-0.90}$ & $-6.50\substack{+0.22\\-0.27}$ \T \B \\
488072604 & 62.073251 & -53.920150 & 0.77290 & 22.129 & $10.07\substack{+0.16\\-0.19}$ & 81.5 & $-7.20\substack{+0.50\\-0.71}$ & $-6.50\substack{+0.23\\-0.25}$ \T \B \\
489521528 & 62.088464 & -53.864945 & 0.24688 & 20.718 & $9.83\substack{+0.09\\-0.08}$ & 125.9 & $-7.43\substack{+0.52\\-0.73}$ & $-6.52\substack{+0.21\\-0.21}$ \T \B \\
488069699 & 62.097258 & -53.886716 & 0.92560 & 22.617 & $9.67\substack{+0.22\\-0.23}$ & 49.6 & $-7.58\substack{+0.61\\-0.91}$ & $-6.52\substack{+0.25\\-0.27}$ \T \B \\
488070602 & 62.030809 & -53.908820 & 0.26729 & 20.096 & $9.86\substack{+0.06\\-0.06}$ & 130.4 & $-7.45\substack{+0.50\\-0.71}$ & $-6.56\substack{+0.20\\-0.21}$ \T \B \\
489523576 & 62.050079 & -53.881679 & 0.94745 & 22.694 & $11.08\substack{+0.07\\-0.08}$ & 107.8 & $-6.73\substack{+0.29\\-0.36}$ & $-6.57\substack{+0.20\\-0.21}$ \T \B \\
MUSE24/488069253 & 62.104998 & -53.902502 & 1.14831 & 23.055 & $9.55\substack{+0.26\\-0.28}$ & 32.3 & $-7.76\substack{+0.66\\-1.01}$ & $-6.59\substack{+0.27\\-0.29}$ \T \B \\
488074678 & 62.146922 & -53.934938 & 0.27044 & 19.713 & $10.34\substack{+0.04\\-0.04}$ & 174.0 & $-7.09\substack{+0.40\\-0.52}$ & $-6.59\substack{+0.20\\-0.20}$ \T \B \\
488074840 & 62.107311 & -53.932685 & 0.73105 & 21.409 & $10.54\substack{+0.09\\-0.11}$ & 123.4 & $-7.00\substack{+0.38\\-0.50}$ & $-6.59\substack{+0.21\\-0.22}$ \T \B \\
488071415 & 62.113750 & -53.913080 & 0.74981 & 22.287 & $9.54\substack{+0.18\\-0.15}$ & 68.6 & $-7.78\substack{+0.61\\-0.89}$ & $-6.60\substack{+0.24\\-0.23}$ \T \B \\
488070947 & 62.158048 & -53.912350 & 0.76818 & 18.274 & $11.02\substack{+0.10\\-0.14}$ & 150.3 & $-6.80\substack{+0.32\\-0.41}$ & $-6.60\substack{+0.21\\-0.23}$ \T \B \\
488067669 & 62.135908 & -53.895219 & 0.46127 & 22.257 & $9.47\substack{+0.18\\-0.28}$ & 98.0 & $-7.85\substack{+0.62\\-1.03}$ & $-6.61\substack{+0.23\\-0.29}$ \T \B \\
488071411 & 62.045447 & -53.913359 & 0.30642 & 21.438 & $9.44\substack{+0.10\\-0.11}$ & 107.0 & $-7.89\substack{+0.58\\-0.89}$ & $-6.63\substack{+0.21\\-0.22}$ \T \B \\
489523301 & 62.097716 & -53.879281 & 0.27278 & 22.768 & $8.75\substack{+0.16\\-0.22}$ & 75.8 & $-8.53\substack{+0.46\\-0.67}$ & $-6.65\substack{+0.23\\-0.26}$ \T \B \\
488072531 & 62.108813 & -53.919489 & 0.75070 & 22.550 & $9.73\substack{+0.17\\-0.20}$ & 80.6 & $-7.68\substack{+0.57\\-0.86}$ & $-6.67\substack{+0.23\\-0.25}$ \T \B \\
488070816 & 62.035444 & -53.909470 & 0.60004 & 22.347 & $9.92\substack{+0.23\\-0.24}$ & 121.6 & $-7.52\substack{+0.58\\-0.82}$ & $-6.68\substack{+0.25\\-0.27}$ \T \B \\
404169047 & 62.136795 & -53.861686 & 0.16020 & 18.916 & $10.04\substack{+0.04\\-0.04}$ & 169.1 & $-7.42\substack{+0.46\\-0.63}$ & $-6.69\substack{+0.20\\-0.20}$ \T \B \\
488070607 & 62.037079 & -53.908383 & 0.59964 & 22.356 & $9.76\substack{+0.18\\-0.21}$ & 117.2 & $-7.72\substack{+0.57\\-0.85}$ & $-6.74\substack{+0.23\\-0.26}$ \T \B \\
488067357 & 62.163804 & -53.896708 & 0.72762 & 21.625 & $10.72\substack{+0.06\\-0.11}$ & 156.1 & $-7.10\substack{+0.36\\-0.46}$ & $-6.77\substack{+0.20\\-0.22}$ \T \B \\
404169634 & 62.112378 & -53.871809 & 0.80252 & 21.675 & $10.34\substack{+0.15\\-0.37}$ & 111.3 & $-7.28\substack{+0.43\\-0.72}$ & $-6.79\substack{+0.22\\-0.34}$ \T \B \\
488074345 & 62.179533 & -53.930268 & 0.43930 & 20.522 & $10.58\substack{+0.06\\-0.12}$ & 218.3 & $-7.23\substack{+0.37\\-0.50}$ & $-6.84\substack{+0.20\\-0.22}$ \T \B \\
489521853 & 62.084697 & -53.867583 & 0.27148 & 21.675 & $9.27\substack{+0.12\\-0.13}$ & 116.9 & $-8.25\substack{+0.60\\-0.92}$ & $-6.84\substack{+0.21\\-0.22}$ \T \B \\
488078736 & 62.114845 & -53.953040 & 0.42826 & 21.346 & $10.35\substack{+0.07\\-0.09}$ & 198.2 & $-7.36\substack{+0.40\\-0.53}$ & $-6.86\substack{+0.21\\-0.21}$ \T \B \\
488066250 & 62.017973 & -53.905064 & 0.57028 & 21.306 & $10.05\substack{+0.10\\-0.14}$ & 154.8 & $-7.59\substack{+0.47\\-0.68}$ & $-6.87\substack{+0.21\\-0.23}$ \T \B \\
488069992 & 62.129682 & -53.882696 & 0.23970 & 21.880 & $8.93\substack{+0.10\\-0.12}$ & 103.8 & $-8.63\substack{+0.53\\-0.78}$ & $-6.91\substack{+0.21\\-0.22}$ \T \B \\
488074838 & 62.144448 & -53.932510 & 0.50871 & 21.484 & $9.97\substack{+0.09\\-0.13}$ & 164.1 & $-7.75\substack{+0.49\\-0.70}$ & $-6.96\substack{+0.21\\-0.22}$ \T \B \\
488078018 & 62.078114 & -53.949484 & 0.51766 & 21.911 & $10.14\substack{+0.09\\-0.17}$ & 180.4 & $-7.60\substack{+0.45\\-0.67}$ & $-6.96\substack{+0.21\\-0.24}$ \T \B \\
488073172 & 62.142324 & -53.923373 & 0.39342 & 21.554 & $9.52\substack{+0.12\\-0.12}$ & 138.8 & $-8.18\substack{+0.58\\-0.87}$ & $-6.98\substack{+0.21\\-0.22}$ \T \B \\
488077424 & 62.142854 & -53.946644 & 0.39071 & 21.170 & $10.14\substack{+0.07\\-0.09}$ & 201.7 & $-7.65\substack{+0.44\\-0.60}$ & $-7.01\substack{+0.21\\-0.21}$ \T \B \\
488077861 & 62.077259 & -53.948516 & 0.51744 & 21.566 & $10.04\substack{+0.10\\-0.12}$ & 177.3 & $-7.75\substack{+0.48\\-0.66}$ & $-7.01\substack{+0.21\\-0.22}$ \T \B \\
488075668 & 62.133767 & -53.937477 & 0.42807 & 20.447 & $9.77\substack{+0.07\\-0.06}$ & 163.6 & $-8.01\substack{+0.52\\-0.75}$ & $-7.04\substack{+0.21\\-0.21}$ \T \B \\
488067945 & 62.019121 & -53.902892 & 0.57171 & 21.720 & $9.77\substack{+0.10\\-0.13}$ & 151.6 & $-8.02\substack{+0.53\\-0.79}$ & $-7.04\substack{+0.21\\-0.23}$ \T \B \\
488072443 & 62.054344 & -53.919213 & 0.27226 & 22.046 & $8.75\substack{+0.14\\-0.13}$ & 103.4 & $-8.94\substack{+0.45\\-0.58}$ & $-7.06\substack{+0.22\\-0.22}$ \T \B \\
488068654 & 62.143374 & -53.894424 & 0.81351 & 22.007 & $10.04\substack{+0.19\\-0.23}$ & 114.0 & $-7.79\substack{+0.53\\-0.77}$ & $-7.06\substack{+0.24\\-0.27}$ \T \B \\
488071272 & 62.133772 & -53.912201 & 0.15809 & 22.207 & $8.60\substack{+0.13\\-0.16}$ & 102.1 & $-9.09\substack{+0.29\\-0.36}$ & $-7.07\substack{+0.22\\-0.23}$ \T \B \\
488072984 & 62.146444 & -53.921724 & 0.60815 & 22.144 & $9.68\substack{+0.13\\-0.18}$ & 142.5 & $-8.13\substack{+0.56\\-0.86}$ & $-7.08\substack{+0.22\\-0.24}$ \T \B \\
488068672 & 62.140255 & -53.894233 & 0.16682 & 21.453 & $8.67\substack{+0.13\\-0.14}$ & 107.7 & $-9.04\substack{+0.37\\-0.47}$ & $-7.09\substack{+0.22\\-0.23}$ \T \B \\
488073629 & 62.125366 & -53.926127 & 0.63364 & 22.258 & $9.43\substack{+0.16\\-0.16}$ & 120.1 & $-8.38\substack{+0.61\\-0.92}$ & $-7.10\substack{+0.23\\-0.23}$ \T \B \\
488073466 & 62.165107 & -53.925645 & 0.93736 & 20.916 & $11.24\substack{+0.05\\-0.07}$ & 183.5 & $-7.21\substack{+0.26\\-0.31}$ & $-7.12\substack{+0.20\\-0.21}$ \T \B \\
488073766 & 62.050530 & -53.926881 & 0.75824 & 22.336 & $9.92\substack{+0.21\\-0.34}$ & 128.8 & $-8.00\substack{+0.57\\-0.95}$ & $-7.17\substack{+0.25\\-0.33}$ \T \B \\
488070459 & 62.171977 & -53.907068 & 0.37773 & 21.361 & $9.65\substack{+0.11\\-0.12}$ & 174.9 & $-8.25\substack{+0.56\\-0.83}$ & $-7.17\substack{+0.21\\-0.22}$ \T \B \\
488070459 & 62.171977 & -53.907068 & 0.37779 & 21.361 & $9.65\substack{+0.11\\-0.12}$ & 174.9 & $-8.25\substack{+0.56\\-0.83}$ & $-7.17\substack{+0.21\\-0.22}$ \T \B \\
488072024 & 62.037238 & -53.917791 & 0.37796 & 22.807 & $9.08\substack{+0.18\\-0.26}$ & 129.9 & $-8.78\substack{+0.62\\-1.01}$ & $-7.19\substack{+0.24\\-0.28}$ \T \B \\
488070460 & 62.171213 & -53.906968 & 0.79720 & 22.595 & $10.56\substack{+0.08\\-0.08}$ & 173.2 & $-7.60\substack{+0.38\\-0.48}$ & $-7.19\substack{+0.21\\-0.21}$ \T \B \\
488072618 & 62.142800 & -53.919871 & 0.77739 & 22.398 & $10.00\substack{+0.17\\-0.35}$ & 132.3 & $-7.96\substack{+0.52\\-0.93}$ & $-7.19\substack{+0.23\\-0.33}$ \T \B \\
488070460 & 62.171213 & -53.906968 & 0.79737 & 22.595 & $10.56\substack{+0.08\\-0.08}$ & 173.2 & $-7.60\substack{+0.38\\-0.48}$ & $-7.19\substack{+0.21\\-0.21}$ \T \B \\
488074030 & 62.163437 & -53.927404 & 0.46567 & 21.400 & $9.79\substack{+0.10\\-0.11}$ & 183.8 & $-8.16\substack{+0.53\\-0.76}$ & $-7.20\substack{+0.21\\-0.22}$ \T \B \\
488070027 & 62.154997 & -53.888704 & 0.81379 & 22.887 & $10.25\substack{+0.18\\-0.26}$ & 142.8 & $-7.74\substack{+0.49\\-0.73}$ & $-7.20\substack{+0.24\\-0.28}$ \T \B \\
404169734 & 62.125315 & -53.873569 & 0.76243 & 22.755 & $9.75\substack{+0.23\\-0.30}$ & 120.3 & $-8.20\substack{+0.61\\-0.96}$ & $-7.21\substack{+0.26\\-0.30}$ \T \B \\
488075485 & 62.112254 & -53.935970 & 0.29471 & 22.103 & $9.08\substack{+0.12\\-0.14}$ & 137.9 & $-8.80\substack{+0.58\\-0.89}$ & $-7.21\substack{+0.22\\-0.23}$ \T \B \\
404169375 & 62.133682 & -53.867494 & 0.16039 & 20.977 & $9.07\substack{+0.11\\-0.12}$ & 148.4 & $-8.82\substack{+0.58\\-0.88}$ & $-7.22\substack{+0.21\\-0.22}$ \T \B \\
MUSE25/488070944 & 62.098868 & -53.909951 & 1.25002 & 23.508 & $9.42\substack{+0.19\\-0.26}$ & 40.4 & $-8.50\substack{+0.63\\-1.02}$ & $-7.22\substack{+0.24\\-0.28}$ \T \B \\
488067796 & 62.025664 & -53.892284 & 0.91979 & 21.768 & $10.53\substack{+0.10\\-0.31}$ & 140.1 & $-7.65\substack{+0.39\\-0.63}$ & $-7.24\substack{+0.21\\-0.31}$ \T \B \\
488073160 & 62.043694 & -53.923652 & 0.91456 & 22.098 & $10.36\substack{+0.18\\-0.20}$ & 130.9 & $-7.75\substack{+0.44\\-0.58}$ & $-7.26\substack{+0.24\\-0.25}$ \T \B \\
404169737 & 62.148449 & -53.873747 & 0.75002 & 22.097 & $10.09\substack{+0.15\\-0.19}$ & 155.0 & $-7.95\substack{+0.49\\-0.70}$ & $-7.26\substack{+0.22\\-0.25}$ \T \B \\
MUSE28/488067056 & 62.107634 & -53.905654 & 1.25400 & 23.100 & $9.45\substack{+0.23\\-0.25}$ & 42.0 & $-8.52\substack{+0.65\\-1.00}$ & $-7.26\substack{+0.25\\-0.28}$ \T \B \\
404169584 & 62.156567 & -53.871330 & 0.80070 & 21.188 & $10.48\substack{+0.29\\-0.13}$ & 174.0 & $-7.70\substack{+0.49\\-0.52}$ & $-7.27\substack{+0.28\\-0.22}$ \T \B \\
405340085 & 62.140943 & -53.880240 & 0.30034 & 21.842 & $8.88\substack{+0.18\\-0.13}$ & 128.4 & $-9.04\substack{+0.55\\-0.74}$ & $-7.27\substack{+0.24\\-0.22}$ \T \B \\
488077259 & 62.092037 & -53.945258 & 0.54971 & 22.234 & $9.56\substack{+0.15\\-0.18}$ & 163.4 & $-8.44\substack{+0.59\\-0.91}$ & $-7.28\substack{+0.22\\-0.24}$ \T \B \\
489521272 & 62.090218 & -53.862527 & 0.63286 & 22.813 & $9.36\substack{+0.23\\-0.25}$ & 134.5 & $-8.64\substack{+0.66\\-1.02}$ & $-7.30\substack{+0.25\\-0.28}$ \T \B \\
489523103 & 62.037949 & -53.878208 & 0.89700 & 21.083 & $10.30\substack{+0.13\\-0.17}$ & 136.0 & $-7.83\substack{+0.43\\-0.58}$ & $-7.31\substack{+0.22\\-0.24}$ \T \B \\
488076578 & 62.107604 & -53.941787 & 0.79782 & 21.477 & $10.19\substack{+0.14\\-0.16}$ & 155.2 & $-7.91\substack{+0.47\\-0.64}$ & $-7.32\substack{+0.22\\-0.23}$ \T \B \\
489521533 & 62.082318 & -53.864622 & 0.67459 & 23.344 & $9.38\substack{+0.26\\-0.28}$ & 128.1 & $-8.65\substack{+0.68\\-1.05}$ & $-7.33\substack{+0.27\\-0.29}$ \T \B \\
489522088 & 62.038672 & -53.869800 & 0.59969 & 22.534 & $9.40\substack{+0.27\\-0.25}$ & 154.2 & $-8.66\substack{+0.68\\-1.02}$ & $-7.36\substack{+0.27\\-0.28}$ \T \B \\
488072301 & 62.160850 & -53.917944 & 0.59760 & 22.185 & $9.47\substack{+0.16\\-0.18}$ & 162.9 & $-8.61\substack{+0.61\\-0.93}$ & $-7.37\substack{+0.23\\-0.24}$ \T \B \\
404169401 & 62.147459 & -53.869008 & 0.75141 & 21.712 & $10.05\substack{+0.24\\-0.28}$ & 164.3 & $-8.10\substack{+0.57\\-0.82}$ & $-7.38\substack{+0.26\\-0.29}$ \T \B \\
404169402 & 62.130404 & -53.867675 & 0.59397 & 23.104 & $9.20\substack{+0.26\\-0.33}$ & 143.7 & $-8.88\substack{+0.68\\-1.12}$ & $-7.40\substack{+0.27\\-0.32}$ \T \B \\
404169758 & 62.146052 & -53.874054 & 0.75095 & 22.586 & $9.85\substack{+0.16\\-0.21}$ & 150.3 & $-8.31\substack{+0.55\\-0.82}$ & $-7.40\substack{+0.23\\-0.25}$ \T \B \\
404169588 & 62.151601 & -53.871104 & 0.56693 & 22.810 & $9.40\substack{+0.18\\-0.19}$ & 166.1 & $-8.73\substack{+0.62\\-0.95}$ & $-7.43\substack{+0.23\\-0.25}$ \T \B \\
488076552 & 62.105955 & -53.941347 & 0.36171 & 22.388 & $9.03\substack{+0.16\\-0.17}$ & 152.8 & $-9.06\substack{+0.59\\-0.89}$ & $-7.43\substack{+0.23\\-0.24}$ \T \B \\
488078159 & 62.047075 & -53.950191 & 0.56953 & 21.811 & $9.71\substack{+0.09\\-0.12}$ & 203.1 & $-8.49\substack{+0.54\\-0.80}$ & $-7.46\substack{+0.21\\-0.22}$ \T \B \\
489522305 & 62.030458 & -53.871796 & 0.91911 & 22.148 & $10.48\substack{+0.19\\-0.38}$ & 162.5 & $-7.90\substack{+0.43\\-0.70}$ & $-7.46\substack{+0.24\\-0.35}$ \T \B \\
488073582 & 62.036870 & -53.925643 & 0.91818 & 21.953 & $10.27\substack{+0.18\\-0.28}$ & 146.6 & $-8.01\substack{+0.45\\-0.66}$ & $-7.48\substack{+0.23\\-0.29}$ \T \B \\
489522522 & 62.023889 & -53.873587 & 0.59935 & 22.742 & $9.35\substack{+0.18\\-0.25}$ & 170.0 & $-8.87\substack{+0.63\\-1.03}$ & $-7.52\substack{+0.24\\-0.28}$ \T \B \\
488075100 & 62.135113 & -53.933925 & 0.42804 & 22.300 & $8.97\substack{+0.07\\-0.07}$ & 154.9 & $-9.22\substack{+0.54\\-0.79}$ & $-7.54\substack{+0.21\\-0.21}$ \T \B \\
404168970 & 62.143547 & -53.858498 & 0.84037 & 22.020 & $10.35\substack{+0.16\\-0.17}$ & 186.9 & $-8.04\substack{+0.43\\-0.57}$ & $-7.55\substack{+0.23\\-0.24}$ \T \B \\
404169534 & 62.142829 & -53.870321 & 0.80082 & 21.848 & $9.86\substack{+0.19\\-0.17}$ & 153.9 & $-8.45\substack{+0.56\\-0.78}$ & $-7.55\substack{+0.24\\-0.24}$ \T \B \\
488073010 & 62.046958 & -53.922500 & 0.27124 & 22.409 & $8.37\substack{+0.16\\-0.14}$ & 123.0 & $-9.78\substack{+0.25\\-0.22}$ & $-7.56\substack{+0.23\\-0.23}$ \T \B \\
488079339 & 62.108903 & -53.956880 & 0.42895 & 21.139 & $9.48\substack{+0.09\\-0.08}$ & 208.9 & $-8.80\substack{+0.58\\-0.86}$ & $-7.56\substack{+0.21\\-0.21}$ \T \B \\
488070430 & 62.027802 & -53.906937 & 0.95625 & 22.490 & $10.13\substack{+0.21\\-0.32}$ & 135.2 & $-8.23\substack{+0.53\\-0.84}$ & $-7.57\substack{+0.24\\-0.31}$ \T \B \\
488071245 & 62.030863 & -53.912062 & 0.77531 & 22.736 & $9.46\substack{+0.25\\-0.19}$ & 133.7 & $-8.84\substack{+0.66\\-0.94}$ & $-7.59\substack{+0.26\\-0.25}$ \T \B \\
488078889 & 62.046282 & -53.955925 & 0.19565 & 20.697 & $9.31\substack{+0.10\\-0.10}$ & 222.4 & $-8.98\substack{+0.59\\-0.90}$ & $-7.60\substack{+0.21\\-0.21}$ \T \B \\
489521965 & 62.045558 & -53.868686 & 0.89759 & 22.513 & $10.04\substack{+0.21\\-0.24}$ & 147.2 & $-8.33\substack{+0.55\\-0.78}$ & $-7.60\substack{+0.25\\-0.27}$ \T \B \\
488065215 & 62.027412 & -53.885104 & 1.10526 & 22.372 & $10.66\substack{+0.26\\-0.37}$ & 143.9 & $-7.99\substack{+0.45\\-0.67}$ & $-7.63\substack{+0.27\\-0.34}$ \T \B \\
488075166 & 62.149266 & -53.934182 & 0.50880 & 22.887 & $9.14\substack{+0.19\\-0.24}$ & 175.5 & $-9.18\substack{+0.63\\-1.00}$ & $-7.64\substack{+0.24\\-0.27}$ \T \B \\
488068535 & 62.149540 & -53.895034 & 0.77707 & 22.702 & $9.28\substack{+0.24\\-0.22}$ & 126.6 & $-9.06\substack{+0.66\\-0.99}$ & $-7.65\substack{+0.26\\-0.26}$ \T \B \\
488073904 & 62.151510 & -53.927139 & 0.27214 & 22.822 & $8.74\substack{+0.12\\-0.16}$ & 162.5 & $-9.55\substack{+0.43\\-0.58}$ & $-7.66\substack{+0.22\\-0.23}$ \T \B \\
488068835 & 62.019738 & -53.894319 & 0.91905 & 22.531 & $10.05\substack{+0.21\\-0.26}$ & 151.3 & $-8.40\substack{+0.54\\-0.80}$ & $-7.68\substack{+0.25\\-0.28}$ \T \B \\
405340125 & 62.147848 & -53.880963 & 0.82674 & 22.738 & $9.58\substack{+0.23\\-0.27}$ & 139.6 & $-8.83\substack{+0.64\\-0.99}$ & $-7.69\substack{+0.26\\-0.29}$ \T \B \\
488076126 & 62.098545 & -53.939165 & 0.29568 & 22.286 & $8.45\substack{+0.12\\-0.11}$ & 142.4 & $-9.86\substack{+0.19\\-0.18}$ & $-7.71\substack{+0.21\\-0.22}$ \T \B \\
488074972 & 62.158897 & -53.933231 & 0.27187 & 22.261 & $8.93\substack{+0.11\\-0.13}$ & 188.4 & $-9.43\substack{+0.54\\-0.79}$ & $-7.71\substack{+0.21\\-0.22}$ \T \B \\
488073464 & 62.128601 & -53.924891 & 1.08324 & 22.899 & $10.15\substack{+0.25\\-0.33}$ & 121.1 & $-8.35\substack{+0.57\\-0.85}$ & $-7.72\substack{+0.26\\-0.32}$ \T \B \\
488072862 & 62.014278 & -53.921997 & 0.44369 & 22.954 & $8.99\substack{+0.17\\-0.23}$ & 180.0 & $-9.41\substack{+0.59\\-0.92}$ & $-7.73\substack{+0.23\\-0.26}$ \T \B \\
488071425 & 62.019165 & -53.913520 & 0.77498 & 22.912 & $9.57\substack{+0.18\\-0.24}$ & 158.9 & $-8.89\substack{+0.60\\-0.96}$ & $-7.74\substack{+0.23\\-0.27}$ \T \B \\
488077585 & 62.091222 & -53.947042 & 0.85319 & 23.034 & $9.94\substack{+0.17\\-0.20}$ & 169.8 & $-8.57\substack{+0.54\\-0.77}$ & $-7.75\substack{+0.23\\-0.25}$ \T \B \\
404169398 & 62.129582 & -53.867590 & 1.04896 & 22.750 & $10.28\substack{+0.30\\-0.36}$ & 142.9 & $-8.28\substack{+0.51\\-0.72}$ & $-7.75\substack{+0.29\\-0.34}$ \T \B \\
488076873 & 62.062246 & -53.943414 & 0.51666 & 22.956 & $8.90\substack{+0.28\\-0.27}$ & 167.7 & $-9.51\substack{+0.64\\-0.89}$ & $-7.76\substack{+0.28\\-0.28}$ \T \B \\
404169833 & 62.165627 & -53.875815 & 0.61452 & 22.767 & $9.16\substack{+0.22\\-0.18}$ & 181.6 & $-9.30\substack{+0.65\\-0.95}$ & $-7.78\substack{+0.25\\-0.24}$ \T \B \\
488074337 & 62.173464 & -53.929392 & 0.40472 & 22.616 & $9.08\substack{+0.15\\-0.17}$ & 205.6 & $-9.40\substack{+0.60\\-0.92}$ & $-7.81\substack{+0.22\\-0.24}$ \T \B \\
488067934 & 62.032469 & -53.898319 & 1.34023 & 21.519 & $10.94\substack{+0.05\\-0.04}$ & 123.0 & $-8.04\substack{+0.32\\-0.39}$ & $-7.81\substack{+0.20\\-0.20}$ \T \B \\
488077841 & 62.133208 & -53.947857 & 0.35684 & 22.687 & $8.91\substack{+0.17\\-0.23}$ & 195.1 & $-9.57\substack{+0.56\\-0.85}$ & $-7.83\substack{+0.23\\-0.26}$ \T \B \\
488070960 & 62.031874 & -53.910446 & 1.23026 & 21.226 & $10.58\substack{+0.19\\-0.19}$ & 129.8 & $-8.25\substack{+0.42\\-0.54}$ & $-7.86\substack{+0.24\\-0.25}$ \T \B \\
489521442 & 62.031862 & -53.864100 & 0.90591 & 22.699 & $9.96\substack{+0.24\\-0.26}$ & 179.0 & $-8.74\substack{+0.58\\-0.83}$ & $-7.94\substack{+0.26\\-0.28}$ \T \B \\
488076117 & 62.040622 & -53.939310 & 1.01820 & 22.596 & $10.24\substack{+0.18\\-0.21}$ & 176.9 & $-8.54\substack{+0.49\\-0.67}$ & $-7.99\substack{+0.23\\-0.26}$ \T \B \\
404169065 & 62.156908 & -53.860476 & 0.75027 & 22.948 & $9.44\substack{+0.23\\-0.25}$ & 200.1 & $-9.34\substack{+0.65\\-1.01}$ & $-8.07\substack{+0.25\\-0.28}$ \T \B \\
489520262 & 62.067931 & -53.853616 & 0.91854 & 22.457 & $9.71\substack{+0.25\\-0.23}$ & 173.3 & $-9.12\substack{+0.63\\-0.89}$ & $-8.10\substack{+0.26\\-0.26}$ \T \B \\
404169193 & 62.141088 & -53.862996 & 1.16021 & 22.384 & $10.50\substack{+0.21\\-0.30}$ & 170.9 & $-8.52\substack{+0.44\\-0.63}$ & $-8.10\substack{+0.25\\-0.30}$ \T \B \\
404169489 & 62.136503 & -53.870431 & 0.04355 & 21.430 & $7.60\substack{+0.37\\-0.32}$ & 144.3 & $-11.07\substack{+0.59\\-0.51}$ & $-8.16\substack{+0.32\\-0.31}$ \T \B \\
404169659 & 62.146071 & -53.872477 & 1.14904 & 21.682 & $9.98\substack{+0.17\\-0.22}$ & 153.9 & $-9.09\substack{+0.53\\-0.78}$ & $-8.31\substack{+0.23\\-0.26}$ \T \B \\
488075055 & 62.169444 & -53.933578 & 1.03614 & 22.691 & $10.13\substack{+0.17\\-0.32}$ & 206.8 & $-8.97\substack{+0.50\\-0.84}$ & $-8.32\substack{+0.23\\-0.31}$ \T \B \\
404169176 & 62.145040 & -53.862782 & 1.14723 & 22.928 & $10.20\substack{+0.24\\-0.30}$ & 176.9 & $-8.92\substack{+0.55\\-0.80}$ & $-8.33\substack{+0.26\\-0.30}$ \T \B \\
404169154 & 62.149335 & -53.862340 & 1.14884 & 22.251 & $9.78\substack{+0.20\\-0.24}$ & 184.1 & $-9.66\substack{+0.59\\-0.88}$ & $-8.69\substack{+0.24\\-0.27}$ \T \B \\
488073779 & 62.020985 & -53.927277 & 1.38408 & 22.280 & $10.52\substack{+0.18\\-0.24}$ & 177.2 & $-9.12\substack{+0.42\\-0.58}$ & $-8.70\substack{+0.23\\-0.27}$ \T \B \\
488072310 & 62.161986 & -53.918003 & 1.25287 & 22.603 & $9.08\substack{+0.18\\-0.14}$ & 165.2 & $-10.89\substack{+0.61\\-0.89}$ & $-9.31\substack{+0.23\\-0.23}$ \T \B \\
488072700 & 62.160696 & -53.920293 & 1.25468 & 22.908 & $9.07\substack{+0.25\\-0.24}$ & 166.2 & $-10.93\substack{+0.66\\-0.98}$ & $-9.33\substack{+0.26\\-0.27}$ \T \B \\
488072408 & 62.161866 & -53.918519 & 1.25018 & 22.937 & $8.95\substack{+0.18\\-0.19}$ & 165.7 & $-11.10\substack{+0.58\\-0.86}$ & $-9.40\substack{+0.24\\-0.25}$ \T \B \\
404169985 & 62.153722 & -53.878539 & 1.41999 & 22.613 & $9.23\substack{+0.26\\-0.24}$ & 154.7 & $-11.00\substack{+0.68\\-1.02}$ & $-9.55\substack{+0.27\\-0.27}$ \T \B \\
489520462 & 62.073666 & -53.855405 & 1.17034 & 22.110 & $7.69\substack{+0.45\\-0.45}$ & 164.0 & $-12.93\substack{+0.72\\-0.72}$ & $-10.10\substack{+0.37\\-0.39}$ \T \B \\
489520752 & 62.055153 & -53.858062 & 1.59303 & 22.328 & $9.14\substack{+0.29\\-0.23}$ & 168.1 & $-11.75\substack{+0.70\\-0.99}$ & $-10.21\substack{+0.28\\-0.27}$ \T \B \\
3070263610 & 62.040098 & -53.885262 & 1.32360 & - &- & 119.0 & - & - \T \B \\
3070263850 & 62.075161 & -53.892278 & 0.91583 & - &- & 42.4 & - & - \T \B \\
3070265512 & 62.076105 & -53.885866 & 0.91540 & - &- & 58.9 & - & - \T \B \\
3070264072 & 62.073661 & -53.889176 & 0.76790 & - &- & 52.4 & - & - \T \B \\
MUSE1 & 62.093482 & -53.907716 & 0.23983 & $-$ & $-$ & 28.9 & $-$ & $-$ \T \B \\
MUSE2 & 62.088497 & -53.889554 & 0.28757 & $-$ & $-$ & 37.4 & $-$ & $-$ \T \B \\
MUSE4 & 62.079502 & -53.896486 & 0.55446 & $-$ & $-$ & 26.2 & $-$ & $-$ \T \B \\
MUSE5 & 62.092164 & -53.902560 & 0.59221 & $-$ & $-$ & 10.3 & $-$ & $-$ \T \B \\
MUSE7 & 62.086607 & -53.902732 & 0.59822 & $-$ & $-$ & 13.0 & $-$ & $-$ \T \B \\
MUSE11 & 62.092623 & -53.909149 & 0.64338 & $-$ & $-$ & 33.7 & $-$ & $-$ \T \B \\
MUSE13 & 62.088612 & -53.911555 & 0.76379 & $-$ & $-$ & 42.2 & $-$ & $-$ \T \B \\
MUSE16 & 62.075319 & -53.889095 & 0.76863 & $-$ & $-$ & 50.4 & $-$ & $-$ \T \B \\
MUSE17 & 62.095373 & -53.907544 & 0.82980 & $-$ & $-$ & 29.5 & $-$ & $-$ \T \B \\
MUSE18 & 62.069991 & -53.898778 & 0.85458 & $-$ & $-$ & 43.5 & $-$ & $-$ \T \B \\
MUSE23 & 62.082424 & -53.891043 & 1.13208 & $-$ & $-$ & 36.1 & $-$ & $-$ \T \B \\
MUSE27 & 62.084658 & -53.911326 & 1.25152 & $-$ & $-$ & 42.9 & $-$ & $-$ \T \B \\
\hline
\multicolumn{9}{c}{\DEStwo} \T \B \\ 
169190452 & 309.539131 & -40.115839 & 0.22917 & 18.213 & $10.76\substack{+0.04\\-0.04}$ & 107.9 & $-6.39\substack{+0.35\\-0.43}$ & $-6.08\substack{+0.20\\-0.20}$ \T \B \\ 
169192596 & 309.492815 & -40.140131 & 0.23003 & 21.389 & $9.28\substack{+0.11\\-0.18}$ & 52.3 & $-7.61\substack{+0.60\\-0.96}$ & $-6.21\substack{+0.21\\-0.24}$ \T \B \\ 
169191228 & 309.535586 & -40.122881 & 0.22900 & 20.025 & $10.13\substack{+0.07\\-0.11}$ & 83.9 & $-6.86\substack{+0.45\\-0.62}$ & $-6.21\substack{+0.21\\-0.22}$ \T \B \\ 
169189459 & 309.539963 & -40.103707 & 0.22827 & 17.848 & $11.00\substack{+0.04\\-0.04}$ & 143.5 & $-6.47\substack{+0.30\\-0.37}$ & $-6.27\substack{+0.20\\-0.20}$ \T \B \\ 
169192249 & 309.537431 & -40.135642 & 0.22864 & 21.077 & $9.60\substack{+0.08\\-0.09}$ & 71.9 & $-7.50\substack{+0.56\\-0.82}$ & $-6.38\substack{+0.21\\-0.21}$ \T \B \\ 
169192351 & 309.523347 & -40.137071 & 0.42874 & 20.494 & $10.12\substack{+0.10\\-0.12}$ & 32.9 & $-7.05\substack{+0.46\\-0.63}$ & $-6.39\substack{+0.21\\-0.22}$ \T \B \\ 
169191897 & 309.517763 & -40.130806 & 0.34445 & 22.210 & $8.69\substack{+0.13\\-0.12}$ & 28.5 & $-8.56\substack{+0.40\\-0.49}$ & $-6.63\substack{+0.22\\-0.22}$ \T \B \\ 
169193594 & 309.514195 & -40.155037 & 0.27820 & 21.126 & $9.31\substack{+0.11\\-0.13}$ & 65.3 & $-8.19\substack{+0.60\\-0.92}$ & $-6.81\substack{+0.21\\-0.22}$ \T \B \\ 
169194640 & 309.471495 & -40.170293 & 0.23647 & 19.289 & $10.42\substack{+0.06\\-0.06}$ & 162.4 & $-7.38\substack{+0.39\\-0.51}$ & $-6.91\substack{+0.20\\-0.21}$ \T \B \\ 
169192636 & 309.498710 & -40.140676 & 0.42710 & 22.669 & $9.45\substack{+0.16\\-0.26}$ & 37.3 & $-8.27\substack{+0.61\\-1.01}$ & $-7.02\substack{+0.23\\-0.28}$ \T \B \\ 
169193902 & 309.487074 & -40.160804 & 0.42725 & 18.676 & $11.35\substack{+0.03\\-0.03}$ & 108.6 & $-7.09\substack{+0.24\\-0.27}$ & $-7.04\substack{+0.20\\-0.20}$ \T \B \\ 
169191786 & 309.521701 & -40.129094 & 0.34157 & 22.937 & $8.60\substack{+0.23\\-0.18}$ & 40.3 & $-9.15\substack{+0.43\\-0.40}$ & $-7.13\substack{+0.26\\-0.24}$ \T \B \\ 
169191437 & 309.540911 & -40.125437 & 0.22716 & 21.577 & $8.65\substack{+0.10\\-0.10}$ & 91.4 & $-9.34\substack{+0.33\\-0.39}$ & $-7.37\substack{+0.21\\-0.21}$ \T \B \\ 
169193432 & 309.557474 & -40.152652 & 0.34079 & 19.475 & $10.47\substack{+0.07\\-0.07}$ & 138.8 & $-7.83\substack{+0.39\\-0.50}$ & $-7.39\substack{+0.21\\-0.21}$ \T \B \\ 
169194065 & 309.562909 & -40.162398 & 0.34506 & 19.299 & $10.78\substack{+0.07\\-0.05}$ & 168.7 & $-7.75\substack{+0.35\\-0.43}$ & $-7.45\substack{+0.20\\-0.20}$ \T \B \\ 
169191326 & 309.496715 & -40.122896 & 0.34271 & 22.345 & $8.96\substack{+0.18\\-0.19}$ & 64.9 & $-9.20\substack{+0.58\\-0.86}$ & $-7.50\substack{+0.23\\-0.25}$ \T \B \\ 
169189145 & 309.516470 & -40.095073 & 0.20084 & 19.786 & $9.31\substack{+0.11\\-0.11}$ & 151.7 & $-8.89\substack{+0.60\\-0.91}$ & $-7.51\substack{+0.21\\-0.22}$ \T \B \\ 
169192814 & 309.520646 & -40.143046 & 0.52902 & 22.046 & $9.57\substack{+0.20\\-0.22}$ & 33.5 & $-8.75\substack{+0.62\\-0.94}$ & $-7.60\substack{+0.24\\-0.26}$ \T \B \\ 
169194067 & 309.482408 & -40.160695 & 0.07580 & 20.890 & $8.19\substack{+0.26\\-0.19}$ & 116.7 & $-10.15\substack{+0.41\\-0.31}$ & $-7.77\substack{+0.27\\-0.25}$ \T \B \\ 
169193455 & 309.496328 & -40.151391 & 0.36829 & 22.005 & $8.79\substack{+0.11\\-0.12}$ & 66.3 & $-9.68\substack{+0.46\\-0.62}$ & $-7.83\substack{+0.21\\-0.22}$ \T \B \\ 
169189914 & 309.518874 & -40.103760 & 0.19742 & 22.878 & $8.00\substack{+0.20\\-0.17}$ & 121.5 & $-10.71\substack{+0.32\\-0.27}$ & $-8.16\substack{+0.24\\-0.24}$ \T \B \\ 
169194531 & 309.503092 & -40.167505 & 0.42654 & 21.622 & $9.83\substack{+0.10\\-0.12}$ & 112.1 & $-9.10\substack{+0.52\\-0.75}$ & $-8.18\substack{+0.21\\-0.22}$ \T \B \\ 
169189067 & 309.542813 & -40.092954 & 0.28598 & 21.217 & $9.10\substack{+0.14\\-0.12}$ & 180.7 & $-9.92\substack{+0.60\\-0.89}$ & $-8.35\substack{+0.22\\-0.22}$ \T \B \\ 
169193237 & 309.561418 & -40.148516 & 0.34018 & 22.225 & $9.14\substack{+0.12\\-0.14}$ & 143.8 & $-9.92\substack{+0.59\\-0.91}$ & $-8.39\substack{+0.22\\-0.23}$ \T \B \\ 
169193555 & 309.568673 & -40.153285 & 0.34218 & 21.564 & $9.39\substack{+0.12\\-0.14}$ & 168.2 & $-9.75\substack{+0.59\\-0.92}$ & $-8.43\substack{+0.21\\-0.23}$ \T \B \\ 
169191966 & 309.541745 & -40.132233 & 0.60086 & 20.513 & $10.76\substack{+0.07\\-0.23}$ & 85.3 & $-9.01\substack{+0.35\\-0.52}$ & $-8.70\substack{+0.21\\-0.26}$ \T \B \\ 
169189826 & 309.535769 & -40.102557 & 0.47794 & 22.031 & $9.99\substack{+0.08\\-0.09}$ & 141.1 & $-9.52\substack{+0.48\\-0.67}$ & $-8.75\substack{+0.21\\-0.21}$ \T \B \\ 
169194495 & 309.511978 & -40.166836 & 1.13033 & 22.524 & $10.29\substack{+0.24\\-0.27}$ & 107.3 & $-9.34\substack{+0.48\\-0.64}$ & $-8.82\substack{+0.26\\-0.29}$ \T \B \\ 
169193901 & 309.570179 & -40.158238 & 0.47405 & 21.269 & $10.18\substack{+0.09\\-0.11}$ & 178.9 & $-9.50\substack{+0.44\\-0.60}$ & $-8.89\substack{+0.21\\-0.22}$ \T \B \\ 
169189023 & 309.491739 & -40.092652 & 0.54487 & 20.724 & $10.80\substack{+0.03\\-0.03}$ & 168.6 & $-9.26\substack{+0.34\\-0.42}$ & $-8.97\substack{+0.20\\-0.20}$ \T \B \\ 
169193078 & 309.564077 & -40.146099 & 0.34071 & 22.869 & $8.37\substack{+0.14\\-0.14}$ & 148.7 & $-11.21\substack{+0.23\\-0.23}$ & $-8.99\substack{+0.22\\-0.23}$ \T \B \\ 
169191992 & 309.457026 & -40.131800 & 0.51834 & 21.469 & $9.85\substack{+0.11\\-0.13}$ & 150.8 & $-10.17\substack{+0.52\\-0.75}$ & $-9.27\substack{+0.21\\-0.22}$ \T \B \\ 
169195181 & 309.510546 & -40.177017 & 1.17836 & 22.655 & $9.61\substack{+0.15\\-0.17}$ & 144.0 & $-10.62\substack{+0.58\\-0.88}$ & $-9.51\substack{+0.23\\-0.24}$ \T \B \\ 
169192635 & 309.485842 & -40.140711 & 1.35058 & 20.463 & $7.58\substack{+0.37\\-0.41}$ & 71.5 & $-12.53\substack{+0.60\\-0.66}$ & $-9.61\substack{+0.32\\-0.37}$ \T \B \\ 
169191596 & 309.458841 & -40.126588 & 0.48342 & 22.409 & $8.93\substack{+0.18\\-0.17}$ & 149.4 & $-11.35\substack{+0.57\\-0.82}$ & $-9.63\substack{+0.23\\-0.24}$ \T \B \\ 
169190526 & 309.539140 & -40.111919 & 0.91987 & 21.537 & $10.65\substack{+0.18\\-0.33}$ & 118.4 & $-10.39\substack{+0.41\\-0.63}$ & $-10.03\substack{+0.24\\-0.32}$ \T \B \\ 
169196183 & 309.518797 & -40.190901 & 0.97564 & 21.295 & $10.77\substack{+0.24\\-0.32}$ & 195.0 & $-10.38\substack{+0.43\\-0.60}$ & $-10.08\substack{+0.26\\-0.31}$ \T \B \\ 
169193763 & 309.500311 & -40.155916 & 0.67100 & 22.476 & $9.75\substack{+0.27\\-0.19}$ & 74.5 & $-11.24\substack{+0.65\\-0.84}$ & $-10.25\substack{+0.28\\-0.25}$ \T \B \\ 
169189572 & 309.526023 & -40.099386 & 0.62258 & 20.045 & $9.74\substack{+0.07\\-0.05}$ & 141.4 & $-11.35\substack{+0.53\\-0.75}$ & $-10.35\substack{+0.20\\-0.20}$ \T \B \\ 
169192546 & 309.545721 & -40.139586 & 0.67484 & 21.345 & $10.03\substack{+0.11\\-0.16}$ & 95.0 & $-11.16\substack{+0.48\\-0.70}$ & $-10.43\substack{+0.21\\-0.23}$ \T \B \\ 
169194605 & 309.556455 & -40.168268 & 0.60064 & 22.811 & $9.43\substack{+0.21\\-0.27}$ & 167.4 & $-11.80\substack{+0.64\\-1.02}$ & $-10.52\substack{+0.25\\-0.28}$ \T \B \\ 
169194113 & 309.498489 & -40.161174 & 0.87838 & 22.655 & $10.00\substack{+0.22\\-0.26}$ & 93.9 & $-11.50\substack{+0.56\\-0.81}$ & $-10.74\substack{+0.25\\-0.28}$ \T \B \\ 
169195495 & 309.501974 & -40.181155 & 0.92178 & 21.862 & $9.98\substack{+0.19\\-0.16}$ & 161.0 & $-11.68\substack{+0.54\\-0.72}$ & $-10.89\substack{+0.24\\-0.23}$ \T \B \\ 
169190809 & 309.511333 & -40.116436 & 0.81855 & 20.801 & $10.56\substack{+0.18\\-0.24}$ & 74.1 & $-11.89\substack{+0.42\\-0.58}$ & $-11.50\substack{+0.24\\-0.27}$ \T \B \\ 
169193629 & 309.483800 & -40.153801 & 0.82247 & 21.200 & $10.41\substack{+0.16\\-0.16}$ & 97.0 & $-12.27\substack{+0.42\\-0.55}$ & $-11.80\substack{+0.23\\-0.23}$ \T \B \\ 
169196130 & 309.490867 & -40.190128 & 0.81949 & 21.775 & $10.90\substack{+0.04\\-0.05}$ & 199.3 & $-12.75\substack{+0.32\\-0.40}$ & $-12.50\substack{+0.20\\-0.20}$ \T \B \\ 
169195905 & 309.520673 & -40.186885 & 0.82054 & 21.349 & $10.42\substack{+0.15\\-0.20}$ & 181.3 & $-13.15\substack{+0.42\\-0.57}$ & $-12.69\substack{+0.23\\-0.25}$ \T \B \\ 
14 & 309.515660 & -40.134512 & 0.77392 & 21.462 & $10.55\substack{+0.08\\-0.12}$ & 14.9 & $-14.27\substack{+0.38\\-0.50}$ & $-13.87\substack{+0.21\\-0.22}$ \T \B \\ 
13 & 309.515880 & -40.135034 & 0.77492 & 20.761 & $11.32\substack{+0.05\\-0.06}$ & 14.3 & $-14.00\substack{+0.24\\-0.29}$ & $-13.94\substack{+0.20\\-0.20}$ \T \B \\ 
169194395 & 309.544629 & -40.165182 & 0.76076 & 22.386 & $10.04\substack{+0.32\\-0.18}$ & 136.6 & $-14.95\substack{+0.66\\-0.71}$ & $-14.22\substack{+0.30\\-0.24}$ \T \B \\ 
169194256 & 309.543818 & -40.163338 & 0.76067 & 22.326 & $9.84\substack{+0.19\\-0.27}$ & 130.2 & $-15.19\substack{+0.57\\-0.89}$ & $-14.29\substack{+0.24\\-0.29}$ \T \B \\ 
169190626 & 309.551365 & -40.113228 & 0.78249 & 22.480 & $9.87\substack{+0.21\\-0.30}$ & 139.5 & $-17.17\substack{+0.58\\-0.92}$ & $-16.28\substack{+0.25\\-0.30}$ \T \B \\ 
15 & 309.516878 & -40.135319 & 0.77619 & 22.884 & $10.18\substack{+0.13\\-0.20}$ & 16.3 & $-17.18\substack{+0.46\\-0.69}$ & $-16.58\substack{+0.22\\-0.25}$ \T \B \\ 
169192350 & 309.514393 & -40.137815 & -9.00000 & - & - & 8.8 & - & - \T \B \\
\hline
\end{longtable}
}

\newcommand{\tabphotozgroup}
{
\begin{table*}
	\caption{Properties of redMaPPer clusters in the field of view of \DEStwo. The columns show the cluster ID, cluster redshift, richness, velocity dispersion (rounded to the nearest 10\,\kms), the group centroid (in RA and Dec), projected distance of the centroid to the lens ($\Delta\theta$), and flexion shift $\log(\Delta_3 x (\rm{arcsec}))$. See \S \ref{sec:photogroup} for further discussion}.
	\label{tab:photogroups}
	\centering
	\begin{minipage}{\linewidth}
		\centering
	\begin{tabular}{lcccccc}
		\hline
\texttt{MEM\_MATCH\_ID} & $\bar{z}_{\rm group}$ & $\lambda$ & $\sigma_{\rm{group}}$ &  RA$_{\rm ctr}$, DEC$_{\rm ctr}$ &  $\Delta \theta$ & $\log_{10}(\Delta_3 x)$  \\
& & & (\kms) & (deg) &  (arcsec) & ($\log_{10}(\rm{arcsec})$) \T \B \\
\hline
62659 & $0.221 \pm 0.008$  & $5.1 \pm 1.7$ & $340\substack{+60\\-70}$ & 309.53913, -40.11584 & $ 107.9 $ & $-5.1^{+0.3}_{-0.4}$ \T \B \\ 
138669 & $0.405 \pm 0.017$ & $10.8 \pm 2.0$& $430\substack{+50\\-60}$ & 309.48707, -40.16080 & $108.6$  & $-6.0 \pm 0.2$ \T \B \\ 
\hline
\end{tabular}
\begin{flushleft}
\end{flushleft}
\end{minipage}
\end{table*}
}

\newcommand{\tabgroupmem}{
\begin{longtable}{cccccc}
	\caption{Properties of galaxies in each trial group. The columns display, in order, the coordinates (RA, DEC in degrees; ICRS), redshift $z$, DES $i$-band \texttt{MOF\_CM\_MAG\_CORRECTED} magnitudes (if available; assigned a value of -99 if not), ID, and whether that galaxy passes the iterative group membership algorithm described in \S\ref{sec:groups}. Spectroscopic redshift uncertainties are about 100~\kms, or 0.00033 in redshift. Galaxies with DES Y3 IDs have 9 digits, while galaxies with DES Y1 IDs have 10 digits. The lens galaxy is indicated as such, and galaxies with spectroscopic redshifts from MUSE have IDs that begin with "MUSE"} 
	\label{tab:group_mem} \\
\hline
RA & DEC & $z$ & $i$-band Mag & ID & Group Member? \\
\hline
\multicolumn{6}{c}{\DESone} \T \B \\ 
\multicolumn{6}{c}{Group 1} \T \B \\ 
62.030809 & -53.908820 & 0.26729 & 20.096 & 488070602 & False \\ 
 62.084697 & -53.867583 & 0.27148 & 21.675 & 489521853 & True \\ 
 62.061177 & -53.921532 & 0.27257 & 19.767 & 488072697 & True \\ 
 62.144676 & -53.934498 & 0.27176 & 18.315 & 488074578 & True \\ 
 62.084648 & -53.885204 & 0.27247 & 18.309 & 488070148 & True \\ 
 62.097716 & -53.879281 & 0.27278 & 22.768 & 489523301 & True \\ 
 62.046958 & -53.922500 & 0.27124 & 22.409 & 488073010 & True \\ 
 62.092960 & -53.861224 & 0.27153 & 20.306 & 489521068 & True \\ 
 62.054344 & -53.919213 & 0.27226 & 22.046 & 488072443 & True \\ 
 62.151510 & -53.927139 & 0.27214 & 22.822 & 488073904 & True \\ 
 62.146922 & -53.934938 & 0.27044 & 19.713 & 488074678 & False \\ 
 62.158897 & -53.933231 & 0.27187 & 22.261 & 488074972 & True \\ 
 62.083413 & -53.882657 & 0.27216 & 22.456 & 488069700 & True \\ 
\multicolumn{6}{c}{Group 2} \T \B \\ 
62.049904 & -53.911313 & 0.30581 & 21.203 & 488071089 & True \\ 
 62.045447 & -53.913359 & 0.30642 & 21.438 & 488071411 & True \\ 
 62.080583 & -53.922670 & 0.30741 & 21.184 & 488073021 & True \\ 
 62.093088 & -53.919040 & 0.30909 & 21.573 & 488072433 & True \\ 
 62.096336 & -53.944173 & 0.30792 & 19.259 & 488076699 & True \\ 
 62.020675 & -53.907526 & 0.30566 & 19.692 & 488067185 & True \\ 
 \multicolumn{6}{c}{Group 3} \T \B \\ 
 62.139479 & -53.928501 & 0.42858 & 19.429 & 488073735 & True \\ 
 62.025309 & -53.910519 & 0.42954 & 19.664 & 488070449 & False \\ 
 62.114845 & -53.953040 & 0.42826 & 21.346 & 488078736 & True \\ 
 62.108903 & -53.956880 & 0.42895 & 21.139 & 488079339 & True \\ 
 62.131777 & -53.926465 & 0.42904 & 21.665 & 488073762 & True \\ 
 62.148320 & -53.919275 & 0.42744 & 20.060 & 488072414 & True \\ 
 62.133767 & -53.937477 & 0.42807 & 20.447 & 488075668 & True \\ 
 62.135113 & -53.933925 & 0.42804 & 22.300 & 488075100 & True \\ 
 62.140970 & -53.931443 & 0.43054 & 20.683 & 488074608 & False \\ 
  \multicolumn{6}{c}{Group 4} \T \B \\ 
62.106161 & -53.923866 & 0.57035 & 20.933 & 488073290 & True \\ 
 62.023367 & -53.910705 & 0.57025 & 19.314 & 488066235 & True \\ 
 62.151601 & -53.871104 & 0.56693 & 22.810 & 404169588 & False \\ 
 62.047075 & -53.950191 & 0.56953 & 21.811 & 488078159 & True \\ 
 62.017973 & -53.905064 & 0.57028 & 21.306 & 488066250 & True \\ 
 62.019121 & -53.902892 & 0.57171 & 21.720 & 488067945 & True \\ 
  \multicolumn{6}{c}{Group 5} \T \B \\ 
62.037079 & -53.908383 & 0.59964 & 22.356 & 488070607 & True \\ 
 62.169886 & -53.887360 & 0.59607 & 20.158 & 488070133 & True \\ 
 62.160850 & -53.917944 & 0.59760 & 22.185 & 488072301 & True \\ 
 62.130404 & -53.867675 & 0.59397 & 23.104 & 404169402 & False \\ 
 62.035444 & -53.909470 & 0.60004 & 22.347 & 488070816 & True \\ 
 62.068708 & -53.913477 & 0.59843 & 21.581 & 488071428 & True \\ 
 62.071692 & -53.906245 & 0.59727 & -9999.000 & 488066584 & True \\ 
 62.134647 & -53.901754 & 0.59479 & 21.047 & 488067363 & False \\ 
 62.078380 & -53.881917 & 0.59451 & 19.918 & 489523481 & False \\ 
 62.072770 & -53.910296 & 0.59811 & 19.975 & 488070807 & True \\ 
 62.083912 & -53.903969 & 0.60048 & 22.304 & 488066462 & True \\ 
 62.067875 & -53.910488 & 0.59977 & 21.729 & 488070966 & True \\ 
 62.038672 & -53.869800 & 0.59969 & 22.534 & 489522088 & True \\ 
 62.023889 & -53.873587 & 0.59935 & 22.742 & 489522522 & True \\ 
 62.095575 & -53.898291 & 0.59441 & 21.867 & 488065185 & False \\ 
 62.090417 & -53.899889 & 0.59671 & 19.770 & Lens & True \\ 
 62.092164 & -53.902560 & 0.59221 & -99.000 & MUSE5 & False \\ 
 62.077898 & -53.903878 & 0.59797 & -99.000 & MUSE6 & True \\ 
 62.086607 & -53.902732 & 0.59822 & -99.000 & MUSE7 & True \\ 
 62.103108 & -53.897861 & 0.59832 & -99.000 & MUSE8 & True \\ 
 62.085632 & -53.891330 & 0.59906 & -99.000 & MUSE9 & True \\ 
 62.085862 & -53.910180 & 0.60029 & -99.000 & MUSE10 & True \\ 
  \multicolumn{6}{c}{Group 6} \T \B \\ 
62.148449 & -53.873747 & 0.75002 & 22.097 & 404169737 & True \\ 
 62.113750 & -53.913080 & 0.74981 & 22.287 & 488071415 & False \\ 
 62.147459 & -53.869008 & 0.75141 & 21.712 & 404169401 & True \\ 
 62.156908 & -53.860476 & 0.75027 & 22.948 & 404169065 & False \\ 
 62.146052 & -53.874054 & 0.75095 & 22.586 & 404169758 & True \\ 
 62.108813 & -53.919489 & 0.75070 & 22.550 & 488072531 & False \\ 
 
  \multicolumn{6}{c}{Group 7} \T \B \\ 
62.106067 & -53.909453 & 0.77562 & 21.908 & 488070825 & False \\ 
 62.158048 & -53.912350 & 0.76818 & 18.274 & 488070947 & False \\ 
 62.125315 & -53.873569 & 0.76243 & 22.755 & 404169734 & False \\ 
 62.073251 & -53.920150 & 0.77290 & 22.129 & 488072604 & False \\ 
 62.090965 & -53.901634 & 0.76866 & 20.096 & 488068102 & True \\ 
 62.050530 & -53.926881 & 0.75824 & 22.336 & 488073766 & False \\ 
 62.090541 & -53.904725 & 0.77568 & 22.925 & 488066886 & False \\ 
 62.030863 & -53.912062 & 0.77531 & 22.736 & 488071245 & False \\ 
 62.019165 & -53.913520 & 0.77498 & 22.912 & 488071425 & False \\ 
 62.090243 & -53.903609 & 0.77069 & 22.015 & 488066144 & False \\ 
 62.073661 & -53.889176 & 0.76790 & -99.000 & 3070264072.0 & True \\ 
 62.088612 & -53.911555 & 0.76379 & -99.000 & MUSE13 & False \\ 
 62.099785 & -53.908977 & 0.76669 & -99.000 & MUSE14 & True \\ 
 62.080992 & -53.904565 & 0.76785 & -99.000 & MUSE15 & True \\ 
 62.075319 & -53.889095 & 0.76863 & -99.000 & MUSE16 & True \\ 
  \multicolumn{6}{c}{Group 8} \T \B \\ 
62.142829 & -53.870321 & 0.80082 & 21.848 & 404169534 & True \\ 
 62.107604 & -53.941787 & 0.79782 & 21.477 & 488076578 & True \\ 
 62.171213 & -53.906968 & 0.79737 & 22.595 & 488070460 & True \\ 
 62.156567 & -53.871330 & 0.80070 & 21.188 & 404169584 & True \\ 
 62.112378 & -53.871809 & 0.80252 & 21.675 & 404169634 & True \\ 
 62.171213 & -53.906968 & 0.79720 & 22.595 & 488070460 & True \\ 
  \multicolumn{6}{c}{Group 9} \T \B \\ 
62.036870 & -53.925643 & 0.91818 & 21.953 & 488073582 & True \\ 
 62.043694 & -53.923652 & 0.91456 & 22.098 & 488073160 & True \\ 
 62.025664 & -53.892284 & 0.91979 & 21.768 & 488067796 & True \\ 
 62.030458 & -53.871796 & 0.91911 & 22.148 & 489522305 & True \\ 
 62.019738 & -53.894319 & 0.91905 & 22.531 & 488068835 & True \\ 
 62.067931 & -53.853616 & 0.91854 & 22.457 & 489520262 & True \\ 
 62.075161 & -53.892278 & 0.91583 & -99.000 & 3070263850.0 & True \\ 
 62.076105 & -53.885866 & 0.91540 & -99.000 & 3070265512.0 & True \\ 
 62.088268 & -53.890642 & 0.91504 & -99.000 & MUSE19 & True \\  
  \multicolumn{6}{c}{Group 10} \T \B \\ 
62.161866 & -53.918519 & 1.25018 & 22.937 & 488072408 & True \\ 
 62.160696 & -53.920293 & 1.25468 & 22.908 & 488072700 & True \\ 
 62.161986 & -53.918003 & 1.25287 & 22.603 & 488072310 & True \\ 
 62.098868 & -53.909951 & 1.25002 & -99.000 & MUSE25 & True \\ 
 62.095831 & -53.909607 & 1.25062 & -99.000 & MUSE26 & True \\ 
 62.084658 & -53.911326 & 1.25152 & -99.000 & MUSE27 & True \\ 
 62.107634 & -53.905654 & 1.25400 & -99.000 & MUSE28 & True \\ 
\hline
\multicolumn{6}{c}{\DEStwo} \T \B \\ 
\multicolumn{6}{c}{Group 1} \T \B \\ 
309.492815 & -40.140131 & 0.23003 & 21.389 & 169192596 & True \\
309.539963 & -40.103707 & 0.22827 & 17.848 & 169189459 & True \\
309.539131 & -40.115839 & 0.22917 & 18.213 & 169190452 & True \\
309.535586 & -40.122881 & 0.22900 & 20.025 & 169191228 & True \\
309.540911 & -40.125437 & 0.22716 & 21.577 & 169191437 & True \\
309.537431 & -40.135642 & 0.22864 & 21.077 & 169192249 & True \\
309.511379 & -40.137024 & 0.22829 & $-$    & 169191076/Lens    & True \\ 
\multicolumn{6}{c}{Group 2} \T \B \\ 
309.521701 & -40.129094 & 0.34157 & 22.937 & 169191786 & True \\
309.517763 & -40.130806 & 0.34445 & 22.210 & 169191897 & True \\
309.564077 & -40.146099 & 0.34071 & 22.869 & 169193078 & True \\
309.561418 & -40.148516 & 0.34018 & 22.225 & 169193237 & True \\
309.557474 & -40.152652 & 0.34079 & 19.475 & 169193432 & True \\
309.496715 & -40.122896 & 0.34271 & 22.345 & 169191326 & True \\
309.568673 & -40.153285 & 0.34218 & 21.564 & 169193555 & True \\
309.562909 & -40.162398 & 0.34506 & 19.299 & 169194065 & True \\
\hline

\end{longtable}
}
\newcommand{\figdesonegroupsa}{
\begin{figure*}
	\centering
	\subfigure{\label{fig:0408_0}\includegraphics[width=0.3\textwidth]{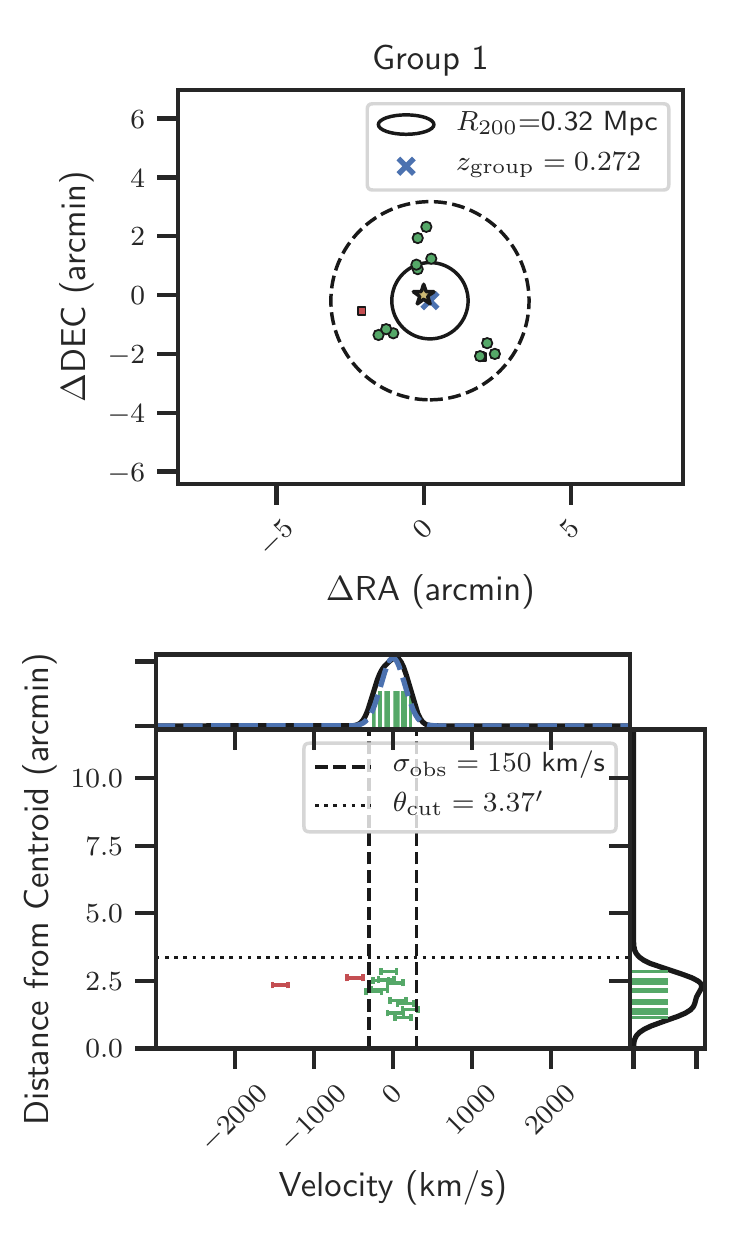}}
	\subfigure{\label{fig:0408_1}\includegraphics[width=0.3\textwidth]{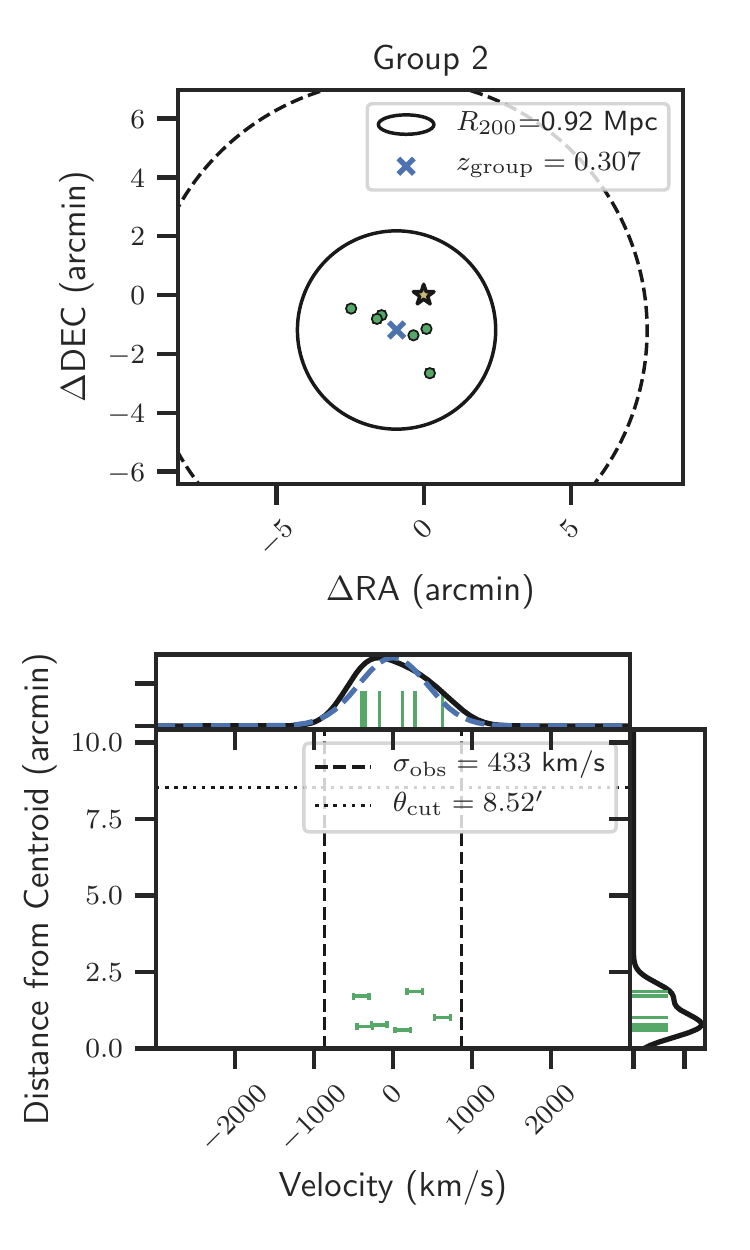}}
	\subfigure{\label{fig:0408_2}\includegraphics[width=0.3\textwidth]{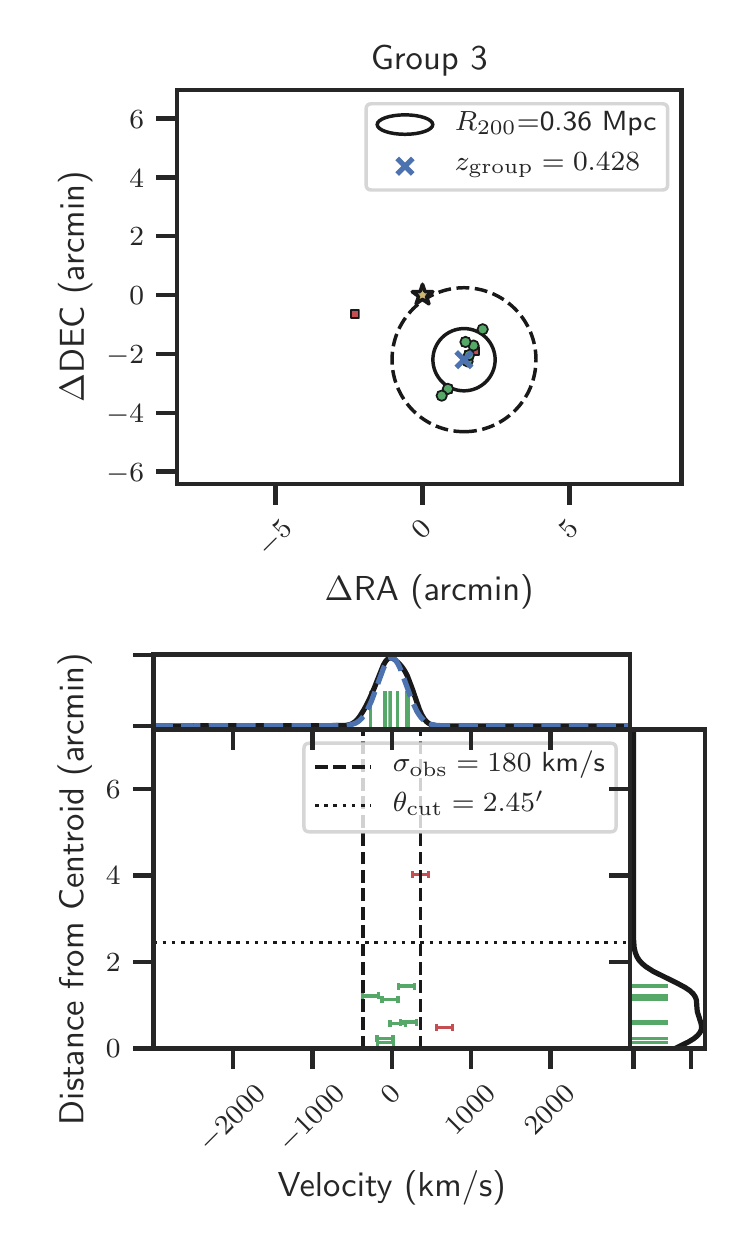}}
	\subfigure{\label{fig:0408_3}\includegraphics[width=0.3\textwidth]{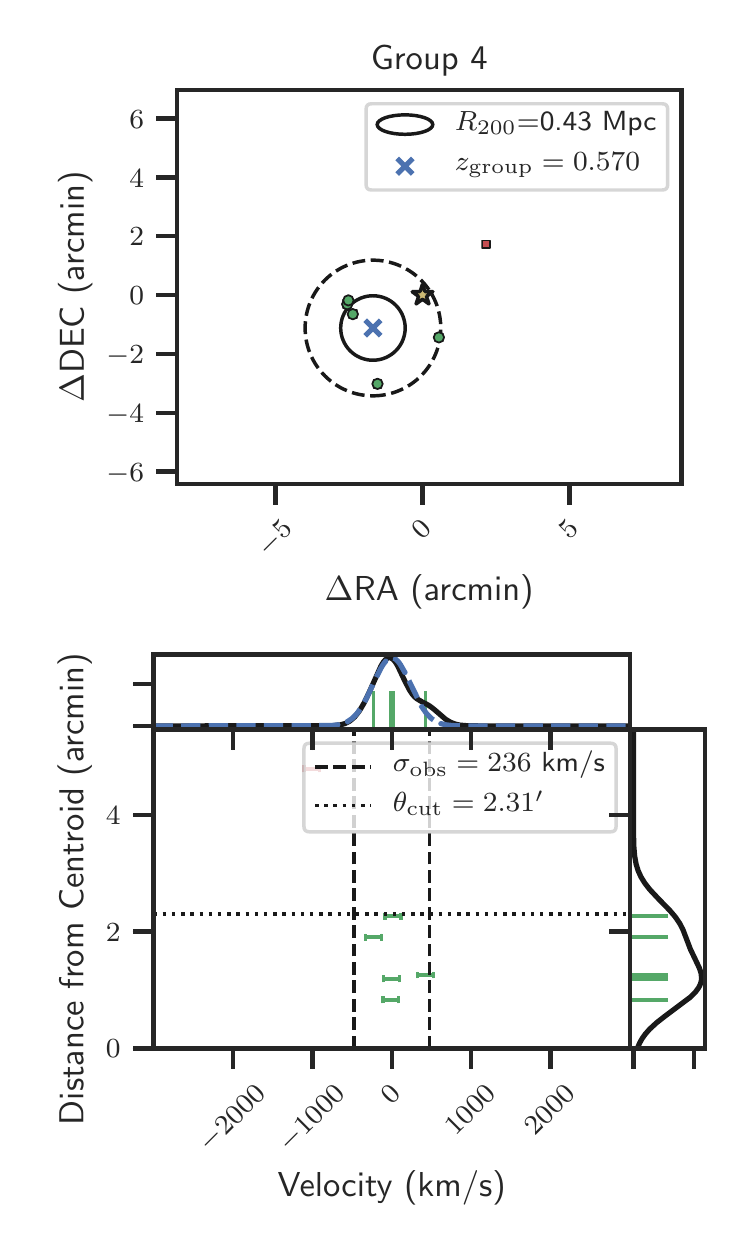}}
	\subfigure{\label{fig:0408_4}\includegraphics[width=0.3\textwidth]{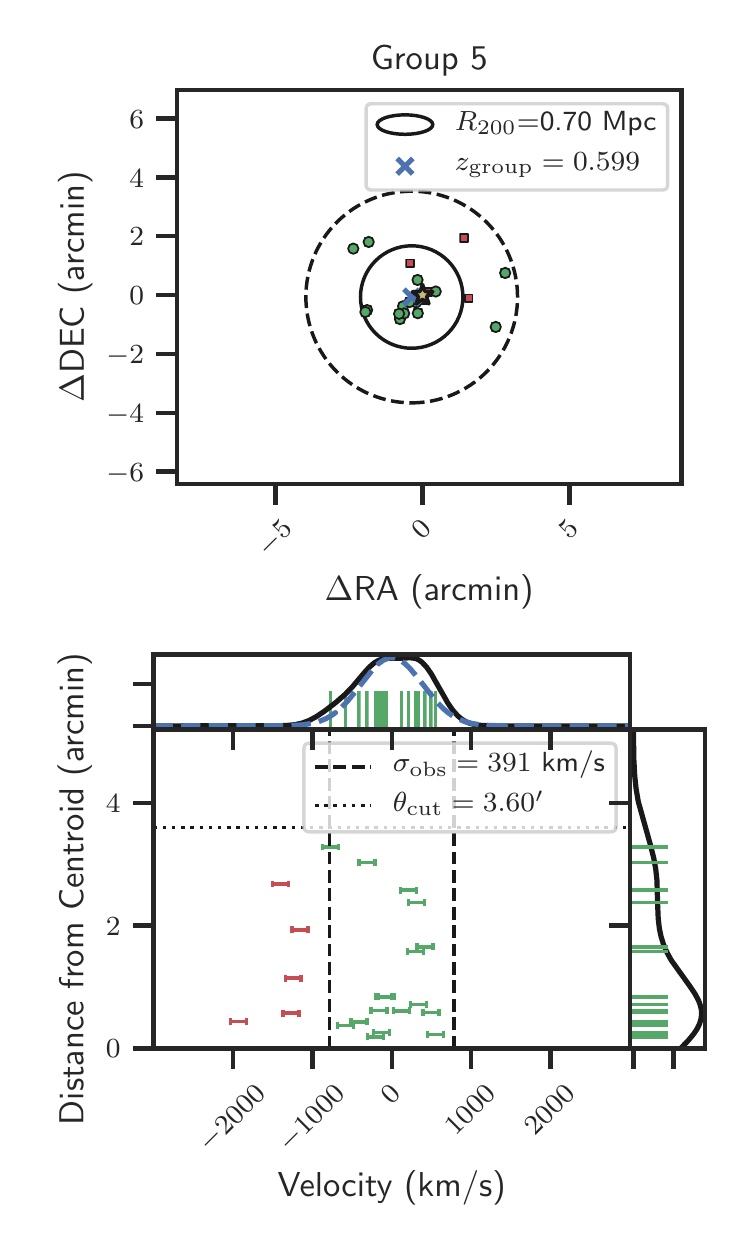}}
	\subfigure{\label{fig:0408_5}\includegraphics[width=0.3\textwidth]{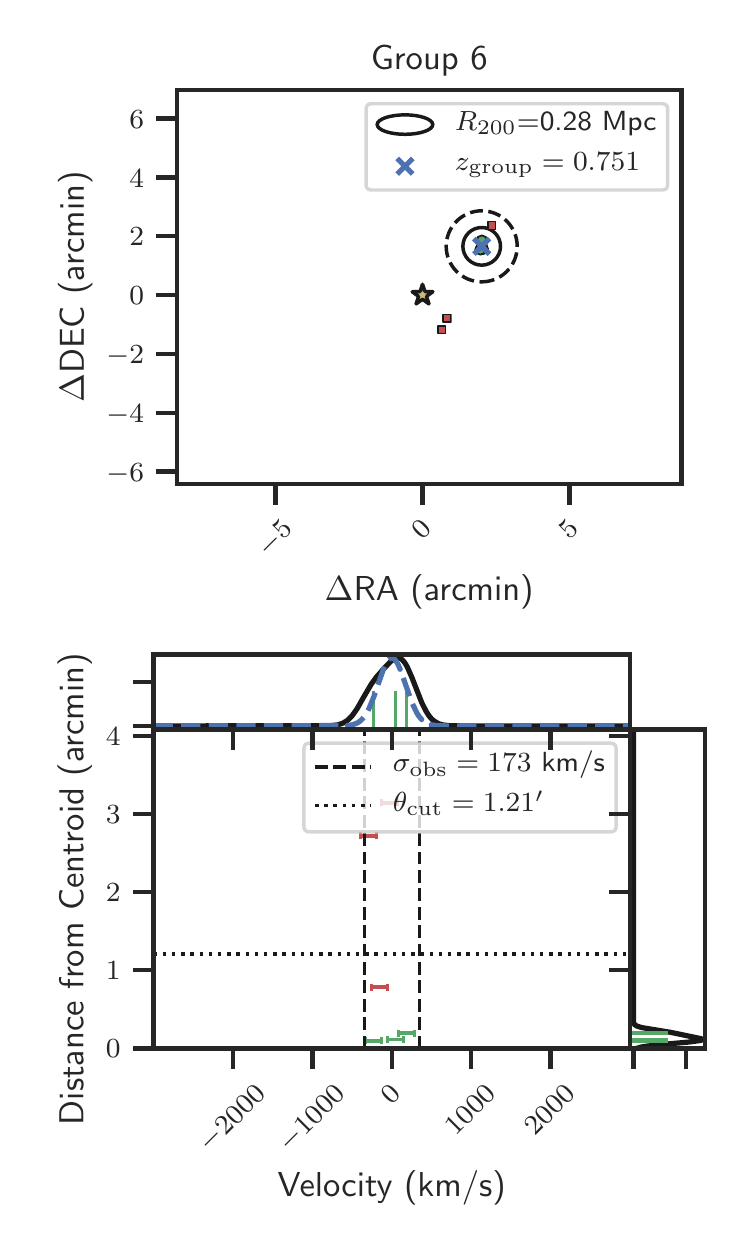}}
	\captcont{\label{fig:0408groups} Galaxy groups identified in the spectroscopic sample of galaxies in the field of view of \DESone. For each group, the first plot (above) shows the positions of the candidate member galaxies associated with that group relative to the lens galaxy, with rejected group members represented as red squares and accepted group members represented as green circles. The lens galaxy \DESone (star) and group centroid (cross) are also displayed. The $R_{200}$ radius of the group is represented by a solid line, while the dashed circle represents the angular separation cut of the group-finding algorithm in its final iteration. The second plot (below) shows the observer-frame velocity of individual member galaxies relative to the group centroid as a function of that galaxy's angular distance from the centroid. Galaxies that passed the iterative algorithm described in \S \ref{sec:groups} are shown in green, while trial galaxy members that were cut through the algorithm are shown in red.  Horizontal error bars represent the measurement error for each galaxy (see \S \ref{sec:redshifts}). The final observer-frame velocity dispersion and angular separation cuts from the group-finding algorithm are presented as dashed and dotted lines respectively. We also show 1-D histograms and rug plots of the velocity and distance distributions of the member galaxies. The 1-D histograms are produced using a Kernel Density Estimate (KDE) with a bandwidth determined using Scott's Rule. In the 1D velocity histogram, the dashed blue line shows a gaussian with width equal to the observer-frame velocity dispersion of the group.}
\end{figure*}}

\newcommand{\figdesonegroupsb}{
\begin{figure*}
	\subfigure{\label{fig:0408_6}\includegraphics[width=0.33\textwidth]{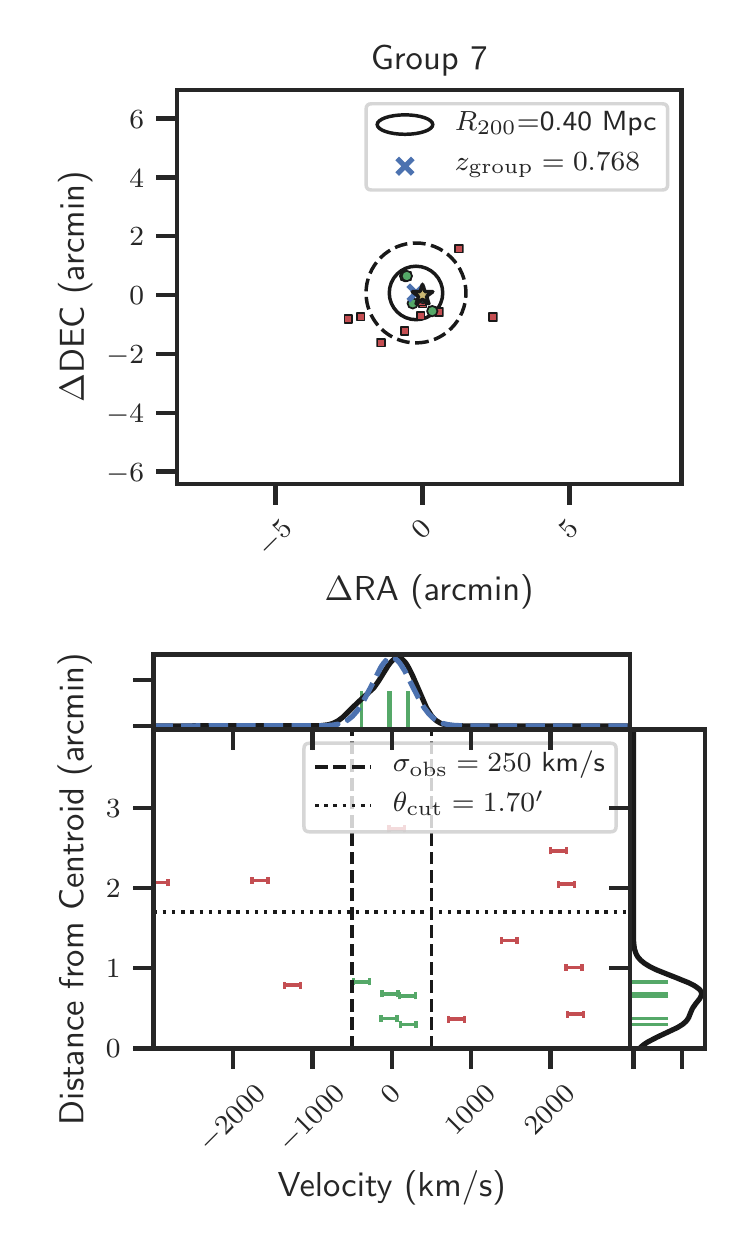}}
	\subfigure{\label{fig:0408_7}\includegraphics[width=0.33\textwidth]{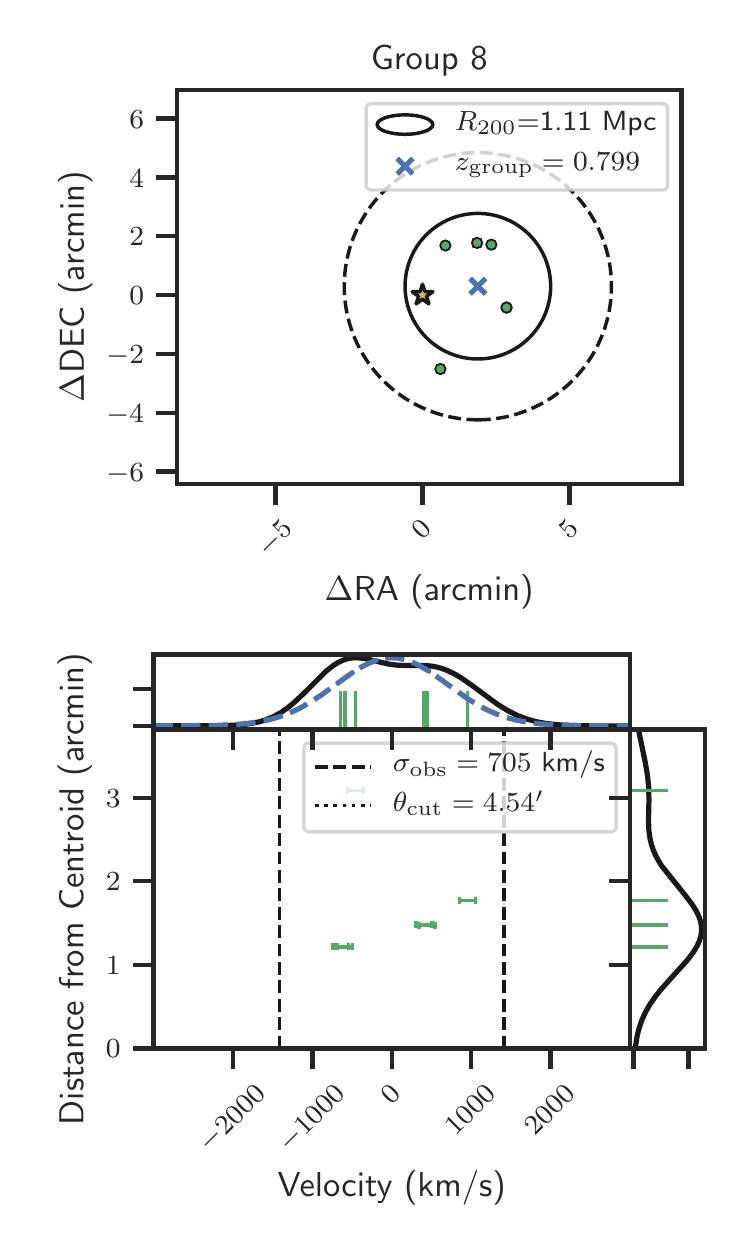}} \\
	\subfigure{\label{fig:0408_8}\includegraphics[width=0.33\textwidth]{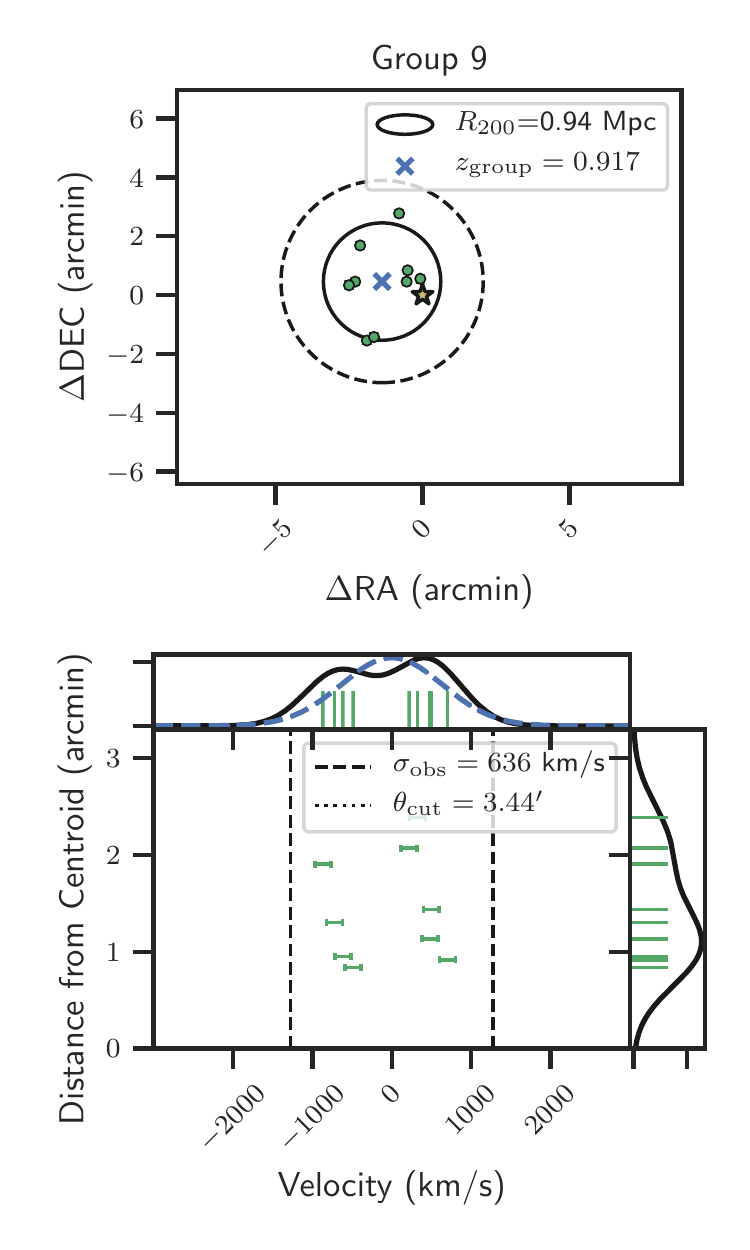}}
	\subfigure{\label{fig:0408_9}\includegraphics[width=0.33\textwidth]{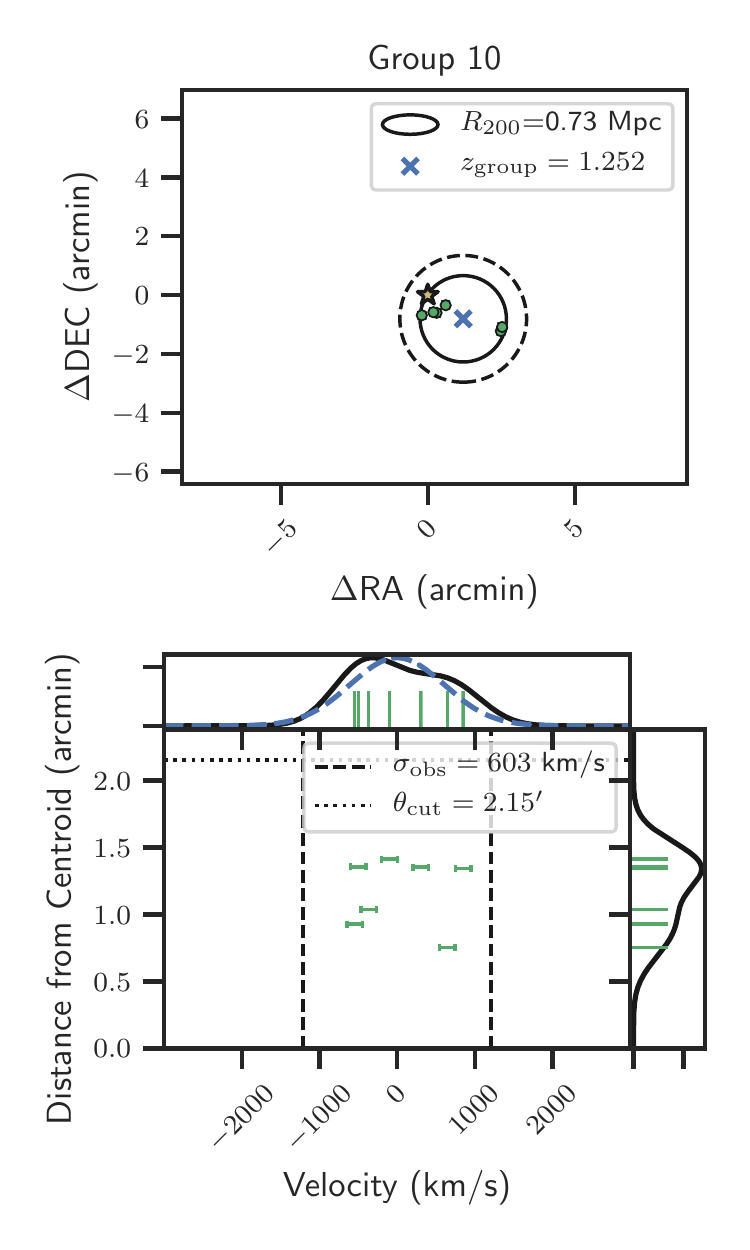}}
	\caption{(continued)}
\end{figure*}}

\newcommand{\figdestwogroups}{
\begin{figure*}
	\subfigure{\label{fig:2038_1}\includegraphics[width=0.33\textwidth]{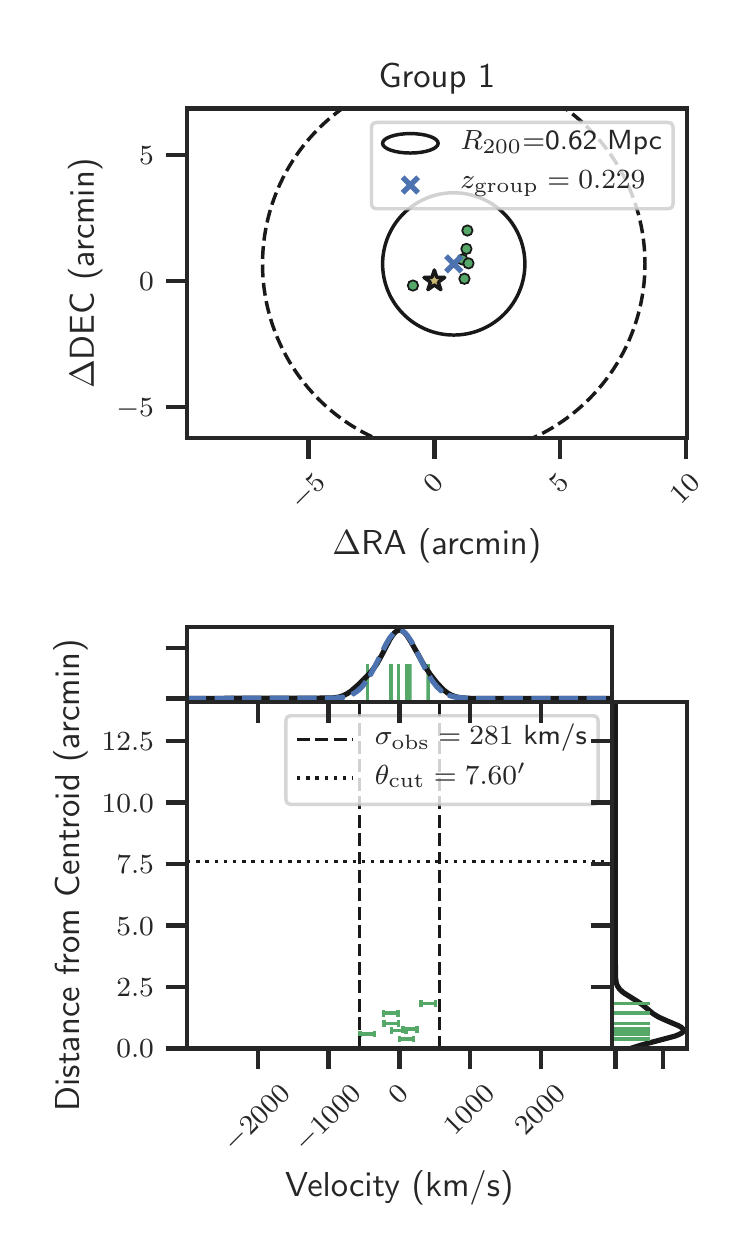}}
	\subfigure{\label{fig:2038_2}\includegraphics[width=0.33\textwidth]{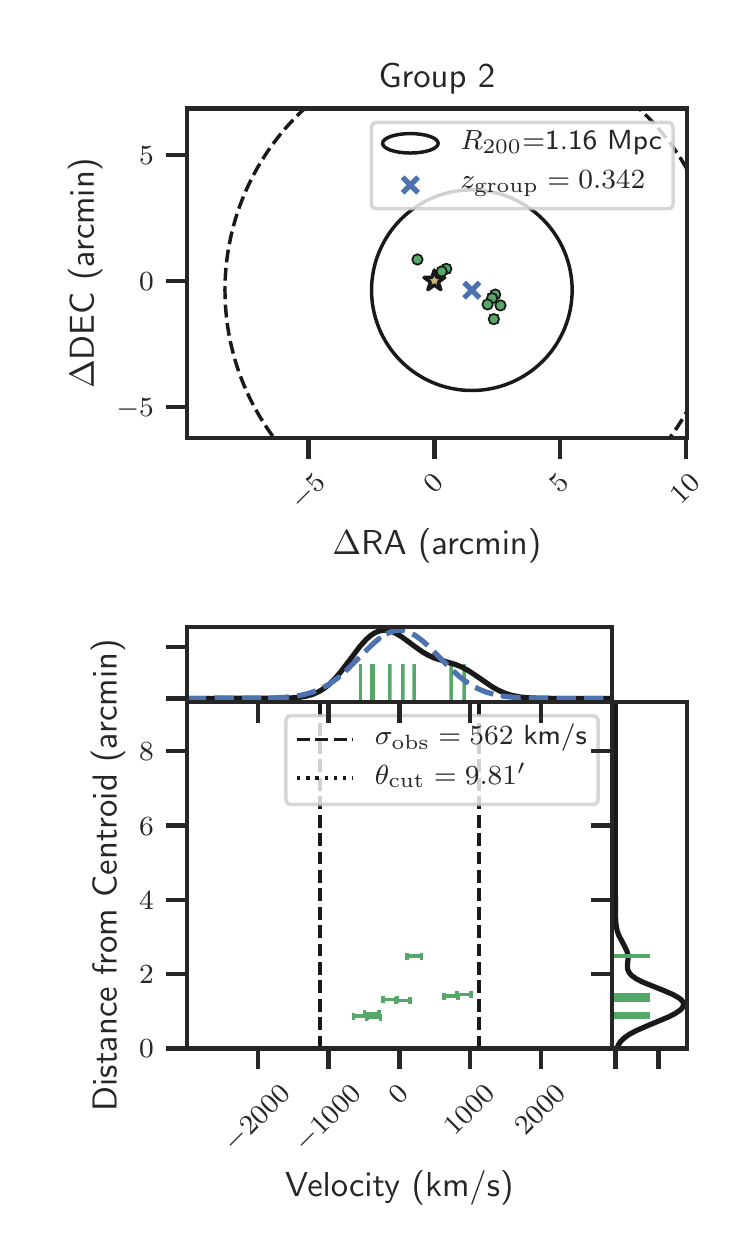}}
	\caption{\label{fig:2038groups}Galaxy groups identified using the spectroscopic sample of galaxies in the environment of \DEStwo. See  Figure \ref{fig:0408groups} for an explanation of the figures.}
\end{figure*}}

\newcommand{\figflexionhistogram}{
\begin{figure}
	\subfigure{\includegraphics[width=0.45\textwidth]{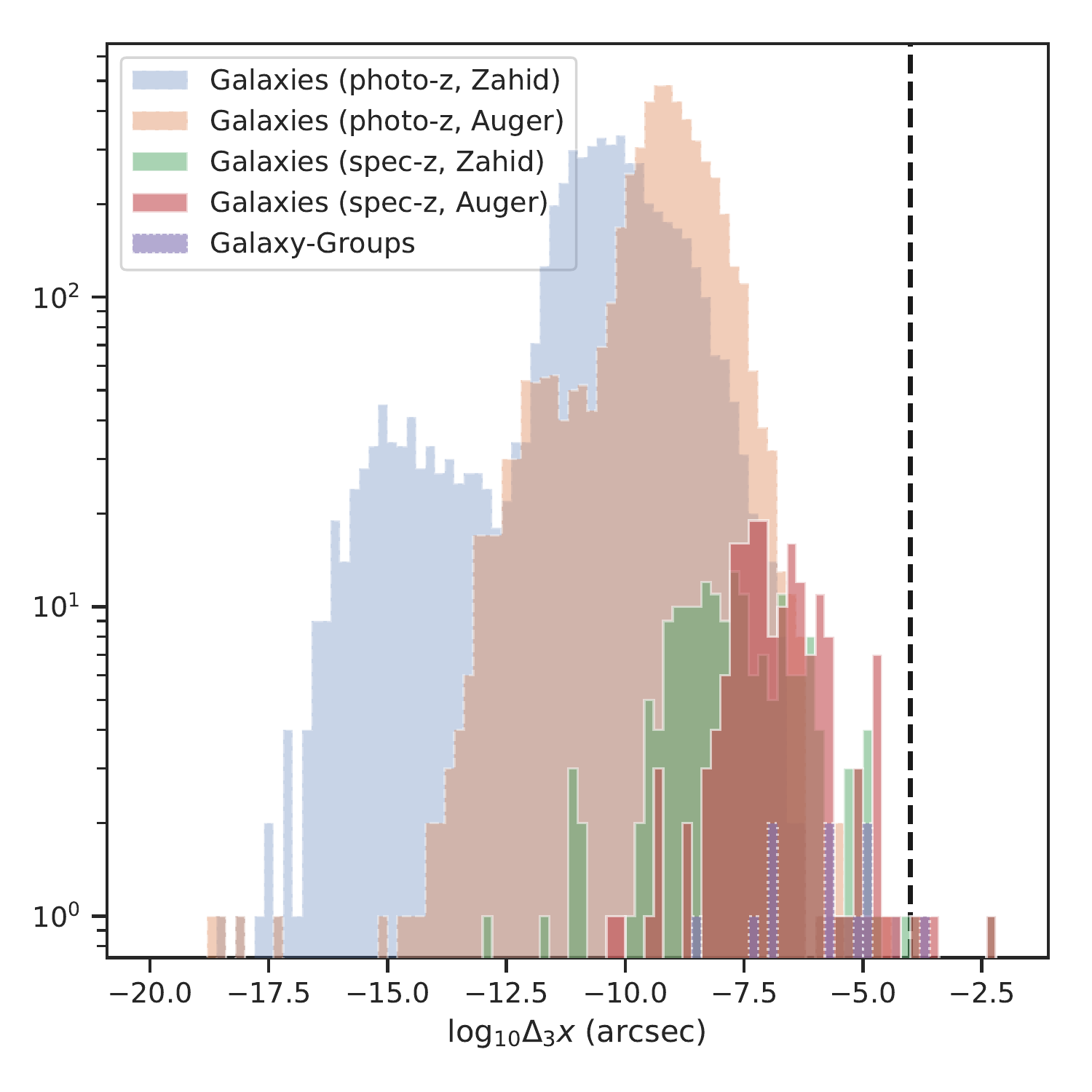}}
	\caption{\label{fig:0408_flexion} Flexion shift histograms for galaxies and galaxy groups in the environment of \DESone.}
	\subfigure{\includegraphics[width=0.45\textwidth]{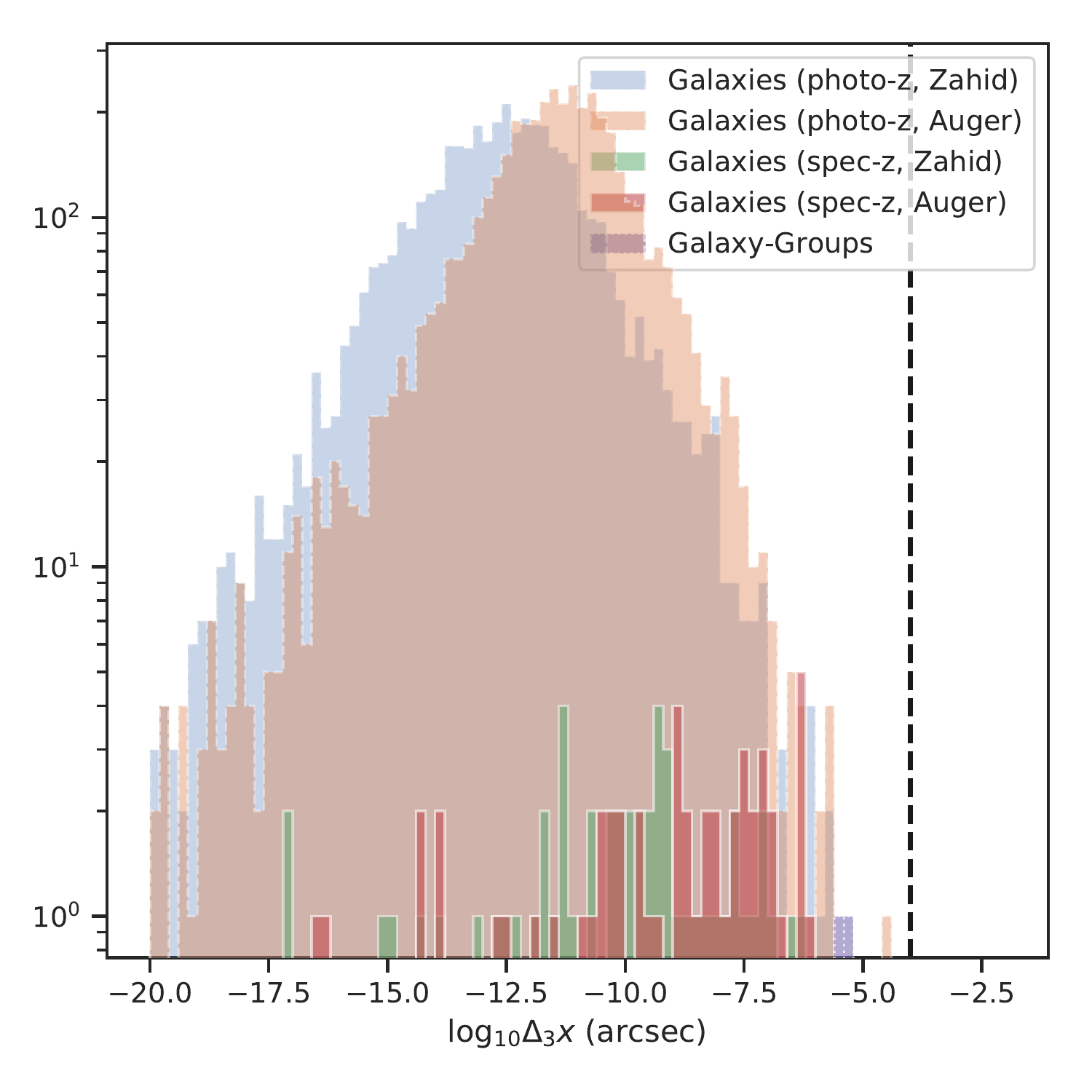}}
	\caption{\label{fig:2038_flexion} Flexion shift histogram for galaxies and galaxy groups in the environment of \DEStwo.}
\end{figure}}

\title[Environment of \DESone\ and \DEStwo]{STRIDES: Spectroscopic and photometric characterization of the environment and effects of mass along the line of sight to the gravitational lenses \DESone and \DEStwo}

\author[E.J. Buckley-Geer et al.]{
E.~J.~Buckley-Geer,$^{1}$\thanks{E-mail: buckley@fnal.gov}
H.~Lin,$^{1}$
C.~E.~Rusu,$^{2}$
J. Poh,$^{3,4}$
A.~Palmese,$^{1}$
A. Agnello,$^{5}$
\newauthor
L. Christensen,$^{5}$
J. Frieman,$^{1,3,4}$
A.~J.~Shajib,$^{6}$
T. Treu,$^{6}$
T.~Collett,$^{7}$
S.~Birrer,$^{6,8}$
\newauthor
T. Anguita,$^{9,10}$
C.~D.~Fassnacht,$^{11}$
G.~Meylan,$^{12}$
S.~Mukherjee,$^{13}$
K.~C.~Wong,$^{14}$
\newauthor
M.~Aguena,$^{15,16}$
S.~Allam,$^{1}$
S.~Avila,$^{17}$
E.~Bertin,$^{18,19}$
S.~Bhargava,$^{20}$
D.~Brooks,$^{21}$
\newauthor
A.~Carnero~Rosell,$^{22}$
M.~Carrasco~Kind,$^{23,24}$
J.~Carretero,$^{25}$
F.~J.~Castander,$^{26,27}$
\newauthor
M.~Costanzi,$^{28,29}$
L.~N.~da Costa,$^{16,30}$
J.~De~Vicente,$^{22}$
S.~Desai,$^{31}$
H.~T.~Diehl,$^{1}$
P.~Doel,$^{21}$
\newauthor
T.~F.~Eifler,$^{32,33}$
S.~Everett,$^{34}$
B.~Flaugher,$^{1}$
P.~Fosalba,$^{26,27}$
J.~Garc\'ia-Bellido,$^{17}$
\newauthor
E.~Gaztanaga,$^{26,27}$
D.~Gruen,$^{35,36,37}$
R.~A.~Gruendl,$^{23,24}$
J.~Gschwend,$^{16,30}$
G.~Gutierrez,$^{1}$
\newauthor
S.~R.~Hinton,$^{38}$
K.~Honscheid,$^{39,40}$
D.~J.~James,$^{41}$
K.~Kuehn,$^{42,43}$
N.~Kuropatkin,$^{1}$
\newauthor
M.~A.~G.~Maia,$^{16,30}$
J.~L.~Marshall,$^{44}$
P.~Melchior,$^{45}$
F.~Menanteau,$^{23,24}$
R.~Miquel,$^{46,25}$
\newauthor
R.~L.~C.~Ogando,$^{16,30}$
F.~Paz-Chinch\'{o}n,$^{23,24}$
A.~A.~Plazas,$^{45}$
E.~Sanchez,$^{22}$
V.~Scarpine,$^{1}$
\newauthor
M.~Schubnell,$^{47}$
S.~Serrano,$^{26,27}$
I.~Sevilla-Noarbe,$^{22}$
M.~Smith,$^{48}$
M.~Soares-Santos,$^{49}$
\newauthor
E.~Suchyta,$^{50}$
M.~E.~C.~Swanson,$^{24}$
G.~Tarle,$^{47}$
D.~L.~Tucker,$^{1}$
and T.~N.~Varga$^{51,52}$
\newauthor
(The DES Collaboration)\\
Affiliations appear at the end of the paper
}

\date{Accepted XXX. Received YYY; in original form ZZZ}

\pubyear{2020}

\begin{document}
\label{firstpage}
\pagerange{\pageref{firstpage}--\pageref{lastpage}}
\maketitle

\begin{abstract}
{In time-delay cosmography, three of the key ingredients are 1) determining the velocity dispersion of the lensing galaxy, 2) identifying galaxies and groups along the line of sight with sufficient proximity and mass to be included in the mass model,
and 3) estimating the external convergence $\kappa_\mathrm{ext}$ from less massive structures that are not included in the mass model.
We present results on all three of these ingredients for two time-delay lensed quasar systems, \DESone and \DEStwo. We use the Gemini, Magellan and VLT telescopes to obtain spectra to both measure the stellar velocity dispersions of the main lensing galaxies and to identify the line-of-sight galaxies in these systems. Next, we identify 10 groups in \DESone and 2 groups in \DEStwo using a group-finding algorithm. We then identify the most significant galaxy and galaxy-group perturbers using the "flexion shift" criterion. We determine the probability distribution function of the external convergence $\kappa_\mathrm{ext}$ for both of these systems based on our spectroscopy and on the DES-only multiband wide-field observations. Using weighted galaxy counts, calibrated based on the Millennium Simulation, we find that \DESone is located in a significantly underdense environment, leading to a tight (width $\sim3\%$), negative-value $\kappa_\mathrm{ext}$ distribution. On the other hand, \DEStwo is located in an environment of close to unit density, and its low source redshift results in a much tighter $\kappa_\mathrm{ext}$ of $\sim1\%$, as long as no external shear constraints are imposed. 
}

\end{abstract}

\begin{keywords}
gravitational lensing: strong -- quasars: individual: \DESone, \DEStwo\ -- galaxies: groups: general
\end{keywords}

\AddToShipoutPictureBG*{%
	\AtPageUpperLeft{%
		\hspace{0.75\paperwidth}%
		\raisebox{-3.5\baselineskip}{%
			\makebox[0pt][l]{\textnormal{DES-2019-0471}}
}}}%

\AddToShipoutPictureBG*{%
	\AtPageUpperLeft{%
		\hspace{0.75\paperwidth}%
		\raisebox{-4.5\baselineskip}{%
			\makebox[0pt][l]{\textnormal{FERMILAB-PUB-20-082-AE-SCD}}
}}}%



\section{Introduction} 
\label{sec:intro}
When a source with a time-varying luminosity such as a quasar or a supernova undergoes strong gravitational lensing, the light reaching the observer from the  multiple images takes different paths and hence different travel times. It was noted by \citet{Refsdal1964} that these time-delays between the images can be used to measure cosmological distances and the Hubble constant $H_0$. 
In recent years, the $H_0$ Lenses in COSMOGRAIL’s Wellspring (H0LiCOW) collaboration has been leading an active time-delay cosmography program to measure $H_0$ using lensed quasars, see \citet{Suyu2017a,Wong2019} and references therein.
For a recent review of the field of time-delay cosmography see \citet{Treu2016a}. The significant improvements in the uncertainty in the measurements of $H_0$ in the last two decades have come from the understanding of the key ingredients required to achieve an accurate measurement.  In particular, three of these ingredients are: 1) determining the velocity dispersion of the lensing galaxy, 2) identifying galaxies and groups along the line of sight that are close enough to the lens and massive enough to be included in the mass model, and 3) estimating the external convergence $\kappa_\mathrm{ext}$ due to less massive structures that are not included explicitly in the mass model.  In this work, we present results on these three ingredients for two time-delay lensed quasar systems, \DESone \citep[source redshift $z_{\rm s}=\,$\zsone, main deflector redshift $z_{\rm d}=\,$\zdone,][]{Lin2017a} and \DEStwo \citep[$z_{\rm s} =\,$\zstwo, $z_{\rm d}=\,$\zdtwo,][]{Agnello2018}, as part of the STRong Lensing Insights into the Dark Energy Survey (\STRIDES) campaign \citep{Treua}, an external collaboration of the Dark Energy Survey. 

It has been known for sometime that including stellar kinematics of the lensing galaxy allows one to break the degeneracies inherent in the mass profile of the lens \citep{TreuKoopmans2002}. To obtain the stellar velocity dispersion of the lens we took spectroscopic observations of the main lensing galaxy in \DESone and \DEStwo.

If the perturbers along the line of sight are not explicitly accounted for in the lens modeling, these perturbations can result in systematic errors of order a few percent in the inferred value of $H_0$. To reduce such systematics, we identify galaxies and galaxy groups in the fields of \DESone and \DEStwo that may significantly impact the lensing potential of the system. These galaxies and galaxy groups will be included in the lens models for both \DESone \citep[][Yidirim et al. in prep]{Shajib2019_H0} and \DEStwo (Wong et al. in prep). 

To identify the most significant perturbing galaxies and galaxy groups, we use the ``\textit{flexion shift}'' diagnostic proposed by \citet{McCully2014, McCully2017}, which has also been used in the line-of-sight analysis of the \HOLI lenses \HEofor \citep{Sluse2017} and \WFItwenty  \citep{Sluse2019}. This diagnostic estimates the difference in lensed image positions caused by the leading order non-tidal (i.e. third-order) perturbation produced by a nearby galaxy or galaxy group. \citet{McCully2017} showed that by explicitly modeling perturbers with flexion shifts larger than the conservative limit of $\Delta_3 x > 10^{-4}$~\arcsec, we can constrain the bias on $H_0$ due to this uncertainty to the percent level.

In addition, we determine for both systems the probability distribution function of the external convergence $\kappa_\mathrm{ext}$ due to less massive structures, which do not need to be explicitly incorporated in the mass modeling, but nonetheless contribute a uniform mass-sheet. Indeed, if unaccounted for, this quantity would bias $H_0$ such that $H_0=H_0^\mathrm{model}(1-\kappa_\mathrm{ext})$ \citep[e.g.,][]{Suyu2010DISSECTING}, for $H_0^\mathrm{model}$ obtained from lens modeling. For the first time we determine $\kappa_\mathrm{ext}$ based on multi-band $(griz)$, wide-field images obtained from the Dark Energy Survey\footnote{\protect\url{https://www.darkenergysurvey.org}} (DES) data. Following previous work \citep{Fassnacht2011,Greene2013,Rusu2017,Rusu2019,Birrer2019,Chen2019}, we measure the under/overdensity of the line of sight towards both lens systems relative to the ``average'' line of sight density throughout the Universe, obtained from the cosmologically representative sample provided by the DES in the form of a set of control fields. Aiming to constrain $\kappa_\mathrm{ext}$ as tightly as possible, as well as to study the effect of different analysis choices, we determine under/overdensities using various combinations of weighting schemes for the galaxy counts, such as the radial distance to the lens/field center and the redshift. Finally, we convert the measured under/overdensities into a $\kappa_\mathrm{ext}$ distribution, using ray-tracing through the Millennium Simulation \citep[MS;][]{Springel2005}. We explore several aperture sizes, and two different photometric redshift algorithms, which we further test through simulations. 

We perform a spectroscopic survey to obtain redshifts of galaxies in the fields of \DESone and \DEStwo. This redshift data is used to identify galaxy groups located in the environment or along the line of sight to these strong lensing systems as well as in the calculation of the under/overdensity of the line of sight towards both lens systems. 

The structure of this paper is as follows. In Section~\ref{sec:data} we describe our photometric and spectroscopic data, and in Section~\ref{sec:redshifts} we present our techniques to measure redshifts and stellar masses. In Section~\ref{sec:velocity_dispersion} we derive the stellar velocity dispersions for the main lensing galaxies in the two systems, and in Section~\ref{sec:galaxy_groups} we describe our technique to identify galaxy groups. In Section~\ref{sec:contribution_environment} we identify the structures which can potentially affect the modeling of the lensing systems. 
In Section~\ref{sec:number_counts} we present our measurement of the relative weighted galaxy count ratios for \DESone and \DEStwo, including accounting for relevant errors. In Section~\ref{sec:kappa_ext} we use ray-tracing through the MS in order to obtain $P(\kappa_\mathrm{ext}$) for the measured ratios, and present our tests for robustness. Finally, we conclude in Section~\ref{sec:conclude}. 

The current work represents one of a series of papers 
from the STRIDES collaboration, which together aim to obtain an accurate and precise estimate of $H_0$ with a blinded approach, from a comprehensive modelling of \DESone and \DEStwo. In particular, lens modeling is performed by two independent modeling teams \cite[][as well as Yildirim et al. in prep, Wong et al. in prep]{Shajib2019_H0}, both of which make use of the stellar velocity dispersion, environment and line-of-sight constraints derived in the present work.
Throughout this paper, we assume a flat $\Lambda$CDM cosmology with $H_0 = 70 $ \kmsMpc, $\Omega_m = 0.3$ for convenience when estimating physical individual galaxy and galaxy group properties (\S\ref{sec:galaxy_groups}-\S\ref{sec:contribution_environment}). However, in the latter part of the analysis, where we determine the under/overdensities of the fields of the lenses and then derive $\kappa_\mathrm{ext}$ distributions using ray-tracing through the MS, we adopt the MS cosmology, $\Omega_{m} = 0.25$, $\Omega_{\lambda} = 0.75$, $h = 0.73$, $\sigma_{8} = 0.9$ for consistency. This is not expected to have a significant effect on the inference of $H_0$ (see Section~\ref{sec:kappa_ext}). We present all magnitudes in the AB system. We define all standard deviations as the semi-difference between the 84th and 16th percentiles.

\section{Data} 
\label{sec:data}
The Dark Energy Survey (DES) is a deep sky survey that was carried out using the Dark Energy Camera \citep[DECam,][]{Flaugher2015} located on the Blanco 4m telescope at the Cerro Tololo Inter-American Observatory in the Chilean Andes. The survey ran from 2013-2019 \citep{Diehl2019} and covered $\sim 5100$ sq. degs of the southern sky in five optical filters ($grizY$). The DES data are processed by the DES Data Management team \citep[DESDM,][]{desdm2018} to produce annual data releases that consist of coadded images and object catalogs. We have used two data sets for the work described here, the first year of DES data which is referred to as DES Year 1 (Y1) and the first three years known as DES Year 3 (Y3). The median single epoch PSF FWHM in the $i$-band is 0.88\arcsec and the coadd magnitude limit in the $i$-band is 23.44. More details of the survey data can be found in \citet{Abbott2018}.

\subsection{Spectroscopic Observations}
\label{subsec:spectroobs}
Spectroscopic observations were carried out using three instruments: (1) the Gemini Multi-Object Spectrograph (GMOS-S) on the Gemini South telescope; (2) the Low Dispersion Survey Spectrograph (LDSS-3) on the Magellan Clay telescope; and (3) the Multi Unit Spectroscopic Explorer (MUSE) on the European Southern Observatory (ESO) Very Large Telescope (VLT) Unit Telescope 4 (UT4). 
Table~\ref{tab:spectro_observations} summarizes details of the spectroscopic data taken using these three telescope+instrument setups.

The GMOS-S observations were taken as part of two programs: (1) a Gemini Large and Long Program (LLP; PI: E.\ Buckley-Geer; program IDs GS-2015B-LP-5 and GS-2017A-LP-5) of spectroscopic follow-up for DES strong lensing systems and for DES photometric redshift (photo-z) calibrations; and (2) a dedicated program (PI: H.\ Lin; program ID GS-2018B-Q-220) to observe line-of-sight galaxy redshifts and lensing galaxy velocity dispersions for our two lensed quasar systems.
These programs observed a total of four GMOS-S multi-object spectroscopy (MOS) masks for \DESone and two masks for \DEStwo, and the data were taken in queue mode on Gemini South during the semesters 2015B, 2017A, and 2018B.  
Nearly all the masks were each observed using both the GMOS-S B600 and R400 gratings, in order to provide spectral coverage over both blue and red wavelength ranges, respectively, spanning approximately 3800\AA$-$7500\AA\ and 5000\AA$-$10500\AA. 
Multiple science exposures were taken to reject cosmic rays, and the grating central wavelength was dithered slightly for different exposures to fill in wavelength gaps due to spatial gaps between the three GMOS-S CCDs. 
Flat field and wavelength calibration exposures were interspersed with the science exposures.

The LDSS-3 observations were taken as part of a semester 2018A Magellan program (PI: J.\ Frieman) to obtain line-of-sight galaxy spectroscopy for \DESone and two other lensed quad quasar systems.
Four LDSS-3 MOS masks were observed for \DESone over the two nights 2018 January 21,22 UT.    
Each mask was observed using the LDSS-3 VPH-All grism, with wavelength coverage of about 3800\AA$-$10500\AA.
Multiple science exposures were taken to reject cosmic rays.
Flat field and wavelength calibration exposures were taken immediately before and after the sequence of science exposures.

The VLT MUSE observations were taken as part of an ESO program (0102.A-0600(E), PI: A.\ Agnello) to do integral field spectroscopy of \DESone and its surrounding field.
The MUSE observations were done in wide-field mode with adaptive optics corrections and were carried out over the two nights 2019 Jan 11 and 13. 
The final MUSE data cube covered an area of $92\arcsec \times 95\arcsec$, and the wavelength coverage spanned 4700\AA$-$5803\AA\ and 5966\AA$-$9350\AA.
Additional spectroscopic analysis of the MUSE data and further details of the observations and data processing are given in \cite{Shajib2019_H0}.

\begin{table}
\caption{Spectroscopic observations for the \DESone and \DEStwo systems.}
\label{tab:spectro_observations}
\begin{tabular}{lccc}
\hline
\multicolumn{2}{c}{Telescope/Instrument} & & \\
 & Grating/ & & Exposure \\
Mask/Cube & Grism & UT Date & time (sec) \\
\hline
\hline
\multicolumn{4}{|c|}{\DESone} \\
\hline
\hline
\multicolumn{2}{c}{Gemini South/GMOS-S} & & \\
(1) DESJ0408-5354\_42 & B600 & 2015 Dec 09 & $4\times 900$ \\
(2) DESJ0408-5354\_42 & R400 & 2015 Dec 09 & $4\times 900$ \\
(3) DESJ0408-5354\_45 & B600 & 2017 Apr 26$-$27 & $5\times 900$ \\
(4) DESJ0408-5354\_45 & R400 & 2017 Mar 30$-$31 & $4\times 900$ \\
(5) DESJ0408-5354\_A & B600 & 2018 Dec 04 & $4\times 900$ \\
(6) DESJ0408-5354\_A2 & R400 & 2019 Feb 04 & $6\times 1000$ \\
(7) DESJ0408-5354\_B & B600 & 2018 Dec 09 & $4\times 900$ \\
(8) DESJ0408-5354\_B & R400 & 2018 Dec 09 & $4\times 900$ \\
\hline
\multicolumn{2}{c}{Magellan Clay/LDSS-3} & & \\
(9) des0408a & VPH-All & 2018 Jan 21 & $7\times 780$ \\
(10) des0408b & VPH-All & 2018 Jan 21 & $6\times 780$ \\
(11) des0408c & VPH-All & 2018 Jan 22 & $7\times 780$ \\
(12) des0408d & VPH-All & 2018 Jan 22 & $6\times 780$ \\
\hline
\multicolumn{2}{c}{VLT UT4/MUSE} & & \\
(13) MUSE & & 2019 Jan 11,13 & 14400 \\
\hline
\hline
\multicolumn{4}{|c|}{\DEStwo} \\
\hline
\hline
\multicolumn{2}{c}{Gemini South/GMOS-S} & & \\
(14) DESJ2038-4008\_A & B600 & 2018 Nov 06 & $4\times 900$ \\
(15) DESJ2038-4008\_A & R400 & 2018 Nov 07 & $4\times 900$ \\
(16) DESJ2038-4008\_B & B600 & 2018 Nov 07 & $4\times 900$ \\
\end{tabular}
\end{table}

\subsection{Spectroscopic target selection}
\label{subsec:targets}

Galaxy targets for the Gemini and Magellan masks were selected using DES photometry.  
The exact selection criteria changed somewhat with time, due to improvements in DES photometric measurements, star-galaxy separation, and object catalogs.  
Specifically, three sets of selection criteria were used for the masks in Table~\ref{tab:spectro_observations}, listed below in order from earliest to latest in time. \\

{\noindent \em (A) Gemini South masks (1)-(4):} Galaxies were selected from the DES Year 1 (Y1) ``Y1A1 COADD'' catalog \citep{drlica-wagner2018}, using the $i$-band {\tt SExtractor AUTO} magnitude \citep{Bertin1996} cuts $20 \leq {\tt MAG\_AUTO\_I} < 22.5$.  
No Milky Way extinction corrections were applied to the magnitudes before selection. 
Star-galaxy separation used the {\tt SExtractor SPREAD\_MODEL} classifier \citep{Desai2012}, also in the $i$-band: ${\tt SPREAD\_MODEL\_I} > 0.002$.\\

{\noindent \em (B) Magellan masks (9)-(12):} Galaxy targets were now selected from the deeper DES Year 3 (Y3) data set, specifically from the ``Y3 GOLD'' (Sevilla et al. in prep) version 1.0 catalog, using the $i$-band magnitude cuts $18 \leq {\tt MAG\_AUTO\_I} < 23$, in particular extending the faint magnitude limit fainter to aim for a deeper sample. 
Again no Milky Way extinction corrections were applied to the magnitudes. 
Star-galaxy separation used the same $i$-band cut: ${\tt SPREAD\_MODEL\_I} > 0.002$.\\

{\noindent \em (C) Gemini South masks (5)-(8) and (14)-(16):}
These masks were the latest to be designed and therefore used improved selection methods compared to the masks in (A) and (B).
Galaxies were again selected from the ``Y3 GOLD'' data set, but using the most current version 2.2 catalog. 
We also changed the selection magnitudes from {\tt AUTO} to {\tt MOF} \citep[][Sevilla et al. in prep]{drlica-wagner2018} magnitudes, where the latter provide improved photometry, based on simultaneous multiepoch, multiband, and multiobject fits.  
The adopted $i$-band cuts were $15 \leq {\tt MOF\_CM\_MAG\_CORRECTED\_I} < 23$, where these MOF magnitudes also included Milky Way extinction corrections and several other sub-percent photometric zeropoint corrections (Sevilla et al. in prep).
Finally, we also updated the star-galaxy separation classifier to {\tt EXTENDED\_CLASS\_MASH\_MOF} (Sevilla et al. in prep), specifically using the cuts ${\tt EXTENDED\_CLASS\_MASH\_MOF} = 3$ (indicating ``high confidence galaxies'') or 2 (indicating ``mostly galaxies'').
For $i \leq 22.5$, this classification should yield a galaxy efficiency $> 98.5\%$ and a stellar contamination $< 1\%$ (Sevilla et al. in prep).\\

Each Magellan LDSS-3 and Gemini South GMOS-S mask included galaxy targets distributed over about a $5.5\arcmin \times 5.5\arcmin$ sky area centered on each lensed quasar system.  
Generally one to three slits on each mask were manually designed to target objects in or close to the lensed quasar system, e.g., to measure the redshift or velocity dispersion of the main lensing galaxy in each system, or to observe close nearby neighbor galaxies of the systems.
The remaining targets were selected automatically by {\tt GMMPS}\footnote{\url{https://gmmps-documentation.readthedocs.io/en/latest/}} or {\tt maskgen}\footnote{\url{https://code.obs.carnegiescience.edu/maskgen}}, the respective GMOS-S or LDSS-3 mask design software, both of which designed masks to maximize the number of targets observed.
The potential set of galaxies that could be targeted was subject to the selection criteria described above.
In addition, at the time each mask was designed, we removed from the initial target list any galaxies which already had high-confidence redshifts from previous observations, or which were already targeted on companion masks designed for the same semester's (Gemini) observing queue or (Magellan) observing run. 
Moreover, for the Magellan LDSS-3 targets (only), we assigned targeting priorities as inputs to {\tt maskgen}, depending on the $i$-band magnitude and on the radius from the \DESone quad system.
Specifically, for galaxies with radius $\leq 3\arcmin$, we assigned priorities based on {\tt MAG\_AUTO\_I} (as in criteria (B) above), with highest priority given to bright galaxies ${\tt MAG\_AUTO\_I} < 19.5$, next priority to galaxies otherwise brighter than ${\tt MAG\_AUTO\_I} = 22$, and lowest priority to galaxies otherwise brighter than ${\tt MAG\_AUTO\_I} = 23$. 
Finally, for galaxies with radius $> 3\arcmin$, we assigned lower priorities than for galaxies $\leq 3\arcmin$, and these priorities were tied linearly to {\tt MAG\_AUTO\_I} (with brighter galaxies at higher priority).

\subsection{Data for the determination of the line-of-sight under/overdensities}

In order to determine the line-of-sight under/overdensities for \DESone and \DEStwo we need a catalog of galaxy properties that includes magnitudes and photometric redshifts.  We have used the catalogs from the Year 3 Gold version 2.2 catalog.

\begin{figure*}
 \subfigure{\includegraphics[width=\columnwidth]{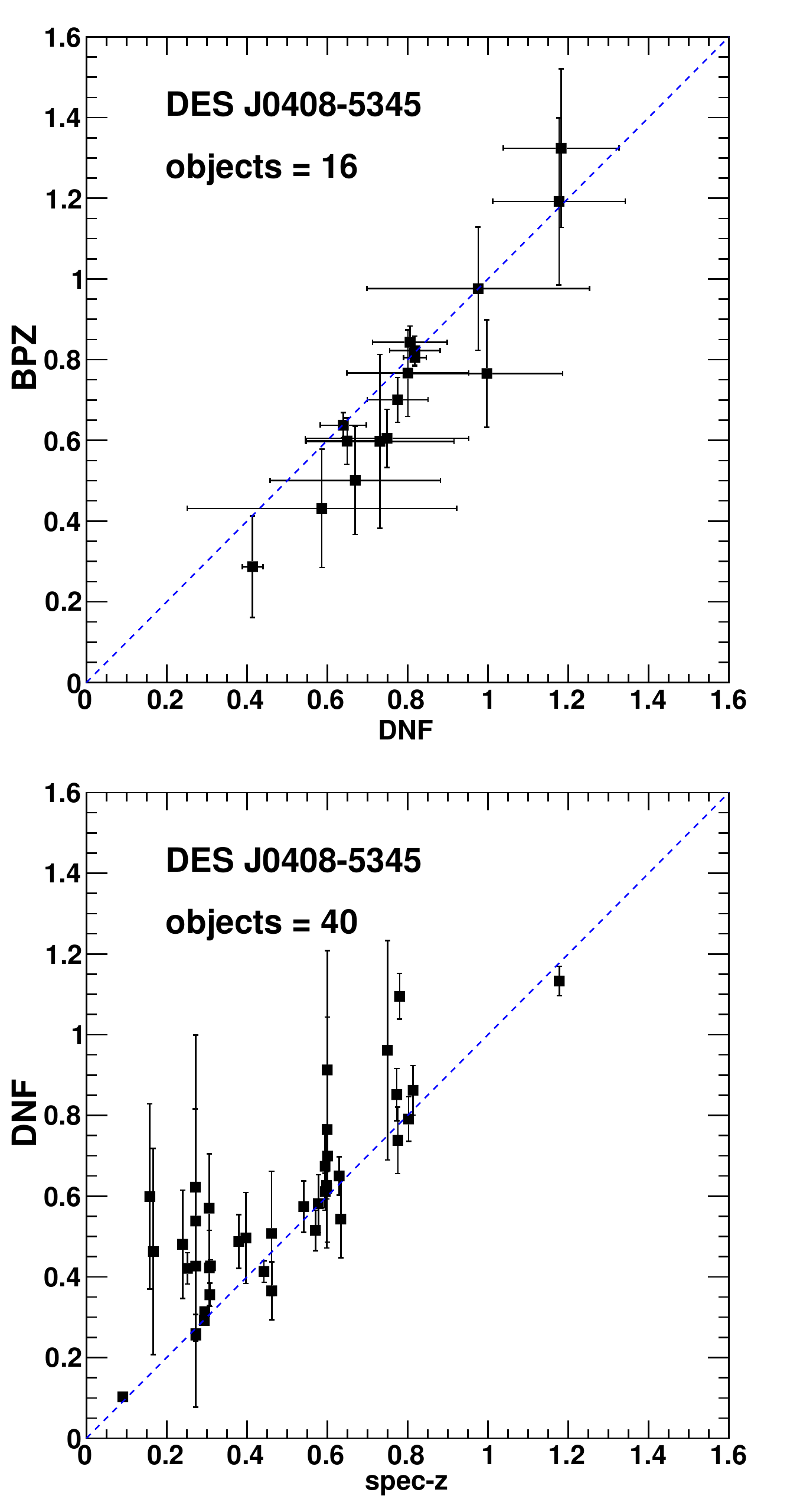}}
 \subfigure{\includegraphics[width=\columnwidth]{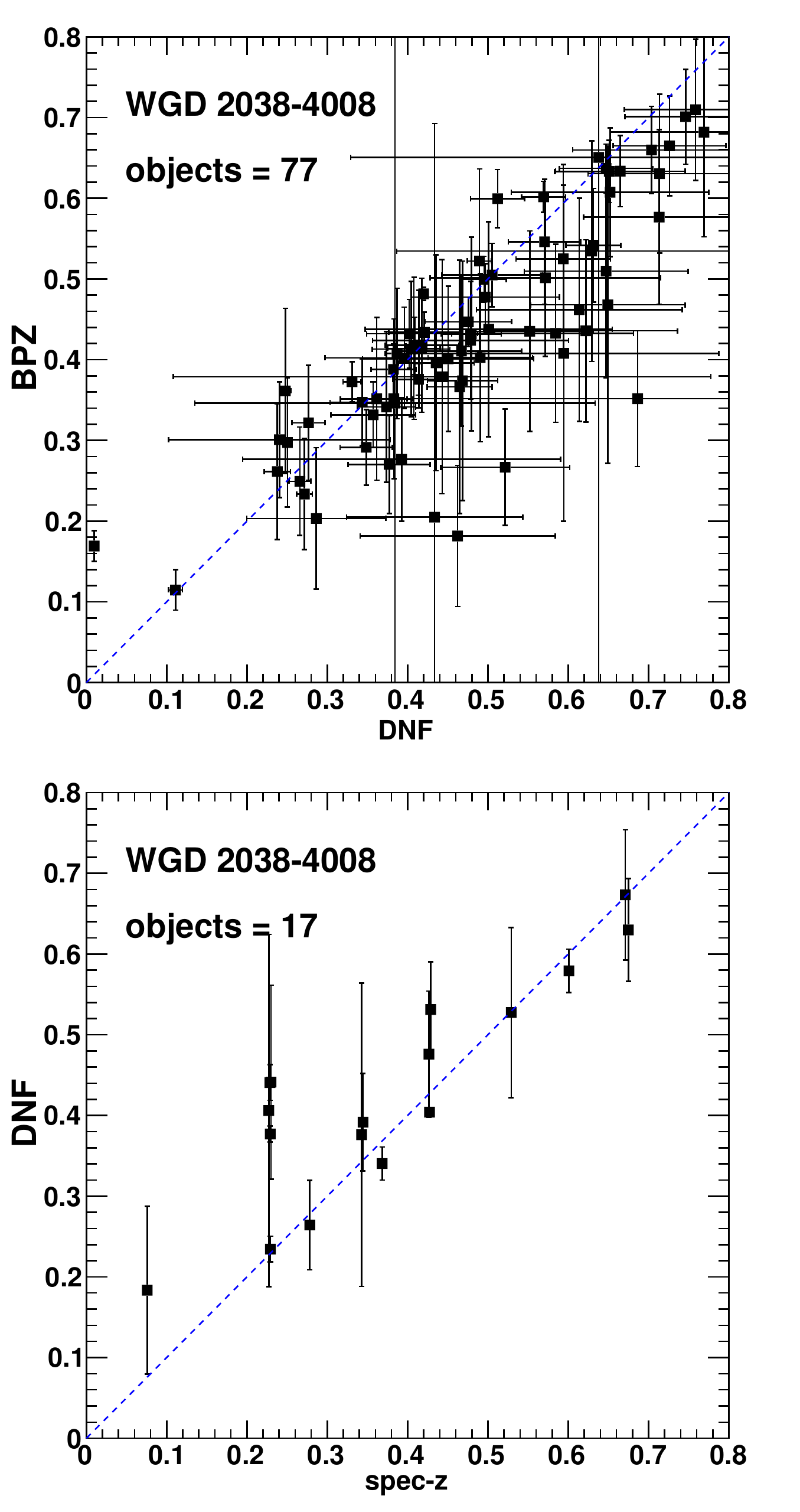}}
 \caption{Upper figures: Comparison of the DNF and BPZ photometric redshifts for galaxies with no spectroscopic redshift which have $i < 22.5$ within 120 arcsec of the lens center. The dashed line indicates BPZ = DNF. Left figure \DESone, right figure \DEStwo. Lower figures: Comparison of spectroscopic and DNF photometric redshifts for galaxies with $i < 22.5$ within 120 arcsec of the lens center. The dashed line indicates specz = DNF photo-z. Left figure \DESone, right figure \DEStwo.}
 \label{fig:specz-dnf-0408-2038}
\end{figure*}

We have used the photometric redshifts computed using the DNF machine-learning algorithm described in \citet{devicente2016}. We also repeat the analysis using the photometric redshifts derived with the Bayesian Photometric Redshift \citep[BPZ:][]{Benitez2000,hoyle2018} method. If an object has a spectroscopic redshift from Gemini or Magellan, as described below in \S\ref{subsec:specz}, then we use that redshift instead of the photometric redshift.  In Figure~\ref{fig:specz-dnf-0408-2038} we show the comparison of the photometric redshifts from the DNF algorithm to the spectroscopic redshifts for objects with $i < 22.5$. The comparison of the DNF and BPZ photo-z's for the objects that only have photometric redshifts is shown also in Figure~\ref{fig:specz-dnf-0408-2038} for both \DESone and \DEStwo.  We observe that there is no obvious mismatch between the two algorithms so we have used results from both of them in the subsequent analysis. We note that the photometric redshifts are computed using the photometry in the $griz$ filters only, as \citet{hoyle2018} have found that the $Y-$band adds little to no predictive power. 

\subsubsection{The Control Field: DES}
\begin{figure}
 \includegraphics[width=\columnwidth]{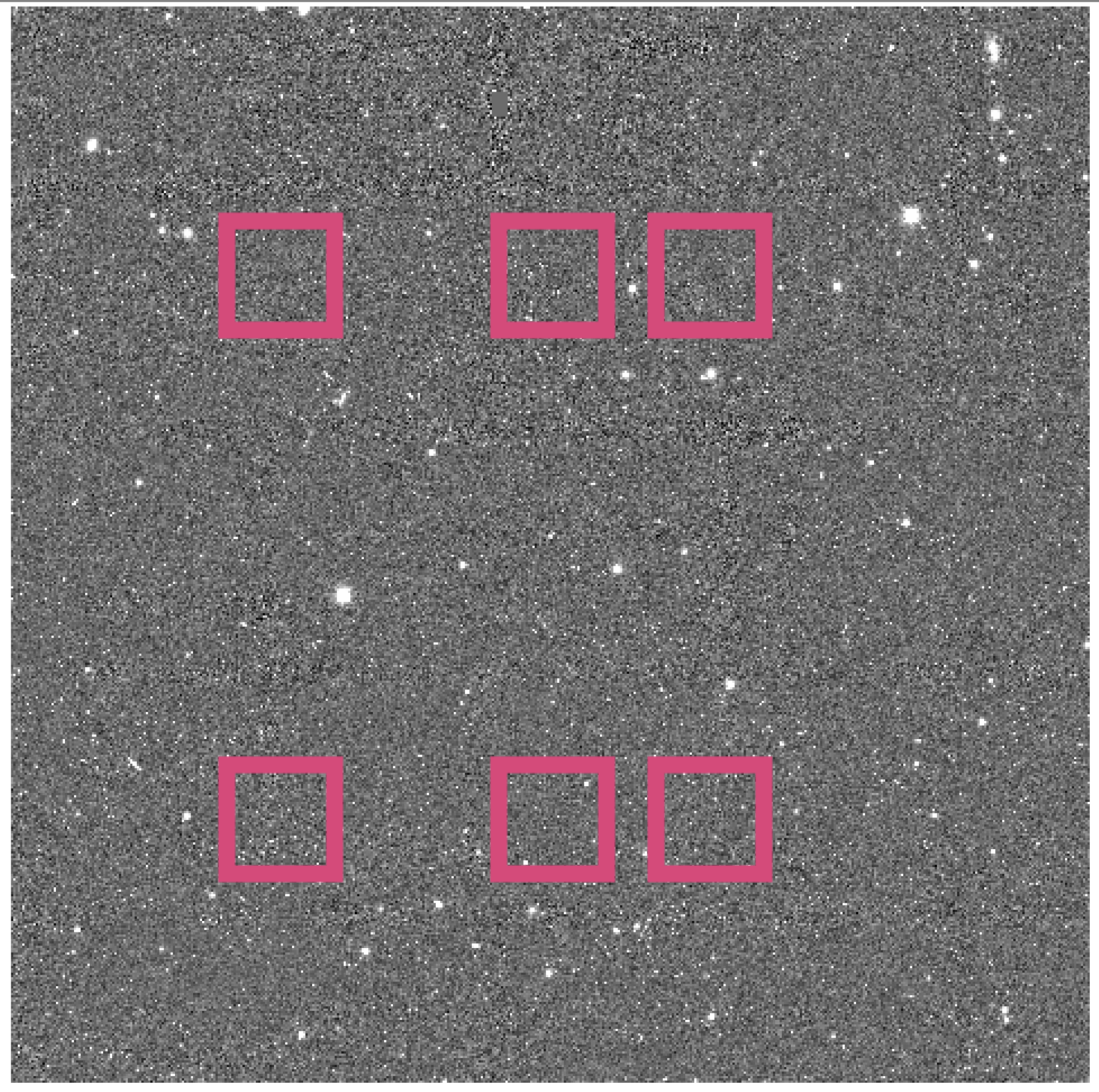}
 \caption{A typical DES tile showing the location of the control fields. Each magenta box is $1000\times 1000$ pixels.}
 \label{fig:control-fields}
\end{figure}

To to apply the weighted number counts technique we need a control field against which to determine an under/overdensity, the constraint we will eventually use to determine $P(\kappa_\mathrm{ext})$. As both lensing systems are within the DES footprint we have chosen to use the full DES survey footprint of 5100 sq. deg to provide our control field. This differs from our approach in the past, where we have used control fields observed with the {\it Hubble Space Telescope (HST)} \citep{Suyu2010,Fassnacht2011,Suyu2013TWOCOSMOLOGY,Greene2013}, and from CFHTLenS \citep{Heymans2012},  \citep[e.g.,][]{Rusu2017,Rusu2019,Birrer2019,Chen2019}. The choice of control fields from DES itself, as opposed to other large-scale cosmological surveys, is optimal. This choice avoids potential biases to which our technique of obtaining $\kappa_\mathrm{ext}$ from weighted number count ratios may be sensitive, such as due to mismatches in image resolution, depth and star-galaxy classification between the lens (target) fields and the control fields \citep[e.g.,][]{Rusu2017}. The full Year 6 DES survey footprint consists of 10169 tiles, each $10,000 \times 10,000$ pixels or $0.53$ sq deg in area. We take a sample of tiles from the center of the footprint that are far from the survey edges and also eliminate tiles that contain very bright stars or very large galaxies. This results in  a total of 5402 tiles of which we select 843 from across the survey footprint. For each tile we select six fields each of $1000\times 1000$ pixels. This gives us a total of 5094 control fields spread over the sky and covering $\approx 27$ sq deg. The location of these six fields are shown in Figure~\ref{fig:control-fields} for one of the DES tiles. This should allow us to account for sample variance.  

\subsubsection{The galaxy samples for the target and control fields}
\label{subsec:galaxy_samples}
\begin{figure}
 \subfigure{\includegraphics[width=\columnwidth]{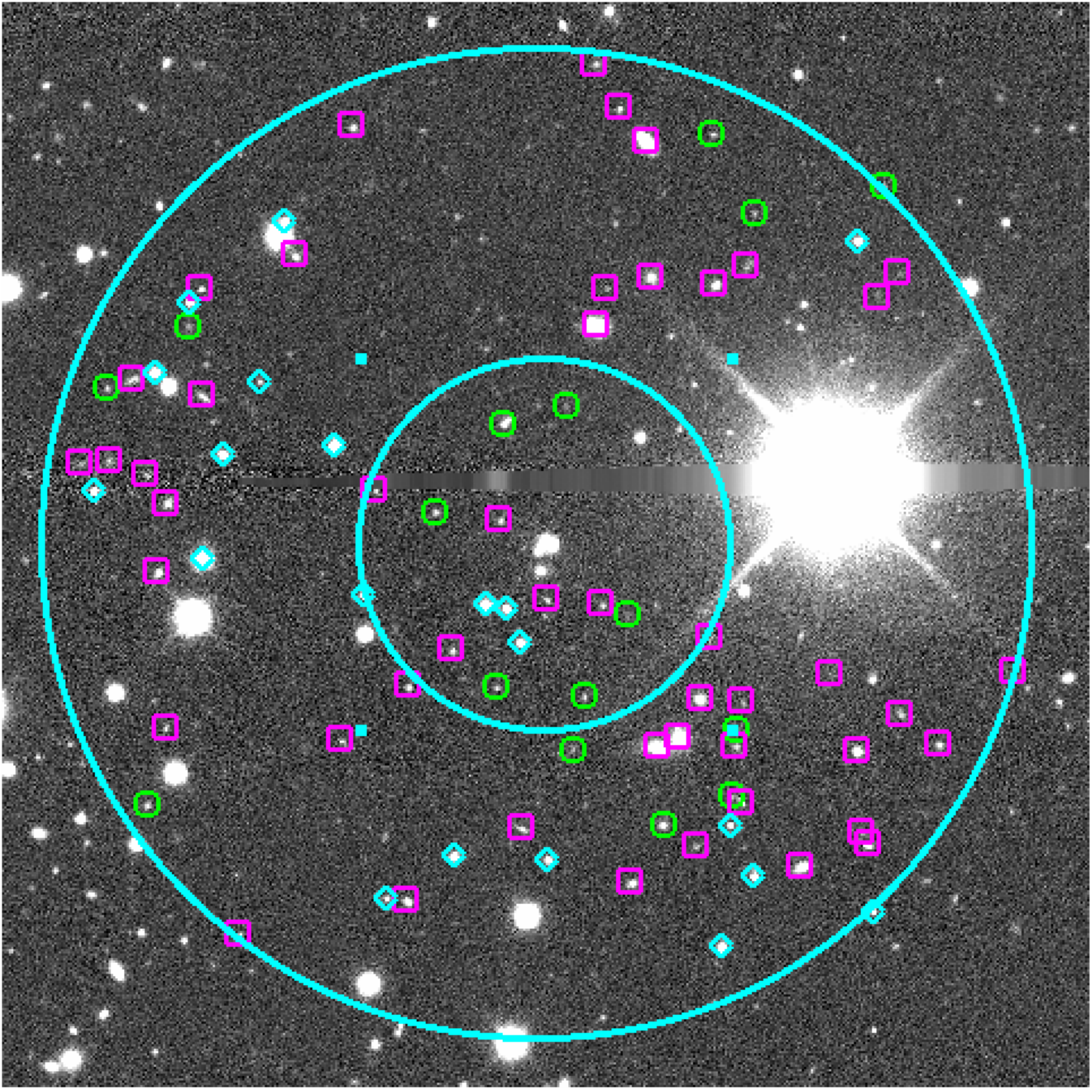}}
 \subfigure{\includegraphics[width=\columnwidth]{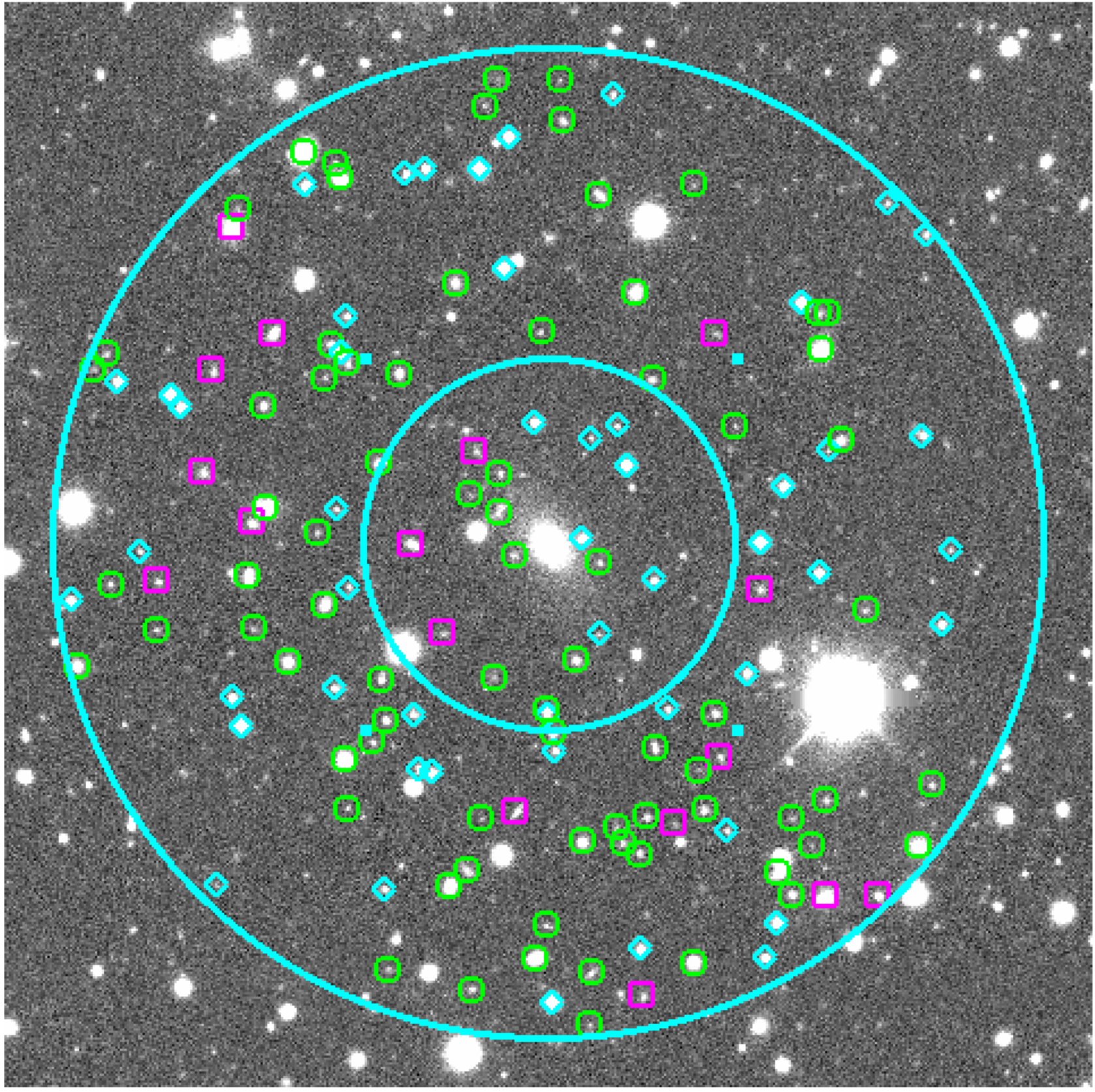}}
 \caption{The $1000 \times 1000$ pixel field of view for \DESone (upper figure) and \DEStwo (lower figure). North is up and East is left. The $i < 22.5$ galaxies inside the $120\arcsec$ radius aperture are indicated by magenta squares for the objects with a spectroscopic redshift and green circles for the objects with no spectroscopic redshift. Stars are indicated by the cyan diamonds. The two concentric cyan circles indicate the apertures of $120\arcsec$ and $45\arcsec$ radius respectively.}
 \label{fig:des0408-2038-field}
\end{figure}

For the target fields we make an initial selection of all objects from the Y3 Gold catalog that fall within a $4'\times 4'$ box centered on the lens. For the control fields we make the same selection but using the center of the control field. We select all objects that satisfy ${\tt FLAGS\_GOLD} = 0$ and ${\tt EXTENDED\_CLASS\_SOF} \geq 2$, which selects galaxies\footnote{When comparing with the available {\it HST} data, we found that five objects in the \DEStwo field are erroneously classified by this pipeline, mainly in the form of stars wrongly classified as galaxies. As some of these are close enough to the lens to bias the inference described below, we were careful to correct the classification.}. We are using the Single-Object Fitting (SOF) magnitudes that are computed using a simplified version of the Multi-Object Fitting (MOF) algorithm described in section 6.3 of \citet{drlica-wagner2018}. We select all objects with $i$-band magnitude $0 < {\tt SOF\_CM\_MAG\_CORRECTED\_I} < 22.5$. The faint-end limiting magnitude, which is the same one used in \citet{Rusu2019} based on the analysis in \citet{Sluse2019}, also for the purpose of constraining $H_0$, ensures that the galaxy classification is reliable, and that the galaxy catalog is complete. It is also deep enough \citep{Collett2013} to keep biases on $\kappa_\mathrm{ext}$ due to depth significantly below the 1\% level. We are using the photometric redshifts that were calculated using the SOF magnitudes, namely ${\tt DNF\_ZMEAN\_SOF}$ and ${\tt BPZ\_ZMEAN\_SOF}$. We also require that the redshift ${\tt DNF\_ZMEAN\_SOF}$ or ${\tt BPZ\_ZMEAN\_SOF}$ of the objects satisfy $z < z_s$, where $z_s$ is the source redshift, and that their distance $\Delta r$ from the center of the field is less than $120\arcsec$. Figure~\ref{fig:des0408-2038-field} shows the objects that pass these selection criteria in a $1000 \times 1000$ pixel field around \DESone and \DEStwo.

\section{Redshifts and stellar masses} 
\label{sec:redshifts}

In this section we first describe our spectroscopic redshift measurement procedure, summarize the results, and show the line-of-sight galaxy redshift distributions for each lensing system.
We then describe and plot the spectroscopic redshift completeness for the overall galaxy samples in the two systems.
Finally, we detail the procedures for measuring stellar masses using photometric model fitting for our galaxies.

\subsection{Spectroscopic redshifts}
\label{subsec:specz}

The Gemini and Magellan data were processed to 2D and 1D spectra using the IRAF Gemini and COSMOS \citep{Dressler2011,Oemler2017} reduction packages, respectively.  
Initial redshifts were determined automatically (without visual inspection) using the IRAF external package {\tt rvsao} \citep{Kurtz1998} to cross correlate the 1D spectra against a set of SDSS galaxy templates.
All processed 2D and 1D spectra, along with the corresponding automated redshift results, were visually inspected in order to assign final quality flags to the redshifts.  
If necessary, an automated redshift may be overridden and manually re-measured from the 2D or 1D spectrum.  
Only high-confidence redshifts were included in the subsequent analyses.

\begin{figure*}
  \includegraphics[width=\columnwidth]{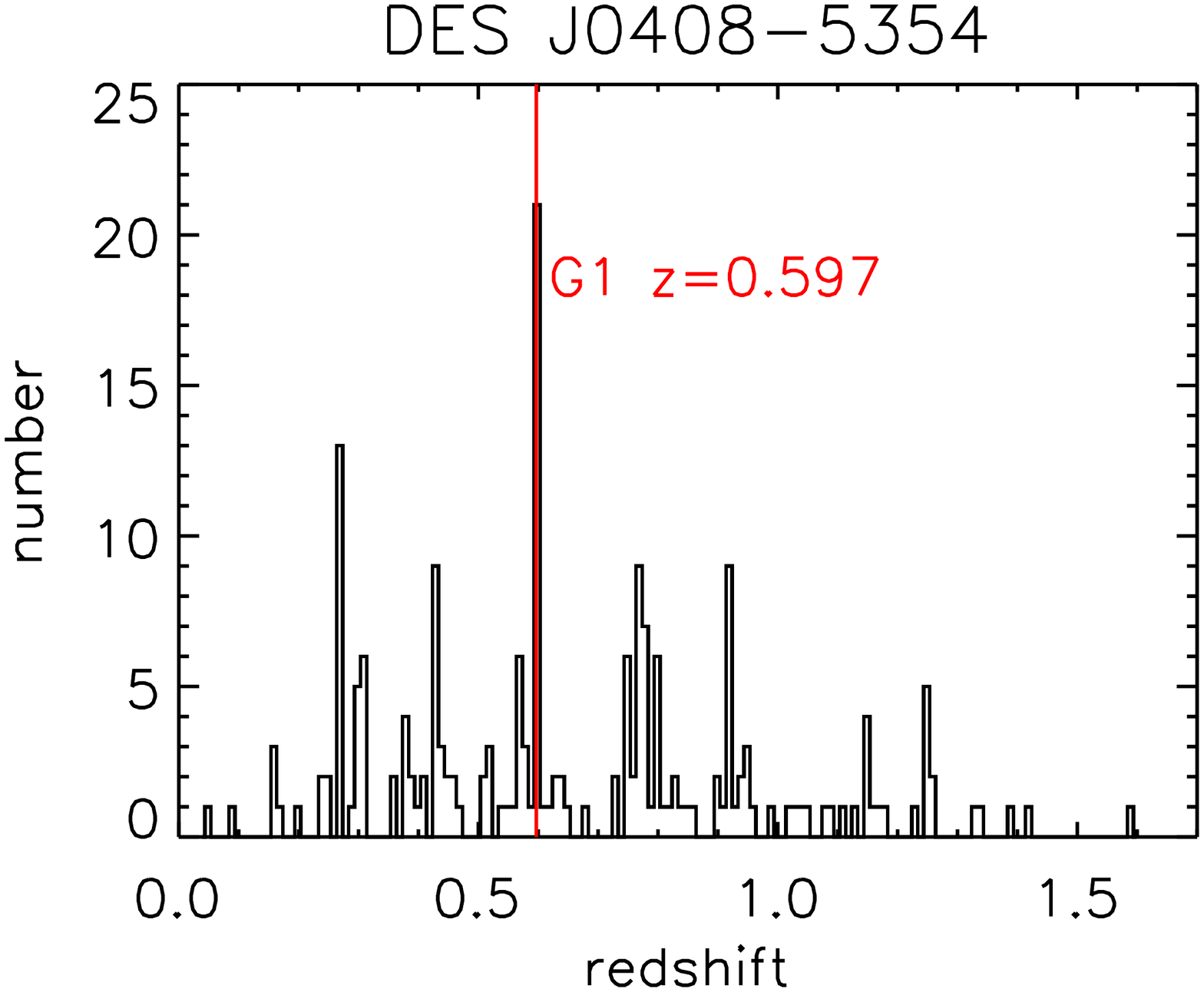}
  \includegraphics[width=\columnwidth]{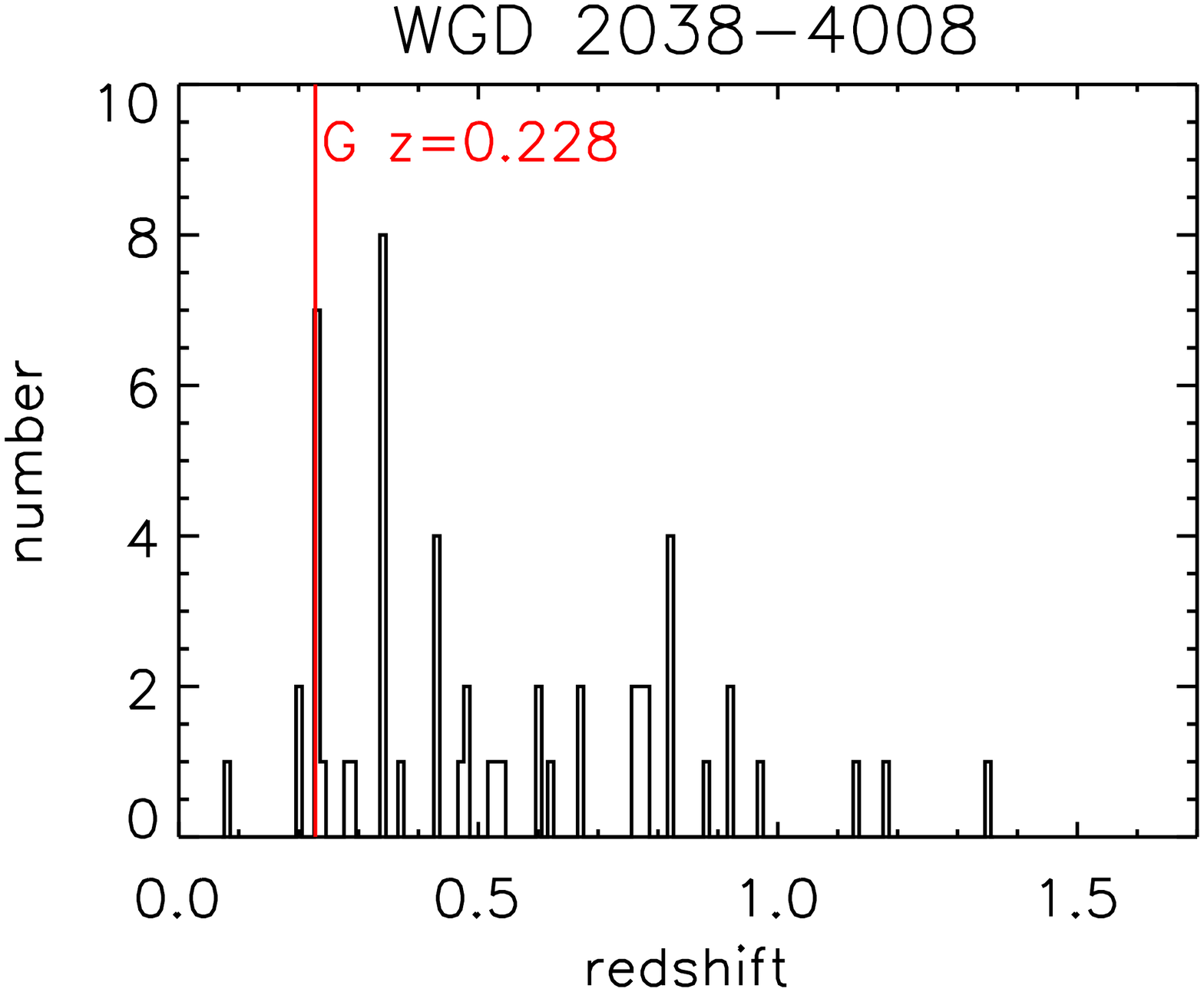}
 \caption{Histograms of line-of-sight galaxy spectroscopic redshifts (\S\ref{subsec:specz}) for the (left) \DESone and (right) \DEStwo systems. In each panel the red vertical line indicates the redshift of the main lensing galaxy in each system.}
 \label{fig:specz_histogram}
\end{figure*}

For \DESone, we obtained 101 high-confidence galaxy redshifts from Gemini South GMOS-S and 70 from Magellan LDSS-3.
From the VLT MUSE data we obtained another 28 redshifts that were not already among the Gemini and Magellan redshifts, thus resulting in a total of 199 high-confidence redshifts for \DESone.
For \DEStwo we obtained a total of 54 high-confidence galaxy redshifts, all obtained from the Gemini South GMOS-S data.
The above counts include the two redshifts of the main lensing galaxies in the systems.
Figure~\ref{fig:specz_histogram} shows histograms of the redshift distributions for each lensing system.

\subsection{Redshift completeness}
\label{subsec:completeness}

We define spectroscopic redshift completeness to be the fraction of DES Y3 Gold (Sevilla et al. in prep) galaxies that have redshifts  (as described in \S\ref{subsec:specz}).
To define our photometric galaxy sample, we use the latest and best available version of the Y3 Gold catalog, version 2.2, and we also use the same magnitude and star-galaxy separation cuts as listed in target selection criteria (C) of \S\ref{subsec:targets} above.

\begin{figure*}
  \includegraphics[width=\columnwidth]{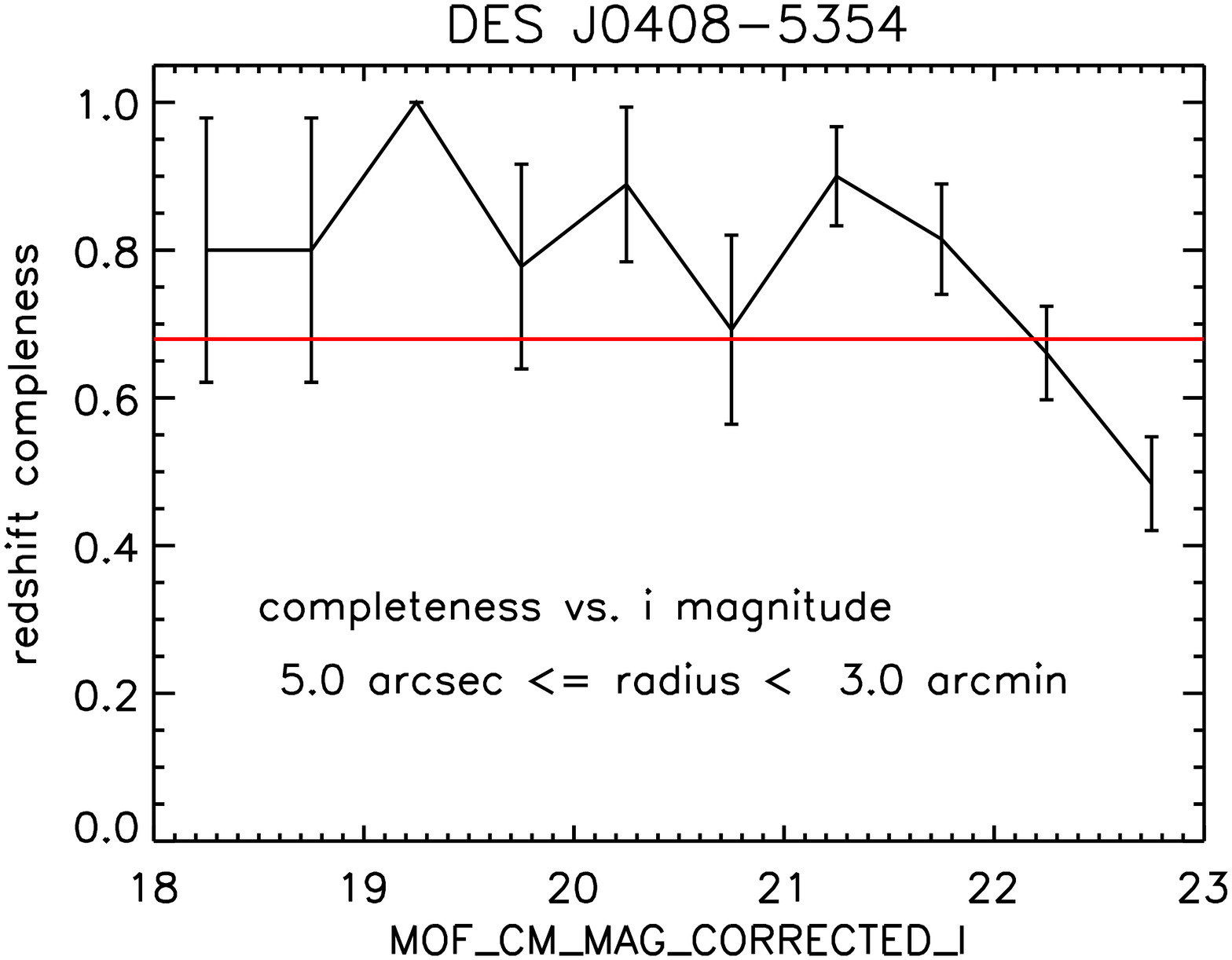}
  \includegraphics[width=\columnwidth]{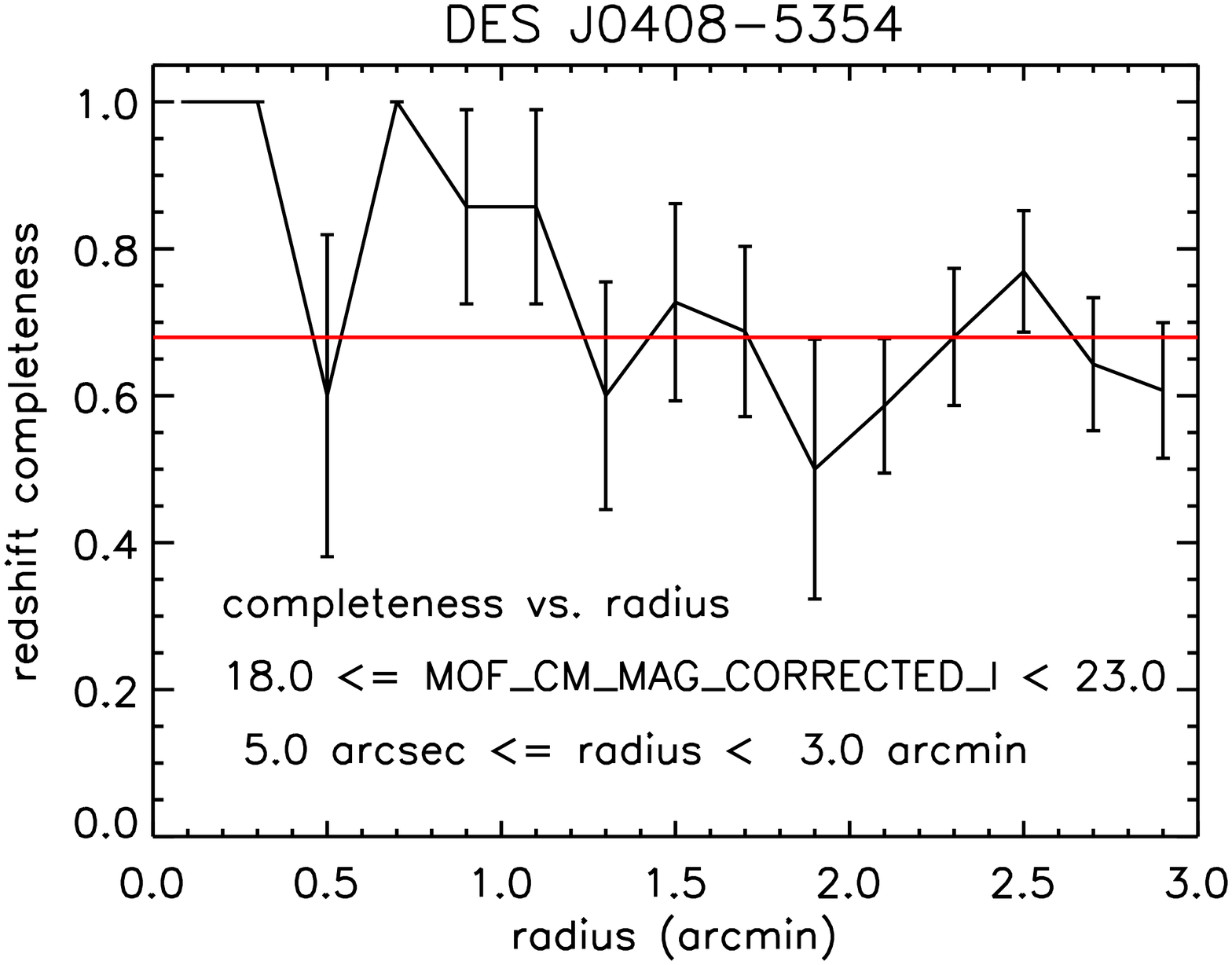}
  \includegraphics[width=\columnwidth]{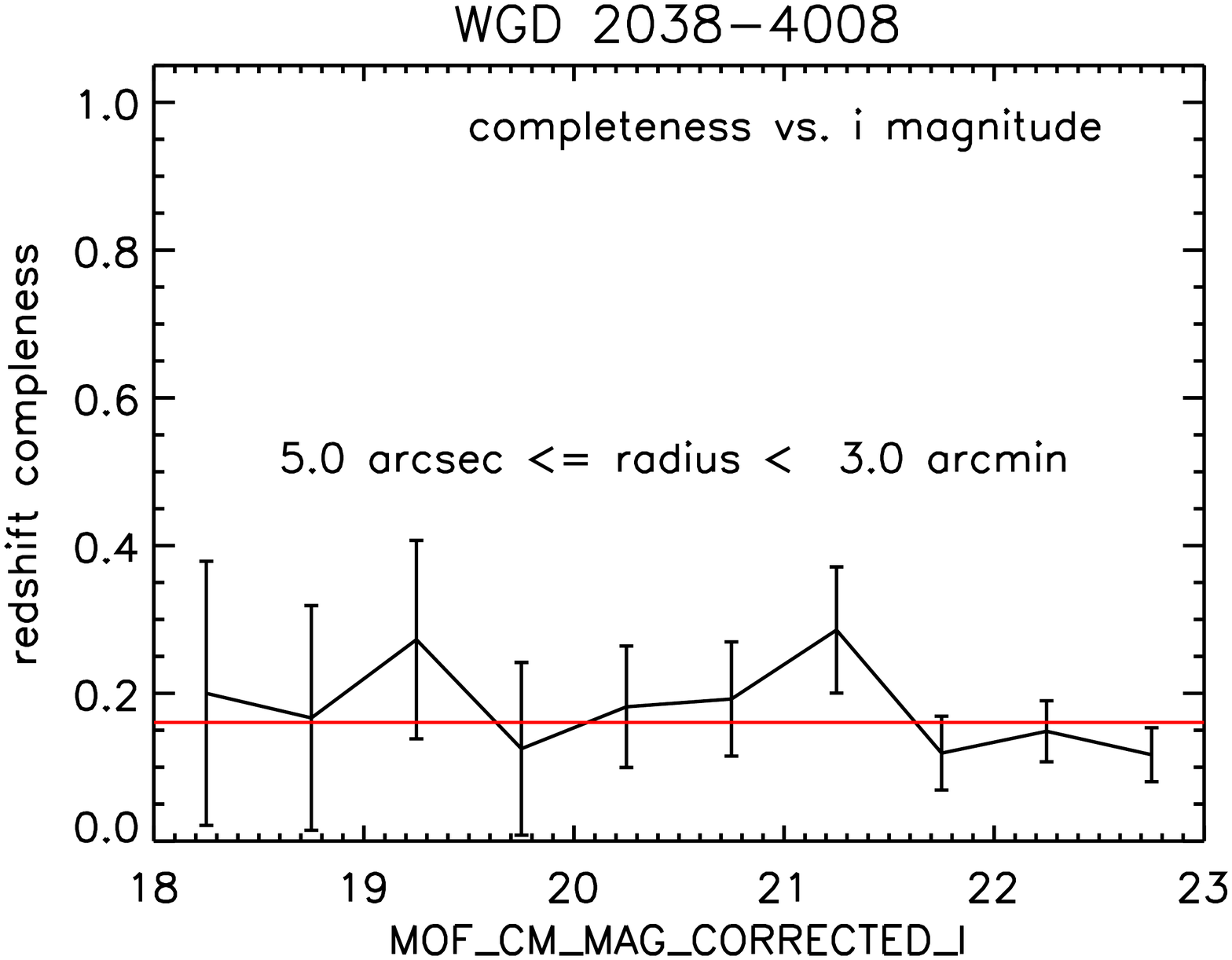}
  \includegraphics[width=\columnwidth]{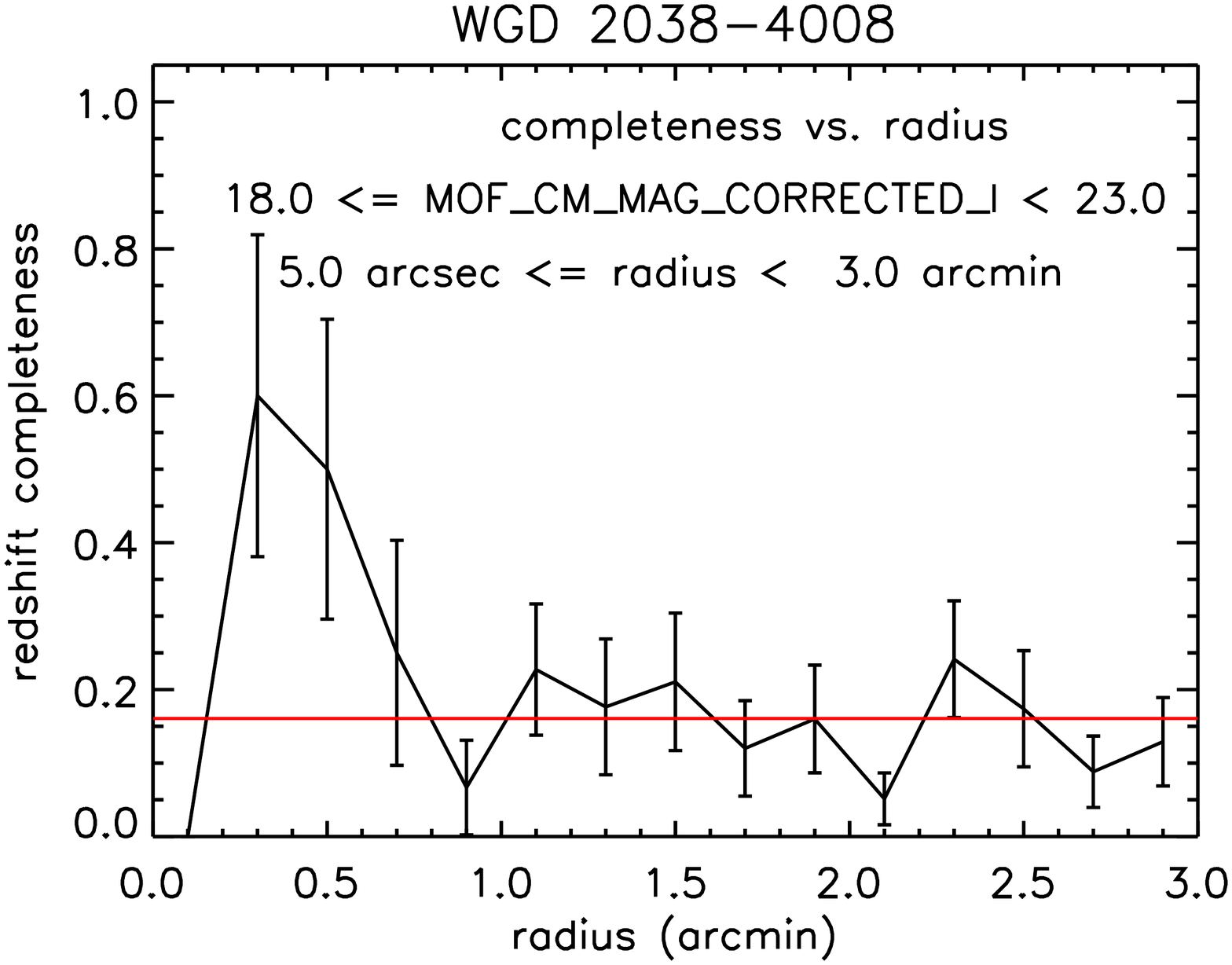}
 \caption{Spectroscopic redshift completeness (defined in \S\ref{subsec:completeness}) for \DESone\ (top panels) and \DEStwo\ (bottom panels). The left panels show redshift completeness vs. $i$-band magnitude, within the radius range 5\arcsec to 3\arcmin\ from the lens galaxy in each system. The right panels show redshift completeness vs.\ radius, within the $i$-band magnitude range 18 to 23. In all panels the red horizontal line indicates the overall redshift completeness within the indicated magnitude and radius ranges for each system.}
 \label{fig:complete}
\end{figure*}

For \DESone, the resulting spectroscopic redshift completeness is 0.68 for $18 <= {\tt MOF\_CM\_MAG\_CORRECTED\_I} < 23$ galaxies and $5\arcsec <=$~radius~$ < 3\arcmin$; see the top panels of Figure~\ref{fig:complete}. 
For \DEStwo, the redshift completeness is 0.16 for the same $i$-band magnitude and radius ranges, as shown in the bottom panels of Figure~\ref{fig:complete}.
Within the plotted magnitude and radius ranges, the redshift completeness stays fairly constant for both systems.
The inner radius cut of $5\arcsec$ is intended to exclude the quasar images from consideration.  
The outer radius cut is set at $3\arcmin$ as the redshift completeness drops very rapidly beyond this radius for either system.

\subsection{Stellar masses}
\label{sec:photozmstar}
 
Stellar masses necessary for the computation of the flexion shift, the criterion used to separate between the structures which need to be accounted for in the lensing model, and those which can be incorporated inside $\kappa_\mathrm{ext}$ (see \S\ref{sec:flexion}) were computed using the galaxy template fitting code Le PHARE \citep{Arnouts1999,Ilbert2006}. In the Le PHARE fits, we either used our spectroscopic redshifts when available, or DES Y3 Gold DNF photo-z's ({\tt DNF\_ZMEAN\_SOF}),
combined with MOF photometry \citep{drlica-wagner2018} from the DES Y3 Gold (Sevilla et al. in prep) version 2.2 catalog.
Specifically, we used the {\tt MOF\_CM\_MAG\_CORRECTED} magnitudes and their associated errors in the $griz$ filters, as these magnitudes included Milky Way extinction and other small photometric corrections (see \ref{subsec:targets}).

The galaxy template set used in the Le PHARE fits were taken from the BC03 \citep{bc03} spectral energy distribution (SED) library.
Specifically, we used a set of 27 BC03 simple stellar population models, computed using the Padova 1994 stellar evolution library \citep[described in][]{bc03} and the \cite{Chabrier2003} initial mass function.
The 27 models consisted of 9 exponentially declining star formation rate (SFR) histories (with decay times $\tau = 0.1, 0.3, 1, 2, 3, 5, 10, 15, 30$ Gyr), each computed at 3 different metallicity values (0.4, 1, and 2.5 times solar).
Each of the 27 models was also computed at different ages (ranging from 0.2 to 13.5 Gyr) and redshifts (up to a maximum redshift of 1.1, in steps of 0.03 in redshift).
No dust extinction was included in the models, but recall the DES magnitudes already included correction for foreground Milky Way extinction.
The stellar masses and uncertainties computed from the Le PHARE fits are tabulated in Table~\ref{tab:flexion_galaxy_all}.

For a few of the close neighbor galaxies of the two lensed quasar systems, the original object deblending and resulting photometry in the DES Y3 catalog were clearly incorrect upon visual inspection of 
the images.
In particular, for \DESone, the neighbor galaxy with ID number 488066768 (or name G5, in Table~\ref{tab:flexion_galaxy} or Table~\ref{tab:flexion_galaxy_all}) had magnitudes that were too bright, and for \DEStwo, the three neighbor galaxies with ID numbers 13, 14, and 15 (Table~\ref{tab:flexion_galaxy_all}) were originally merged into a single object.
To correct these photometry problems, we used the galaxy image fitting code GALFIT \citep{Peng2010} to redo the galaxy model fitting and photometry for these objects, based on the DES Y3 coadd images in the $griz$ filters.
These GALFIT results were used to compute stellar masses via Le PHARE.

\section{Velocity Dispersions of Lensing Galaxies} 
\label{sec:velocity_dispersion}

The main lensing galaxies in both the \DESone\ and \DEStwo\ systems were specifically targeted for stellar velocity dispersion measurements on a number of the spectroscopic masks listed above in Table~\ref{tab:spectro_observations}.
For these observations we describe below the details of the targeting on the spectroscopic slits, the procedures used to extract the lensing galaxy spectra, and the method employed to measure velocity dispersions.

\subsection{\DESone G1}
\label{sec:veldisp_0408}

The main lensing galaxy G1 in \DESone was measured from four independently observed spectra: two from Magellan LDSS-3, one from Gemini South GMOS-S, and one from VLT MUSE.

G1 was targeted on two of the four Magellan LDSS-3 masks listed in Table\ref{tab:spectro_observations}, ``des0408a'' and ``des0408b''; hereafter these two masks will be denoted ``Magellan a'' and ``Magellan b''. 
The slit setup for G1 was the same on both masks, while the remaining targets were different on the two masks.  
Specifically, the slit was oriented so that it included both G1 and quasar image B \citep[the naming convention for the lensing galaxy G1 and the lensed quasar images A, B, and D are shown in][]{Lin2017a}.
In addition, some contaminating flux from quasar image A was also visible in the 2D spectrum from the slit.  
To extract the 1D spectrum of G1, we first fit the spatial profiles along the slit (at each wavelength) of G1, B, and A by two Moffat profiles and a Gaussian profile, respectively.
Moffat profiles were adopted for G1 and B because Gaussian profiles gave worse fits as determined by visual inspection. 
We subtracted off the best-fit spatial profiles for B and A and summed the remaining flux over an extraction window along the slit of $1\arcsec$ (approximately the FWHM of G1's Moffat profile) or $2\arcsec$ (about the extent of G1's profile) in order to extract G1's 1D spectrum.

G1 was targeted on one of the Gemini South GMOS-S masks in Table~\ref{tab:spectro_observations}, ``DESJ0408-5354\_A2'' (R400 grating), hereafter denoted ``Gemini A2,'' on a slit which also included quasar image D.  
The procedure to extract G1's 1D spectrum was entirely analogous to  that described above for the Magellan data.  
To fit the spatial profiles of G1 and D, Moffat profiles were again found to be better than Gaussians. 

In the VLT MUSE data cube, the quasar and source light components were first modeled and removed, and the region near C and G2 was masked out.  
The remaining light from G1 was then extracted to a 1D spectrum by summing the light over a $1\arcsec \times 1\arcsec$ box or a $2.2\arcsec \times 2.2\arcsec$ box.

The velocity dispersion of G1 was then measured from the above data using the ULySS \citep{Koleva2009} galaxy spectral modeling package. 
In the rest wavelength range 4800\AA \ to 5500\AA, including the H$\beta$, Mg, and CaFe features, the G1 spectra were fit to \cite{Vazdekis2010} stellar population models, which used the MILES stellar library \citep{Miles2006} and the \cite{Salpeter1955} initial mass function. 
The wavelength dependent line spread function (LSF) in the Magellan and Gemini-South data were determined from the widths of the arc lamp lines in the respective wavelength calibration spectra, while the LSF of the MUSE data were taken from the fits given in \S3.1 of \cite{Guerou2017}.
Because the Gemini-South and in particular Magellan LSFs were noticeably non-Gaussian, we modified the ULySS package so that it could make use of an empirical LSF, instead of an analytic Gaussian (or low-order Gauss-Hermite) LSF.  
The resulting velocity dispersion measurements and associated statistical errors are given in Table~\ref{tab:velocity_dispersion}, showing good agreement among the results from the four independent data sets.
Plots of the 1D spectra and best-fit models for the $1\arcsec$ extraction window cases are shown in Figure~\ref{fig:velocity_dispersion}.

\subsection{\DEStwo G}
\label{sec:veldisp_2038}

\begin{figure*}
  \includegraphics[width=\textwidth]{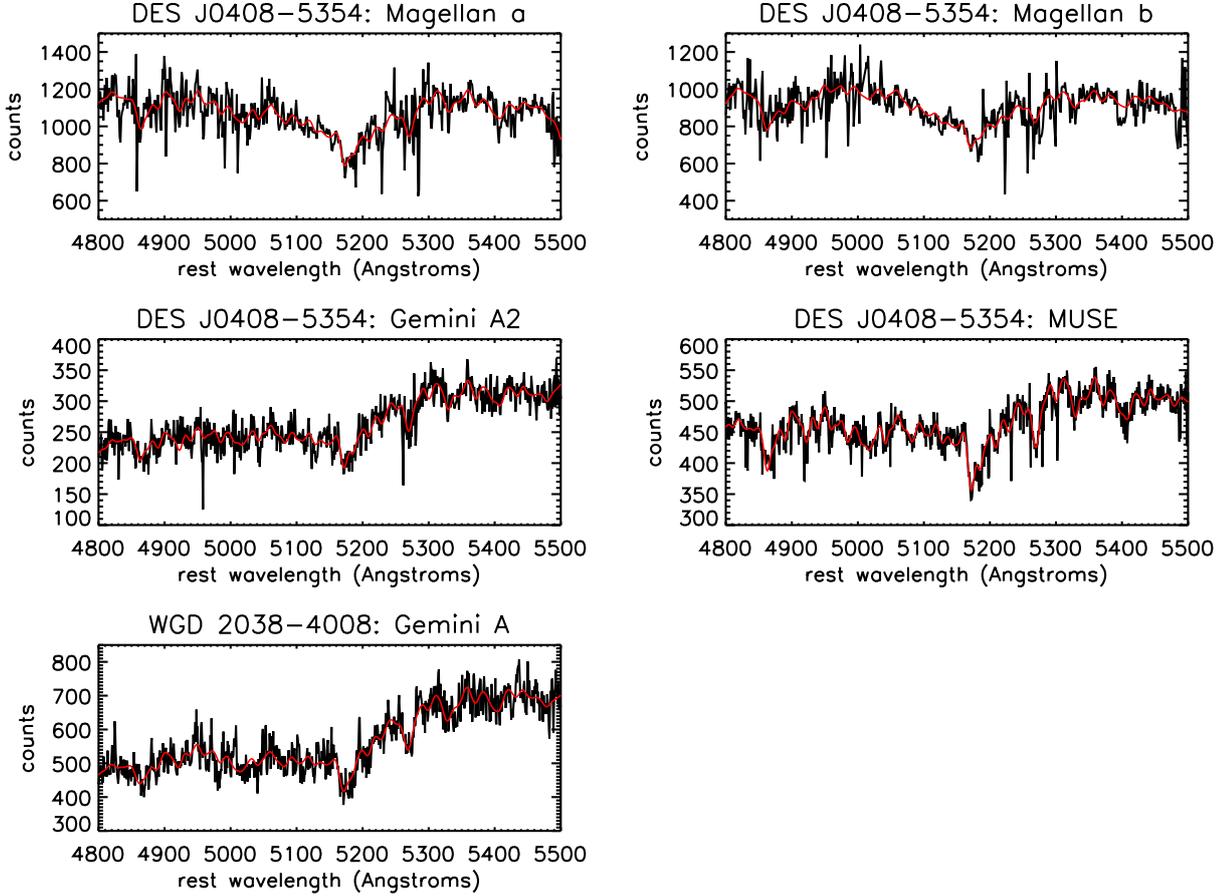}
 \caption{The 1D spectra and fits involved in the velocity dispersion measurements of the main lensing galaxies G1 in the \DESone system (top and middle panels) and G in the \DEStwo system (bottom panel), as described in \S\ref{sec:velocity_dispersion} and listed in Table~\ref{tab:velocity_dispersion}. The black curves in the plots show the observed data, in units of counts vs.\ rest-frame wavelengths in Angstroms, while the red curves show the best-fit models (details in \S\ref{sec:velocity_dispersion}). All the data and fits shown are for the case of a $1\arcsec$ extraction window (see \S\ref{sec:velocity_dispersion} and Table~\ref{tab:velocity_dispersion}).}
 \label{fig:velocity_dispersion}
\end{figure*}

The main lensing galaxy G in \DEStwo was measured from one spectrum observed using Gemini South GMOS-S, specifically on mask (14) listed in Table~\ref{tab:spectro_observations}: ``DESJ2038-4008\_A'' (B600 grating), hereafter denoted ``Gemini A'' for simplicity. 
One slit on this mask targeted galaxy G together with the quasar images C and D \citep[the naming convention for the lensing galaxy and the lensed quasar images are shown in Figure~1 of][]{Agnello2018}.
To extract the 1D spectrum of G, we used the same method described above in \S\ref{sec:veldisp_0408}.
Specifically, we fit the spatial profiles of G, C, and D with three Moffat profiles, subtracted off the best-fit profiles of C and D, and then summed the remaining flux over extraction windows of $1\arcsec$ or $2\arcsec$ along the length of the slit.
The velocity dispersion of G was then measured with ULySS, using the same rest wavelength range and the same stellar population models as above in \S\ref{sec:veldisp_0408}.
The resulting velocity dispersion measurements and statistical errors are given in Table~\ref{tab:velocity_dispersion}.
The 1D spectrum and best-fit model for the $1\arcsec$ extraction window case are shown in Figure~\ref{fig:velocity_dispersion}.

\begin{table}
\caption{Velocity dispersion results for the main lensing galaxies in the \DESone and \DEStwo systems. Details of the measurements are given in \S\ref{sec:velocity_dispersion}.}
\label{tab:velocity_dispersion}
\begin{tabular}{llcccc}
\hline
 & Velocity & Slit & Extraction  & Seeing & Moffat \\
Data & dispersion & width & window & FWHM & index  \\
set & (km/s) & (arcsec) & (arcsec) & (arcsec) & $\beta$ \\
\hline
\hline
\multicolumn{6}{|c|}{\DESone} \\
\hline
\hline
Magellan a & $230 \pm 37$ & 1.0 & 1.0 & 0.68 & 2.97 \\
Magellan b & $236 \pm 42$ & 1.0 & 1.0 & 0.76 & 3.20 \\
Gemini A2 & $220 \pm 21$ & 0.75 & 1.0 & 0.52 & 3.06 \\
MUSE & $227 \pm 9$ & 1.0 & 1.0 & 0.61 & 1.55 \\
\hline
Magellan a & $209 \pm 37$ & 1.0 & 2.0 & 0.68 & 2.97 \\
Magellan b & $230 \pm 47$ & 1.0 & 2.0 & 0.76 & 3.20 \\
Gemini A2 & $261 \pm 21$ & 0.75 & 2.0 & 0.52 & 3.06 \\
MUSE & $227 \pm 9$ & 2.2 & 2.2 & 0.61 & 1.55 \\
\hline
\hline
\multicolumn{6}{|c|}{\DEStwo} \\
\hline
\hline
Gemini A & $296 \pm 19$ & 0.75 & 1.0 & 0.90 & 1.74 \\
Gemini A & $303 \pm 24$ & 0.75 & 2.0 & 0.90 & 1.74 \\
\end{tabular}
\end{table}

\section{Galaxy Group identification}
\label{sec:galaxy_groups}

\subsection{Galaxy Group Identification Algorithm}
\label{sec:groups}

For galaxy-group identification, we employed the same algorithm used in the spectroscopic analysis of the fields of \HOLI lenses \HEofor \citep{Sluse2017} and \WFItwenty \citep{Sluse2019}, which is based on the group-finding algorithms of \citet{Wilman2005} and \citet{Ammons2014}. \citet{Wilson2016} uses a similar method, the results of which were used in the analysis of the \HOLI lens \PGeleven \citep{Chen2019}. We summarize the method here, and refer interested readers to \citet{Sluse2017} for a more complete description and explanation of parameter choices in this algorithm. 

The first step towards identifying galaxy groups involves searching for candidate groups in the spectroscopic redshift distribution of the surveyed galaxy catalog. We begin by constructing a redshift histogram with bins of width 2000 \kms. We identify redshift bins with 5 or more members as candidate groups. To ensure that candidate group members are not split across two bins due to an arbitrary choice of bin boundaries, we construct a second redshift histogram with the bins shifted by half a width of a bin (1000 \kms), and count all non-duplicate redshift peaks from both histograms as candidate groups. We include all other galaxies that are within 1500 \kms of a candidate group member in that candidate group. 

Once we have identified the candidate groups, we use a biweight location estimator \citep{Beers1990} to calculate the mean (group) redshift of each candidate group. The group centroid is also calculated from the positions of the candidate group members. Since \cite{Sluse2019} found that using a luminosity-weighted scheme to calculate the centroid does not improve the match between the group centroid and brightest galaxy in this method, we do not use luminosity-weighted centroids here. 

Once candidate groups have been identified, they are subjected to an algorithm that iteratively removes outliers in both redshift and angular space until the algorithm converges to a stable solution or a group membership of zero. The latter indicates that the candidate group is not gravitationally linked and is spurious. The algorithm is as follows: 

\begin{enumerate}
    \item We set the initial observer-frame velocity dispersion, \vel to 500 \kms. This value will be revised in subsequent iterations. 
    \item Candidate group members that are further than twice the velocity dispersion away from the group redshift are excluded from the group. This corresponds to the following limit 
    \begin{align}
        \delta z_{\rm max} = n \times \sigma_{\rm obs} /c \label{eq:vel_threshold}
    \end{align}
    where $n=2$. This redshift limit is converted into an angular separation limit
    \begin{align}
        \delta \theta_{\rm max} = \frac{c \times \delta z_{\rm max}}{b(1+z)\, H(z)\, D_{\rm \theta}(z) \label{eq:aspect_ratio}} 
    \end{align}
    where $H(z)$ is the Hubble parameter at redshift $z$ and \angdist is the angular diameter distance from the observer to redshift z. Following \cite{Sluse2017}, we set the aspect ratio $b = 3.5$. Candidate group members that have an angular separation that is larger than $\delta \theta_{\rm max}$ from the group centroid are excluded from the group. 
    \item Once cuts have been made in both redshift and angular separation, we recalculate the group centroid, group redshift and observed-frame velocity dispersion \vel from the remaining candidate group members. We obtain the latter two quantities following this framework:
    \begin{itemize}
        \item If there are more than 10 galaxies remaining, we use the biweight location and scale estimators to calculate the group redshift and velocity dispersion, respectively \citep{Mosteller1977}. 
        \item If there are between 4 and 10 galaxies remaining (inclusive), we use the biweight location to calculate the group redshift and the gapper estimator to calculate the velocity dispersion \citep{Wainer1976, Beers1990}.
        \item If there are fewer than 4 galaxies, we use the mean redshift as the group redshift and the standard deviation as the velocity dispersion. 
    \end{itemize}
\end{enumerate}

Steps (ii) and (iii) are repeated until we reach a stable solution. Galaxies that are members of these identified groups are then used to infer group properties, such as the group redshift, centroid, velocity dispersion, and flexion shift (following the method described in \S \ref{sec:flexion}. The rest-frame velocity dispersions are calculated from the observer-frame velocity dispersions using

\begin{align}
    \sigma_{\rm rest} = \frac{\sigma_{\rm obs}}{1+\Bar{z}_{\rm group}}.
\end{align}

We then estimate uncertainties in the group properties by bootstrapping (i.e. random sampling with replacement) the group members of each group 1000 times. We recalculate the group properties of the resampled groups, and use the bootstrapped distribution in those quantities to estimate their uncertainties.

Because the associated measurement uncertainties of the galaxies in the spectroscopic redshift catalog ($\Delta v_{\rm err} \sim 100$ \kms, see \S \ref{sec:redshifts}) are of order the measured velocity dispersion of many of the identified groups, care must be given to account for these uncertainties. To this end, we forward-model the kinematic datasets to infer the velocity dispersion given measurement uncertainties, following techniques used in dwarf-galaxy studies (e.g. \citet{Koposov2011,Walker2011,Amorisco2012}), where it has been found to be especially relevant for systems with small numbers of discrete kinematic tracers \citep{Martin2018,Laporte2019}. We construct a generative likelihood model for the data and evaluate the posterior probability distribution for the intrinsic velocity dispersion, $\sigma_{\rm int}$. The likelihood function is 

\begin{align}
    \mathcal{L} = \prod_{\rm i} \frac{1}{\sqrt{2 \pi} \sigma_{\rm obs}} \exp \left(- 0.5\left(\frac{v_i - \langle v \rangle}{\sigma_{\rm obs}} \right)^2 \right)
\end{align}
where $\langle v \rangle$ is the mean velocity, the product is over all member galaxies $i$ of the galaxy group, and $\sigma_{\rm obs}^2 = \sigma_{\rm int}^2 + \Delta v_{\rm err}^2 $. We assume $\Delta v_{\rm err} = 100 $ \kms, and a non-informative Jeffreys prior for the intrinsic velocity dispersion $\sigma_{\rm int}$ (i.e. $p(\sigma_{\rm int}) \propto 1/\sigma_{\rm int}$) over the range 1 to 1000 \kms. We also assume a uniform prior for the mean group velocity $\langle v \rangle$ over the range -500 to 500 \kms and treat it as a nuisance parameter. We then sample the posterior PDF using the \texttt{emcee} affine invariant Markov Chain Monte Carlo sampler \citep{Foreman2013,Goodman2010}. We then report the median and 68th percentile confidence intervals of the posterior PDF for $\sigma_{\rm int}$ in Table \ref{tab:groups}. For groups where the posterior PDF for $\sigma_{\rm int}$ peaks near zero and the lower bounds are not well-constrained, we report only the 68th percentile upper limits.

\subsection{Identified Groups in the Environment of \DESone}
\label{subsec:DES_group_results_one}
\tabdesgroups

\figdesonegroupsa
\figdesonegroupsb

We applied the galaxy-group identification algorithm to the combined catalog of 199 galaxies with high-confidence redshifts in the field of \DESone described in \S\ref{sec:redshifts}. We identified 10 galaxy groups comprising of 76 galaxies from this spectroscopic sample, which we label Group 1-10 in order of increasing group redshift. Their properties are summarized in  Table \ref{tab:groups}, and Figure \ref{fig:0408groups} shows, for each identified galaxy group, the positions of both accepted and rejected trial member galaxies of that group in right ascension and declination, as well as the distances and velocities relative to the converged group centroid. 

The largest galaxy group identified in this spectroscopic sample is Group 5, which contains 17 member galaxies, including the lens galaxy of \DESone. The centroid of this group is close to \DESone ($26\substack{+16\\-12}$ arcsec), which is also the most luminous member galaxy in the group. 

Aside from group 5, the the identified groups are generally small, with no identified group containing more than 11 member galaxies. For Groups 1, 3, 4, 6 and 7, the posterior PDFs for the intrinsic velocity dispersions peak at or near zero, and the lower limits are not well-constrained. For these distributions, we report the upper 68th percentile confidence intervals for these distributions and treat it as the upper limit of the intrinsic velocity dispersion for that group.

For groups 8 and 9, the distribution of member galaxies in velocity space appears to be bimodal, with two separate subgroups separated by $\sim1000$\kms. However, there are not enough member galaxies in that redshift range to successfully separate these two subgroups into separate groups as none of the individual subgroups have more than 5 potential members. 

The choice of parameters used in the group-finding algorithm described in \S \ref{sec:groups} can impact the final membership of each galaxy group. As mentioned in \S \ref{sec:groups}, the choice of fiducial values for the initial observer-frame velocity dispersion \vel = 500 \kms, velocity threshold $n=2$ (Eq. \ref{eq:vel_threshold}), aspect ratio $b=3.5$ (Eq. \ref{eq:aspect_ratio}) follow that of previous group-finding analyses \citep{Sluse2017,Sluse2019,Wilman2005}. We relaxed the parameter $n$ to  $n=3$ to investigate the effect of a more conservative (i.e. more inclusive) choice in the group finding algorithm and found that relaxing the parameters results in groups that contain outlier members, or have bi- or multimodal configurations, all of which are likely to be spurious. 

As an additional sanity check, we inspected the Chandra X-ray images of the field (PI: Pooley; Program 20419; ACIS-S; 25ks). No diffuse emission is detected (Pooley, D \& Gallo, E. 2019, private communication), with an upper limit (90\% CL) of $\sim10^{44}$ erg/s within a 1Mpc radius (with considerable error bars depending on the assumed temperature). The non-detection makes it unlikely that the lens galaxy is a member of a galaxy cluster.

\subsection{Identified Groups in the Environment of \DEStwo}
\label{subsec:DES_group_results_two}
\figdestwogroups

We applied the group-finder to the 54 galaxies with high confidence redshifts in the field of \DEStwo. From this sample, we identified 2 galaxy groups. The results and properties of these galaxy groups are summarized in Table \ref{tab:groups}, and Figure \ref{fig:2038groups} show, for each identified galaxy group, the positions of both accepted and rejected member galaxies of that group, as well as the distances and velocities of individual galaxies relative to the group centroid. Group 1 in \DEStwo's field contain the eponymous lens galaxy of that field. 

\section{Contribution of Environment Galaxies and Galaxy-Groups to the lens structure}
\label{sec:contribution_environment}
\subsection{Flexion Shift Formalism}
\label{sec:flexion}

A major objective of this analysis is to identify galaxies or galaxy groups along the line of sight or in the environment of lensing systems that significantly perturb the lensing potential of that system and therefore require explicit modeling in the cosmological analysis. Specifically, we want to identify structure that cannot be well-approximated by a uniform perturbation of the lens potential at the position of the lensed images (i.e. external convergence/shear). To do that, we use the ``\textit{flexion shift}'' diagnostic proposed by \citet{McCully2017}, given by

\begin{equation}
\Delta_3 x = f(\beta) \, \times \frac{(\theta_{\rm E} \,\theta_{\rm E,p})^2}{\theta^3}, 
\label{eq:flexion}
\end{equation}

\noindent where $\theta_{\rm E}$ and $\theta_{\rm E, p}$ are the Einstein radii of the main lens and perturber respectively, and $\theta$ is the angular separation on the sky between them. The function $f(\beta)$  is 

\begin{align}
     f(\beta) \in \left\{ 
        \begin{array}{ll} 
         (1-\beta)^2 & \text{if perturber is behind the lens} \\
         1 & \text{if perturber is in the foreground} 
        \end{array} \right\},
\end{align}

\noindent where
\begin{equation}
\beta = \frac{D_{\rm{dp}} D_{\rm{os}}}{D_{\rm {op}} D_{\rm{ds}}},
\end{equation}

\noindent is a combination of angular diameter distances involving the observer (o), deflector (d), perturber (p), and source (s), where the subscripts $D_{ij} = D(z_i, z_j)$ indicate the angular diameter distance between redshifts $z_1$ and $z_2$.

This diagnostic provides a simple quantity to estimate the difference in lensed image positions caused by the leading order non-tidal (i.e. third-order) perturbation produced from a perturber. \citet{McCully2017} showed that by explicitly modeling perturbers with flexion shifts larger than the conservative limit of $\Delta_3 x > 10^{-4} \arcsec$, we can constrain the bias on $H_0$ due to this uncertainty to the percent level. We explain how the Einstein radii, $\theta_{\rm E,p}$, as well as the flexion shift uncertainties for each perturber, are determined for galaxies in \S \ref{subsec:flex_galax}, and for galaxy groups in \S \ref{subsec:flex_groups}. 

We calculated the flexion shift for all galaxies in the spectroscopic survey, as well as the flexion shift of all galaxy groups identified from the survey (\S\ref{subsec:flex_groups}). For individual galaxies, we exclude 4 objects that are in DES Y1 Gold catalog but not in the DES Y3 Gold catalog. We also exclude 12 galaxies with spectroscopic redshifts from MUSE because they do not have DES Y3 photometry. In addition, one galaxy in the spectroscopic sample (\texttt{488065214}) was found to have bad MOF magnitudes (\texttt{MOF\_CM\_MAG\_CORRECTED} magnitudes of -9999 for all bands), and two other galaxies (\texttt{488069251, 488066060}) were found to have bad MOF fits, with unrealistically large sizes (\texttt{MOF\_CM\_T} values on the order of $\sim$5000 square arcseconds) and \texttt{MOF\_CM\_MAG\_CORRECTED\_I} magnitudes that are brighter than their \texttt{MAG\_AUTO\_CORRECTED\_I} magnitudes by more than 4 magnitudes (18.638176 and 17.868253 compared to 23.349312 and 22.776083 respectively). For these three galaxies with spurious MOF photometry, we used the \texttt{MAG\_AUTO\_CORRECTED} photometry to calculate the stellar masses of these galaxies instead. 

For completeness, we also calculated the flexion shifts of all likely galaxies with photometric redshift estimates in the DES "Y3 Gold" photometric catalog within 10\arcmin\  of the lens galaxies, excluding galaxies that are in the spectroscopic sample. To do this, we made the following selections to the DES Y3 Gold catalog: First, we selected all objects within 10\arcmin\  of the lens galaxies in the Y3 Gold catalog that satisfied \texttt{FLAGS\_GOLD = 0} and \texttt{EXTENDED\_CLASS\_MASH\_MOF $\geq$ 2}, which selected likely galaxies. From this catalog, we excluded all galaxies that have \texttt{COADD\_OBJECT\_ID}s that matched galaxies already in the spectroscopic sample. We then selected objects with DNF photometric redshifts that satisfied \texttt{DNF\_ZMEAN\_SOF > 0} and \texttt{DNF\_ZSIGMA\_SOF < 10}, which removed several objects with spurious redshifts. Finally, we made the following cuts specific to the field of each lens galaxy: For \DESone, we removed objects with \texttt{COADD\_OBJECT\_ID = 488068193, 488069583, 488067795}, as they are features of the lens system and not galaxies in the environment of the lens. For \DEStwo, we removed objects with \texttt{COADD\_OBJECT\_ID = 169192447, 169193208, 169192589, 169193438} as they are misclassified stars, and reincluded \texttt{169190696}, which is a galaxy misclassified as a star (see footnote in \S \ref{subsec:galaxy_samples}). After applying these selection criteria, we obtained a photometric catalog of 5082 objects within 10\arcmin\  of \DESone and 4438 objects within 10\arcmin\  of \DEStwo. We then performed the same analysis on these objects as on the spectroscopic sample, using \texttt{DNF\_ZMEAN\_SOF} in lieu of spectroscopic redshift when necessary. 

For the lens galaxies, we use the following quantities in our analysis: For \DESone, we use coordinates $\{\mathrm{RA, DEC}\} = \{62.090417, -53.899889\}$, lens redshift $z_{\rm d} = 0.59671$, source redshift $z_{\rm s} = 2.375$, and Einstein radius $\theta_{\rm E} = 1.80$ \citep{Lin2017a,Shajib2019}. For \DEStwo, we use $\{\mathrm{RA, DEC}\} = \{309.511379, -40.137024\}$, $z_{\rm d} = 0.22829$, 
$z_{\rm s} = 0.777$, $\theta_{\rm E} = 1.38$ \citep{Agnello2018,Shajib2019}.

\subsubsection{Individual Galaxies}
\label{subsec:flex_galax}

We follow the general methodology described in \citet{Sluse2019} to estimate the Einstein radii of galaxies. 

First, we inferred the stellar masses of galaxies from DES photometry using the galaxy template fitting code Le PHARE (see \S\ref{sec:photozmstar}). We then use an empirical scaling relation to estimate the line-of-sight central velocity dispersion of the galaxy, $\sigma$. In this work, we use and compare results derived from two different scaling relations, one from \citet{Zahid2016} and another from \citet{Auger2010}. The \citet{Zahid2016} relation was derived from a sample of $\sim 3.7\times10^{5}$ SDSS elliptical galaxies at z < 0.7 with stellar masses in the range $\log_{10}(M_{\star}/M_{\odot}) \in [9.5,11.5]$. The relation is fit with a broken power law given by Eqn 5 of \citet{Zahid2016}, which we rewrite here in logarithmic form:

\begin{align}
    \log_{10}(\sigma) &= \log_{10}(\sigma_b) + \alpha_1\left(\log_{10}(M_{\star}) - \log_{10}(M_b)\right) \textnormal{for } M_{\star} \leq M_b \nonumber \\
    \log_{10}(\sigma) &= \log_{10}(\sigma_b) + \alpha_2\left(\log_{10}(M_{\star}) - \log_{10}(M_b)\right) \textnormal{for } M_{\star} > M_b 
\end{align}
where $\log_{10}(\sigma_b)=2.073$, $\log_{10}(M_b/M_{\odot}) = 10.26$, $\alpha_1 = 0.403$, and $\alpha_2 = 0.293$. 
Since \cite{Zahid2016} found no significant change of the scaling relation at different redshift bins, we assume that the stellar-mass-to-velocity-dispersion scaling relation does not evolve with redshift.  

Alternatively, we also used the scaling relation from \citet{Auger2010}, which was obtained from fitting a sample of 73 elliptical galaxy lenses from the SLACS survey. The best-fit relation is
\begin{align}
    \log_{10}(\sigma) = 0.18 \log_{10}\left(M_{\star}/(10^{11} M_{\odot})\right) + 2.34
\end{align}
where we have opted to use the best fit parameters for the model that includes the intrinsic scatter, which is $0.04\pm0.01$ in the fit. The elliptical galaxies used for this fit are generally more massive compared to the sample used in the \citet{Zahid2016} analysis, and have stellar masses in the range $\log_{10}(M_{\star}/M_{\odot}) \in [10.5,12]$. We also assume that the stellar-mass-to-velocity-dispersion scaling relation does not evolve with redshift.  

Once we obtain line-of-sight velocity dispersion estimates for each galaxy, we convert the velocity dispersion to the Einstein radius, $\theta_{\rm E,p}$, assuming a Singular Isothermal Sphere (SIS) model

\begin{align}
    \label{eq:SIS}
    \theta_{\rm E,p} = 4 \pi \left( \frac{\sigma}{c} \right)^2 \frac{D_{\rm ps}}{D_{\rm os}}
\end{align},
from which the flexion shift can be calculated with Eq.(\ref{eq:flexion}). 

The uncertainties for the flexion shifts are calculated by adding two different sources of uncertainty in quadrature. The first source comes from the uncertainty in the stellar mass estimates from Le PHARE. The second source of uncertainty comes from the intrinsic scatter in the scaling relation between stellar mass and velocity dispersion. For the \citet{Zahid2016} relation, we quantify this uncertainty by taking half the difference in the central 68th percentile limits of the velocity dispersion distribution at a given stellar mass and use that as the uncertainty from the intrinsic scatter of the scaling relation \citep[see Figure 9(A) of][]{Zahid2016}. For the \citet{Auger2010} relation, we use their fit for the intrinsic scatter, which is $\Delta \log_{10}(\sigma) = 0.05$ (taking the more conservative limit). The two sources of uncertainties are added in quadrature, and then propagated forward into an uncertainty in the flexion shift. 

Flexion shift estimates for galaxies in the spectroscopic sample with stellar mass estimates that are significantly outside the mass ranges used to derive the \citet{Zahid2016} and \citet{Auger2010} scaling relations should be treated with caution, as the scaling relations (and errors) are extrapolated. Therefore, flexion shift estimates for galaxies in our sample with stellar masses $\log_{10}(M_{\star}/M_{\odot}) < 9.5$ should be treated with caution, as both the scaling relations from \citet{Zahid2016} and \citet{Auger2010} may not be valid at the lower end of the stellar mass range. However, the validity of this extrapolation does not affect the main results of this study, since the most significant perturbers (i.e. galaxies that contribute the largest flexion shift contributions) tend to be more massive. The 10 galaxies with the largest flexion shift contributions at the lens positions of \DESone and \DEStwo (Table \ref{tab:flexion_galaxy}) are within the stellar mass ranges used to derive at least one of the two scaling relations. 

\subsubsection{Galaxy Groups} 
\label{subsec:flex_groups}

For galaxy-groups, we obtain a probability density function for the Einstein radius by adopting the same SIS approximation described in Eq. (\ref{eq:SIS}), and sampling 1000 values from the posterior PDF of the intrinsic velocity dispersion of the groups identified in \S \ref{sec:groups} as well as from the bootstrapped PDF of the redshift of each group. 

To obtain the flexion shifts of the galaxy groups and corresponding uncertainties, we use Eq. (\ref{eq:flexion}), sampling from the PDF of the Einstein radius as well as the bootstrapped group centroid position.

\tabdesgalflex

\subsection{Flexion shifts for galaxies and galaxy groups in the field of \DESone and \DEStwo} 
\label{subsec:flexion_results}

\figflexionhistogram

We present a table of the properties of the 10 galaxies with the largest flexion shifts at the lens position in the fields of \DESone and \DEStwo. Comparing the results from the two scaling relations, The scaling relation from \citet{Auger2010} is shallower than the fit by \citet{Zahid2016}, but produces larger estimates of the line-of-sight velocity dispersion for galaxies with stellar masses $\log_{10}(M_{\star}/M_{\odot}) \lesssim 11.45$. Since the majority of galaxies in the sample have smaller stellar mass estimates than that, the \citet{Auger2010} scaling relation produces larger flexion shifts than that from the \citet{Zahid2016} for this sample, and can be treated as the more conservative estimate of the two. In figures \ref{fig:0408_flexion} and  \ref{fig:2038_flexion}, we show the distribution of flexion shifts of the spectroscopic and photometric catalogs for the environment of \DESone and \DEStwo, using both the \citet{Zahid2016} and \citet{Auger2010} scaling relations. The $\log_{10}(\Delta_3 x) > -4$ criteria is indicated by the dashed vertical line. 

When calculating flexion shifts for the photometric catalog, we found some objects with spurious \texttt{MOF\_CM\_MAG\_CORRECTED} photometry.  For these objects we instead used stellar masses computed from \texttt{MAG\_AUTO} photometry, including the same Milky Way extinction and other photometric corrections as for the MOF magnitudes.

Using this criterion, for \DESone, there are four galaxies (G3, G4, G5, G6) with flexion shift contributions above $\log_{10}(\Delta_3 x) > -4$ when using the stellar masses from the \citet{Auger2010} scaling relation. These four galaxies are explicitly modeled in the` lens model analysis \citep{Shajib2019_H0}. In addition, the group including the lensing galaxy, group 5, has a flexion shift of $-3.86\substack{+0.97\\-0.72}$ \arcsec. The large uncertainties in the flexion shift of group 5 is due to a combination of the close proximity of the group centroid to \DESone and the uncertainty in the centroid location, with the upper limits of the flexion shift being produced when the group centroid is near the lens galaxy. 

For \DEStwo, we did not identify any galaxies or galaxy groups with a flexion shift $\log_{10}(\Delta_3 x) > -4$. 

\subsection{Photometrically Identified Galaxy-Groups} \label{sec:photogroup}

\subsubsection{RedMaPPer Clusters in the Field of the Lenses}
Due to the low spectroscopic completeness of the survey of \DEStwo, we complemented our spectroscopic group-identification efforts with a search of all photometrically-identified clusters with richness $\lambda > 5$ in the field of view of \DEStwo using the redMaPPer algorithm \citet{Rykoff2014}. We used the sixth release of the redMaPPer cluster catalog on DES Y3A2 data (v6.4.22+2) and found two redMaPPer clusters within 3\arcmin\ of \DEStwo\ (with unique IDs $\texttt{MEM\_MATCH\_ID = 62659, 138669}$). 

One of the clusters ($\texttt{62659}$) has a photometric cluster redshift of $z_{\rm photo} = 0.221 \pm 0.008$ and a richness of $\lambda = 5.1 \pm 1.7$. The photometric cluster redshift of cluster $\texttt{62659}$ is consistent with the spectroscopically-identified Group 1 ($z_{\rm group} = 0.229$), and its algorithmically-identified central galaxy, $\texttt{COADD\_OBJECT\_ID = 169190452}$, is also a member of Group 1 (see Table \ref{tab:group_mem}), suggesting that redMaPPer cluster $\texttt{62659}$ and Group 1 are the same group. Only two of the seven spectroscopically-identified group members in Group 1 ($\texttt{COADD\_OBJECT\_ID = 169190452, 169189459}$) are also members of the redMaPPer cluster $\texttt{62659}$. However, this could be simply due to bad photometric redshift estimates, as 4 of the 5 group members have $\texttt{DNF\_ZMEAN\_SOF}$ photometric redshift estimates that range from $0.38-0.44$, and the lens galaxy (\texttt{169191076}) has a spurious redshift estimate of $0.00977$. 

The second redMaPPER cluster ($\texttt{138669}$) has a cluster redshift of $z_{\rm photo} = 0.405 \pm 0.017$ and a richness of $\lambda = 10.8 \pm 2.0$. None of the galaxies in this cluster share group membership with the spectroscopically-identified Group 2, suggesting that this group is distinct from Group 2. 

For completeness, we also searched for redMaPPer clusters in the field of \DESone, though the spectroscopic completeness of the field of \DESone is much higher than that of \DEStwo. However, we did not find any clusters within 3\arcmin of \DESone. One reason for this paucity is that a nearby region of the lens has been flagged and precluded from redMaPPer analysis due to a bright star in the foreground. 

\subsubsection{Flexion Shifts for redMaPPer Clusters near the lenses}
We performed the same flexion shift calculations on the redMaPPer clusters as the spectroscopic groups, following the procedure outlined in \S \ref{sec:flexion}, using the same lens parameter quantities, and using the same conservative SIS approximation described by Eq. \ref{eq:SIS} for the perturber. We use the scaling relation given by Eq. (17) of \citet{Andreon2010} to convert the cluster richness into a velocity dispersion estimates for the SIS model. From this, we obtain flexion shifts for the redMaPPer clusters. For cluster $\texttt{62659}$, we obtain a flexion shift of $\log_{10}(\Delta_3 x) = -5.1^{+0.3}_{-0.4}$. For cluster $\texttt{138669}$, we obtain a flexion shift of $\log_{10}(\Delta_3 x) = -6.0 \pm 0.2$. The estimated uncertainties come from propagating both the uncertainties in cluster richness, as well as the uncertainties in the scaling relation. The properties of both groups are summarized in Table \ref{tab:photogroups}. Neither redMaPPer groups in the field of \DEStwo exceeds the threshold of $\log_{10}(\Delta_3 x) > -4$. 

\section{Determining line-of-sight under/overdensities using weighted number counts}
\label{sec:number_counts}

\subsection{Description of the technique}
\label{numbercountsdescript}

To determine the line-of-sight under/overdensities we follow the technique described in section 5 of \citet{Rusu2017}. Like the CFHTLens control fields used in \citet{Rusu2017}, the DES fields also have saturated stars and other artifacts that are masked. Each DES coadd tile contains a mask plane that contains the bleed trails for the saturated stars but not the mask for the stars themselves. The masks for the stars and other artifacts such as dead CCD regions are contained in the {\tt mangle} masks \citep{swanson2008,hamilton2004} that are computed by DESDM for each tile. As can be seen from the upper image in Figure~\ref{fig:des0408-2038-field} there is a large saturated star in the \DESone field close to the lens. The $i-$band and $z-$band {\tt mangle} masks defined a very large mask around this star such that the entire $1000 \times 1000$ pixel area around the lens was masked. We therefore chose to use the {\tt mangle} masks for the $r-$band images for  both the target field and the control fields which did not have this problem. We also chose to use the $r$-band mask for \DEStwo for consistency. 
For each field (target and control) we combine the mask plane and the {\tt mangle} mask to obtain the complete mask. We also use a 5.26\arcsec radius mask at the center of the \DESone target field and a 2.63\arcsec radius mask for \DEStwo to remove the lensing galaxy and quasar images from the calculation. We then apply each control field mask to the target field and apply the target field mask to each control field, as described in section 5.1 of \citet{Rusu2017}. Following the example of \citep{Rusu2019}, for \DESone we are manually removing from the target field catalogue four galaxies that are incorporated in the mass models of \citet{Shajib2019_H0}, in order to avoid double-counting their contribution to $\kappa_\mathrm{ext}$.

We compute the median of the weighted counts for the target field $W_{q}^{meds,t} = N_{gal}^{t} \times median(q_{i}^{t})$, where $q_t$ is the chosen weighting scheme, with $i=1,...,N_{gal}^{t}$ and $N_{gal}^{t}$ is the number of galaxies in the aperture. We compute the same quantity for the control field  $W_{q}^{meds,c} = N_{gal}^{c} \times median(q_{i}^{c})$. For each target field and control field combination we calculate the ratio $\zeta_{q} = W_{q}^{meds,t}/W_{q}^{meds,c}$ for each target and control field combination. Our final weighted count for weighting scheme $q$ is the median of this ratio over all target/control field combinations $\overline{\zeta_{q}^{meds}}$.

\begin{figure*}
 \subfigure{\includegraphics[height=6cm]{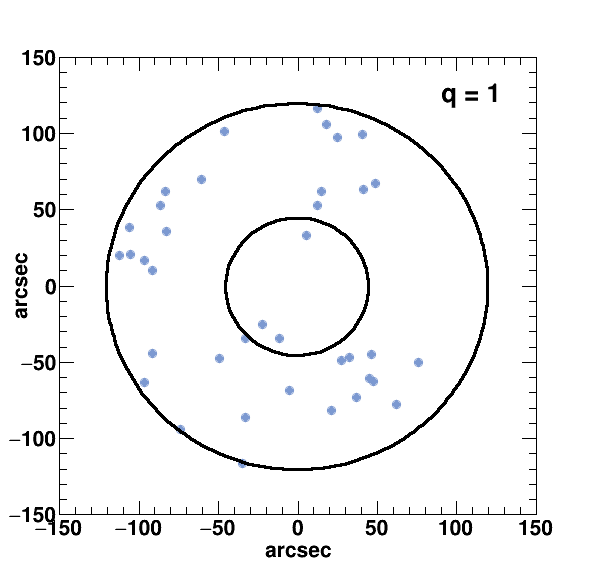}}
 \subfigure{\includegraphics[height=6cm]{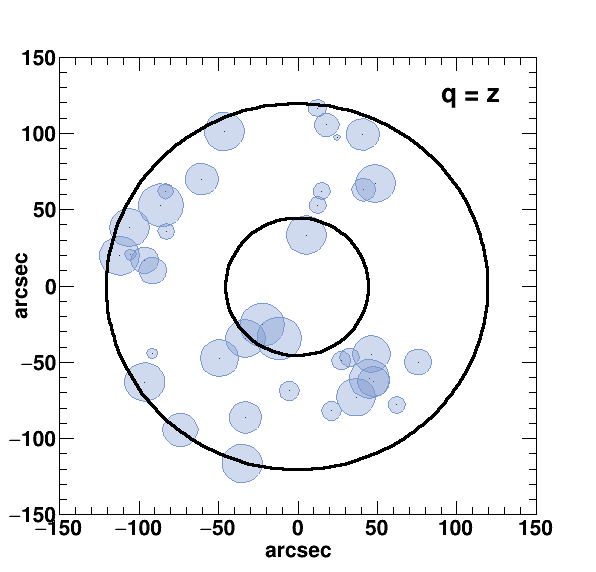}}
 \subfigure{\includegraphics[height=6cm]{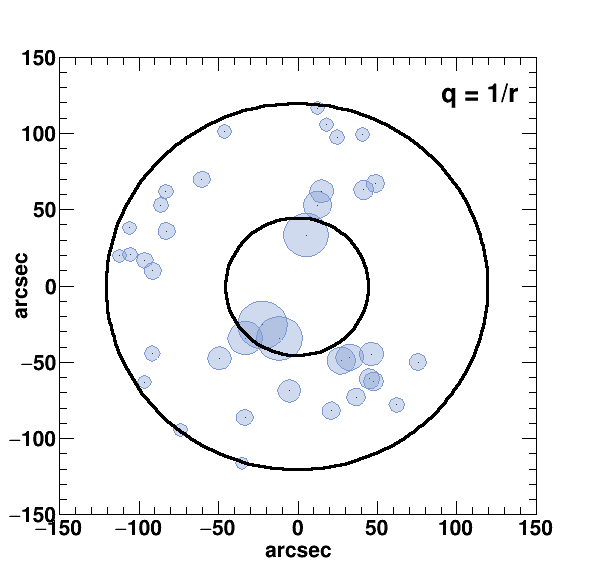}}
 \subfigure{\includegraphics[height=6cm]{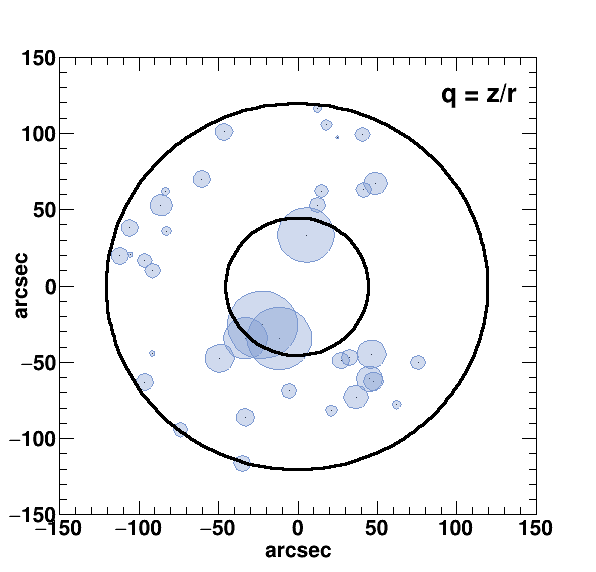}}
 \caption{The relative weights of the galaxies around \DESone for the four weighting schemes $q=1$, $q_{z}= z_{s}\times z_{i} - z_{i}^{2}$, $q_{r} = 1/r$ and $q_{z/r} = (z_{s}\times z_{i} - z_{i}^{2})/r$. The galaxies satisfy $i < 22.5$ and are represented by circles with areas proportional to their weights. The black circles indicate the 120\arcsec and 45\arcsec apertures.}
 \label{fig:average_weights_0408}
\end{figure*}

\begin{figure*}
 \subfigure{\includegraphics[height=6cm]{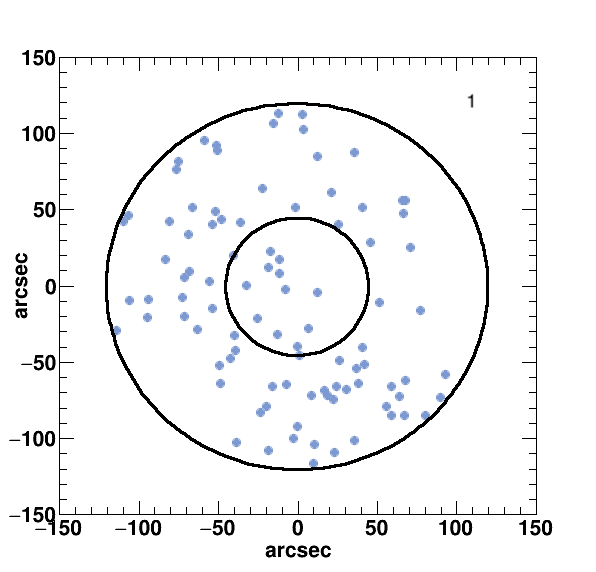}}
 \subfigure{\includegraphics[height=6cm]{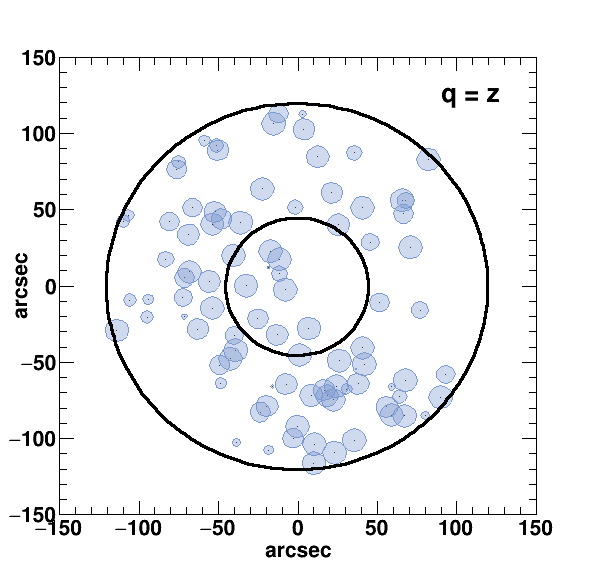}}
 \subfigure{\includegraphics[height=6cm]{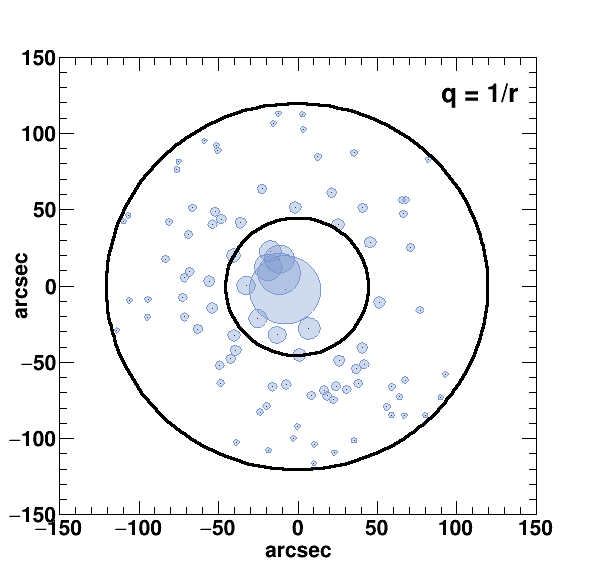}}
 \subfigure{\includegraphics[height=6cm]{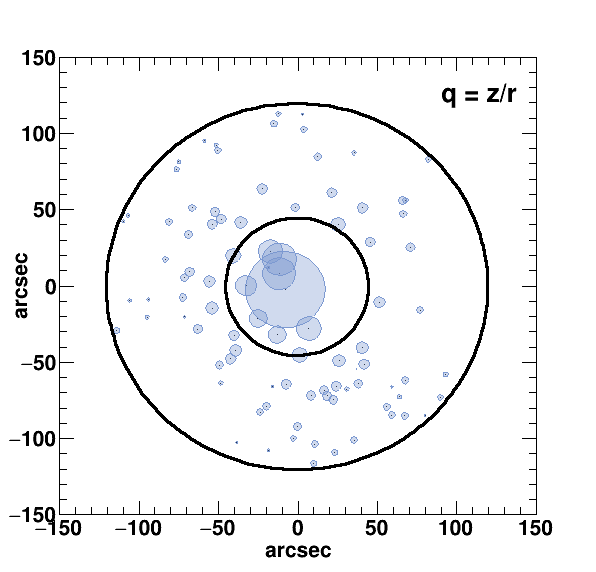}}
 \caption{The relative weights of the galaxies around \DEStwo for the four weighting schemes $q=1$, $q_{z}= z_{s}\times z_{i} - z_{i}^{2}$, $q_{r} = 1/r$ and $q_{z/r} = (z_{s}\times z_{i} - z_{i}^{2})/r$. The galaxies satisfy $i < 22.5$ and are represented by circles with areas proportional to their weights. The black circles indicate the 120\arcsec and 45\arcsec apertures.}
 \label{fig:average_weights_2038}
\end{figure*}

We focus on four weighting schemes we used in \citet{Birrer2019}, $q=1$ which is just the raw counts $N_{gal}^{t}/N_{gal}^{c}$, weighting by redshift $q_{z}= z_{s}\times z_{i} - z_{i}^{2}$, weighting by distance to the lens/center of the field $q_{r} = 1/r$ and weighting by distance to the lens/center of the field and redshift $q_{z/r} = (z_{s}\times z_{i} - z_{i}^{2})/r$. We have used two apertures, one of radius $120\arcsec$ and the other of radius $45\arcsec$.
In Figure~\ref{fig:average_weights_0408} and Figure~\ref{fig:average_weights_2038} we show the relative weights of each galaxy in the \DESone\ and \DEStwo\ fields for $i < 22.5$ and the two apertures $120\arcsec$ and $45\arcsec$. 

\subsection{Resulting distribution for $\zeta_{q}$}

\begin{figure*}
  \includegraphics[width=20cm]{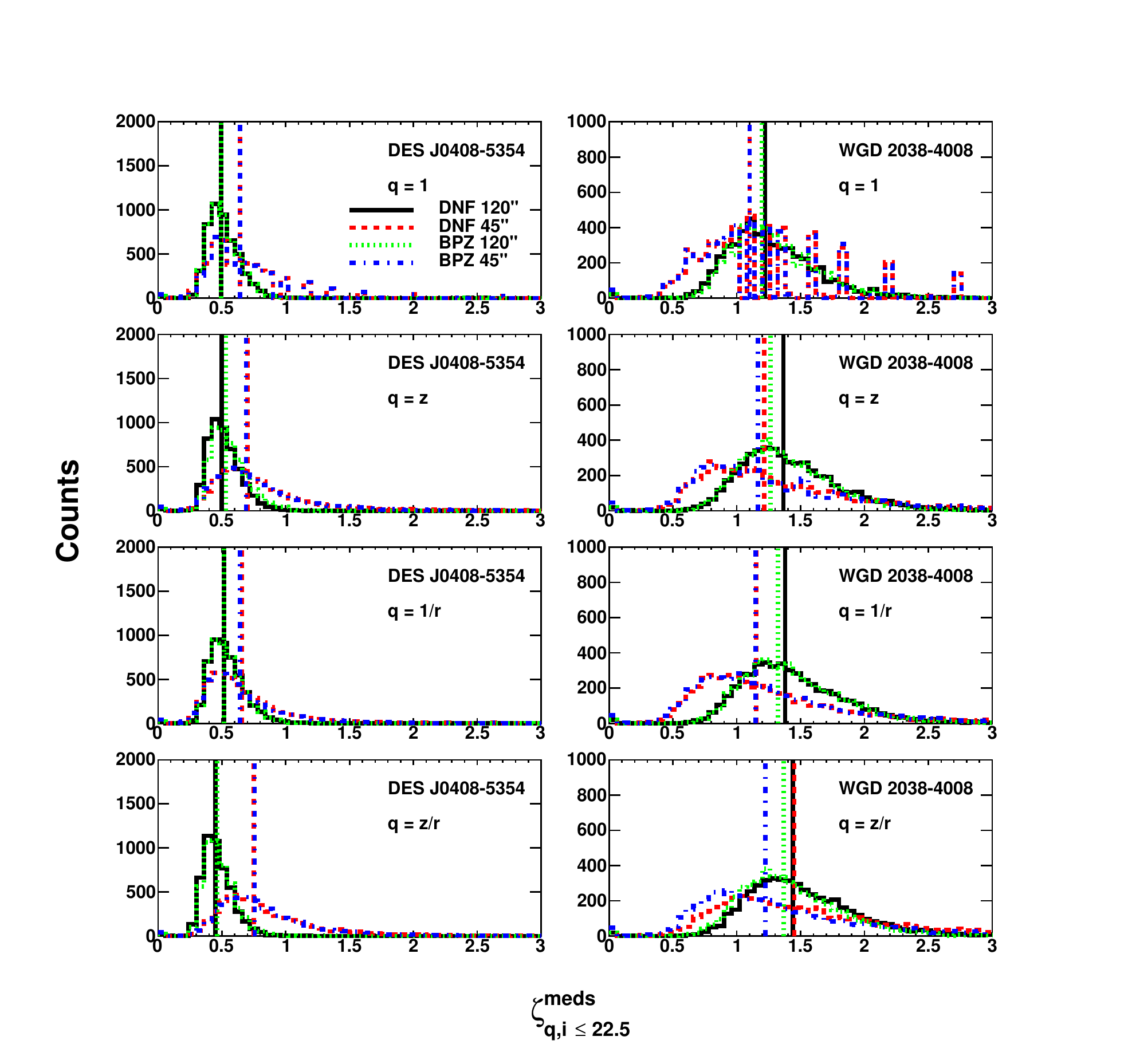}
 \caption{The histograms of the weighted ratios for all $\zeta_{q}^{meds}$ for \DESone and \DEStwo. The vertical lines mark the median of the corresponding distributions. Black solid line - $120\arcsec$ aperture and DNF redshifts. Green dotted line - $120\arcsec$ aperture and BPZ redshifts. Red dashed line - $45\arcsec$ aperture and DNF redshifts. Blue dashed-dotted line - $45\arcsec$ aperture and BPZ redshifts. }
 \label{fig:median_ratios}
\end{figure*}

\begin{figure*}
  \includegraphics[width=15cm]{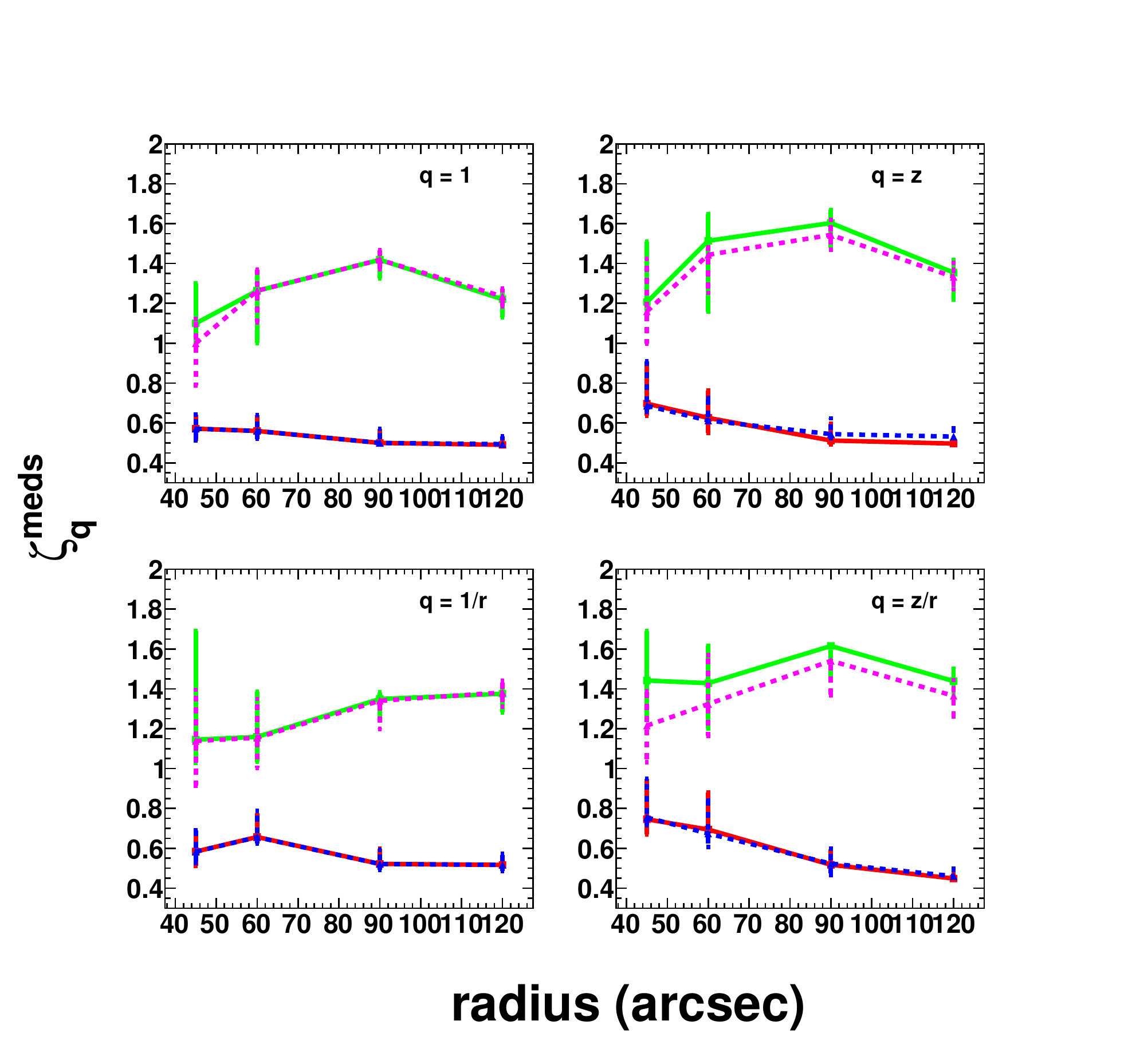}
 \caption{Radial plots of the measured weighted count ratios $\zeta_{q}^{meds}$ calculated for aperture radii of $45\arcsec$, $60\arcsec$, $90\arcsec$ and $120\arcsec$. \DESone: Red - DNF photometric redshifts, Blue - BPZ photometric redshifts. \DEStwo: Green - DNF photometric redshifts, Magenta - BPZ photometric redshifts}
 \label{fig:radial_plot}
\end{figure*}

We present our results for $\overline{\zeta_{q}^{meds}}$ in Table~\ref{tab:ratios}. The uncertainties are derived from taking 20 samplings of the redshift and magnitude errors from a Gaussian PDF distribution corresponding to each galaxy. We show the results for both apertures and for both the DNF and BPZ redshift selections. In Figure~\ref{fig:median_ratios} we show the ratio $\zeta_{q}^{meds}$ for the four weighting schemes for both \DESone and \DEStwo. Figure~\ref{fig:radial_plot} shows a radial plot of the measured over/underdensity for each weight for four different aperture radii: $45\arcsec$, $60\arcsec$, $90\arcsec$ and $120\arcsec$ for both \DESone and \DEStwo. Our analysis shows that the field of \DESone is significantly under-dense (more so than any of the existing H0LiCOW lenses), and this is expected to lead to a tight, negative-value distribution of $\kappa_\mathrm{ext}$ \citep[e.g.,][]{Greene2013}. On the other hand the field of \DEStwo is of about unit density in the $45\arcsec$-aperture, and over-dense in the $120\arcsec$-aperture compared to the random fields.

\renewcommand{\arraystretch}{1.5}
\begin{table*}
 \caption{Weighted galaxy count ratios $\overline{\zeta_{q}}$ for \DESone and \DEStwo.}
 \label{tab:ratios}
 \begin{tabular}{lcccc}
  \hline
  Weight $q$ & $45\arcsec$ & $120\arcsec$ & $45\arcsec$ & $120\arcsec$ \\
  & $i < 22.5$ & $i < 22.5$ & $i < 22.5$ & $i < 22.5$ \\
  & DNF & DNF & BPZ & BPZ \\
  \hline
  \DESone w/ group & & & & \\
  \hline
  1 & $0.643_{-0.06}^{+0.071}$ & $0.495_{-0.01}^{+0.029}$ & $0.643_{-0.06}^{+0.071}$ & $0.495_{-0.01}^{+0.039}$ \\
  $z$ & $0.7_{-0.06}^{+0.207}$ & $0.497_{-0.017}^{+0.029}$ & $0.689_{-0.029}^{+0.223}$ & $0.532_{-0.01}^{+0.042}$\\
  $1/r$ & $0.658_{-0.071}^{+0.095}$ & $0.517_{-0.018}^{+0.048}$ & $0.644_{-0.057}^{+0.108}$ & $0.517_{-0.03}^{+0.056}$\\
  $z/r$ & $0.749_{-0.076}^{+0.204}$ & $0.449_{-0.009}^{+0.033}$ & $0.757_{-0.04}^{+0.193}$ & $0.461_{-0.011}^{+0.038}$ \\
  \hline
  \DESone w/o group & & & & \\
  \hline
  1 & $0.286^{+0.298}_{-0.209}$ & $0.388^{+0.068}_{-0.068}$ & $0.286^{+0.214}_{-0.209}$ & $0.388^{+0.058}_{-0.068}$ \\
  $z$ & $0.389^{+0.376}_{-0.288}$ & $0.373^{+0.100}_{-0.106}$ & 
  $0.394^{+0.278}_{-0.287}$ & $0.388^{+0.093}_{-0.10}$ \\
  $1/r$ & $0.282^{+0.306}_{-0.207}$ & $0.370^{+0.085}_{-0.070}$ & 
  $0.282^{+0.230}_{-0.207}$ & $0.368^{+0.081}_{-0.069}$ \\
  $z/r$ & $0.379^{+0.346}_{-0.278}$ & $0.342^{+0.065}_{-0.079}$ & 
  $0.391^{+0.276}_{-0.286}$ & $0.345^{+0.068}_{-0.076}$ \\
  \hline
  \DEStwo & & & & \\
  \hline
  1 & $0.90_{-0.10}^{+0.20}$ & $1.169_{-0.052}^{+0.039}$ & $0.833_{-0.167}^{+0.556}$ & $1.184_{-0.064}^{+0.066}$ \\
  $z$ & $1.066_{-0.123}^{+0.191}$ & $1.286_{-0.055}^{+0.061}$ & $0.919_{-0.154}^{+0.084}$ & $1.237_{-0.065}^{+0.078}$\\
  $1/r$ & $0.943_{-0.114}^{+0.494}$ & $1.288_{-0.065}^{+0.046}$ & $0.881_{-0.164}^{+0.059}$ & $1.296_{-0.077}^{+0.060}$\\
  $z/r$ & $1.187_{-0.131}^{+0.150}$ & $1.362_{-0.115}^{+0.083}$ & $0.999_{-0.131}^{+0.144}$ & $1.278_{-0.116}^{+0.123}$ \\
  \hline
 \end{tabular}
\end{table*}

\subsection{Computing simulated $\zeta_{q}$ in the MS}

We follow the approached described in \citet{Rusu2017}, in order to implement the same observational constraints to the galaxies in the MS as are relevant to the computation of $\zeta_{q}$ in the DES data. MS is a dark matter-only simulation of the $\Lambda$CDM cosmology, having a periodic box of 500 $h ^{-1}$Mpc on a side with $2160^3 \approx 1.0078 \times 10^{10}$ particles. The simulation run was performed with a modified version of the GADGET-2 code \citet{Springel2005,Lemson2006} and has the spatial resolution limit of $5h ^{-1}$kpc (Plummer-equivalent).  The mass resolution $8.6 \times 10^8\, h^{ -1}M_{\odot}$ and the volume are enough to include a large variety of well-resolved objects from faint quasars to galaxy-clusters. Galaxies can be painted onto these dark matter-only halos using semi-analytical models. Previous H0LiCOW work has employed the models by \citet{DeLucia2007}, but here we are also exploring, for comparison, the newer models of \citet{Henriques2015}. The assignment of these galaxies to halos follows different physical prescriptions, which are adjusted to fit typically low redshift observables. The available catalogues contain synthetic photometry in various bands; we select the $griz$ magnitudes for each galaxy, and sample from them by assigning uncertainties taken from observed DES galaxies, over the same range of magnitude bins. We account for the fraction of galaxies in the target fields that have available spectroscopy, and also for the known DES galaxy-star contamination and incompleteness fractions. The latter fractions are corrected to account for the fraction of the target field apertures covered by {\it HST}, as {\it HST} imaging is assumed to result in the most reliable classification.  

\begin{figure*}
 \includegraphics[width=\textwidth]{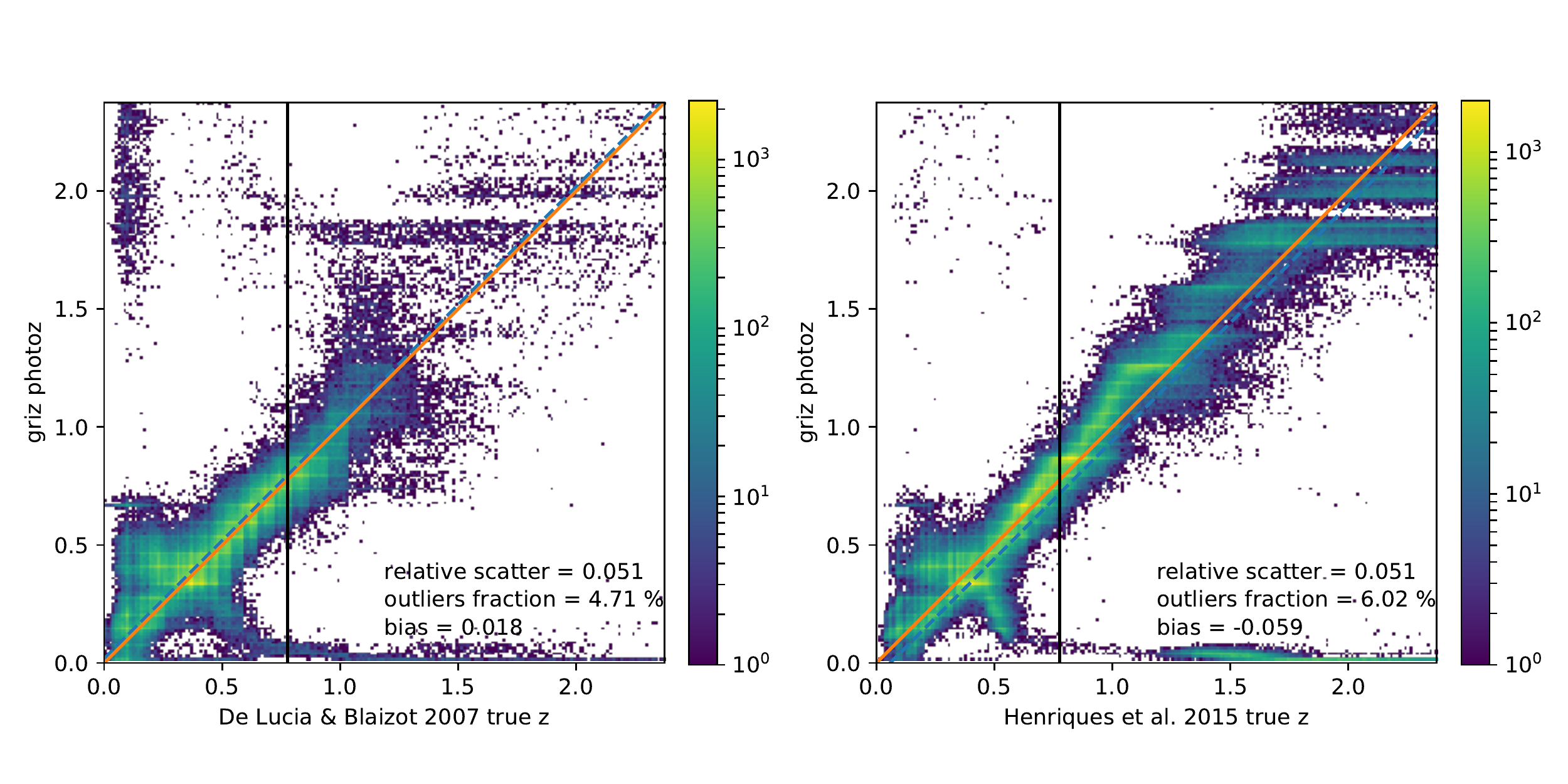}
 \caption{Comparison of catalogue versus photometric redshifts computed for a representative fraction of the MS galaxies painted over the dark matter halos, using the \citet{DeLucia2007} {\it (left)} and \citet{Henriques2015} {\it (right)} semi-analytical galaxy models and resulting synthetic photometry. The photometric redshifts are computed using BPZ, with the errors on the photometry being representative of the DES measurement uncertainties for similar magnitudes. The limiting magnitude used is $i<22.5$ mag. The orange orange solid line marks the true z = photoz identity, and the blue dashed line marks the best linear fit. The plotting range extends to the redshift of \DESone. The vertical black line marks the redshift of \DEStwo. For this lower redshifts, the scatter, fraction of outliers and bias become 0.051 (0.050 for \citet{Henriques2015}), 4.26\% (3.08\%) and 0.023 ($-0.005$), respectively.}
 \label{fig:photozsim}
\end{figure*}

In Figure \ref{fig:photozsim} we plot the resulting comparison of photometric redshifts and catalogue redshifts for a representative sample of the \citet{DeLucia2007}  and \citet{Henriques2015} galaxies, up to the redshift of \DESone (including for the redshift limit of \DEStwo). Our photometric redshifts measured with BPZ have negligible bias up to a redshift of $z\sim1$, above which there is significant scatter, due to the absence of infrared photometry beyond $z-$band. For the \citet{DeLucia2007} models this is a small effect, as there are very few galaxies above this redshift (Figure \ref{fig:SA-Henriques}), but the effect may be more pronounced for the \citet{Henriques2015} models, which predict a significant number of galaxies at large redshift. 

\section{Determining $P(\kappa_\mathrm{ext}$)}
\label{sec:kappa_ext}

Our method of obtaining $P(\kappa_\mathrm{ext}$) relies on selecting lines of sight from the MS which match the observed $\zeta_{q}$ constraints, and constructing the PDF of their associated $\kappa_\mathrm{ext}$ distributions, using the $\kappa_\mathrm{ext}$ maps produced by the ray tracing technique of \citet{Hilbert2009}. This method has been described in detail in \citet{Rusu2017}, and updated in \citet{Birrer2019} and \citet{Rusu2019}. One point we wish to emphasize is that when we computed the relative over/underdensity of the \DESone lens fields in Section \ref{numbercountsdescript} we removed individual galaxy perturbers that were explicitly incorporated into the lens modeling. By doing so, we ensure that these galaxies do not contribute to the $P(\kappa_\mathrm{ext}$) we estimate, and we therefore avoid biasing our estimate high. This is accomplished without the need to alter the input $\kappa_\mathrm{ext}$ maps. 

Furthermore, we have shown in Section \ref{subsec:flexion_results} that the group of galaxies at the redshift of the lens in \DESone contributes a flexion shift close to our threshold for incorporating this structure into the mass models. We therefore compute $P(\kappa_\mathrm{ext}$) for two cases. In the first case, ``w/ group'', we ignore the existence of this structure, and we include the LOS contribution of the constituent galaxies to $P(\kappa_\mathrm{ext}$). In the second case, ``w/o group'', we expect that the structure will be included in the lensing models of \citet{Shajib2019_H0}, Yildirim et al. in prep. and Wong et al. in prep, and we therefore exclude it from the LOS analysis. This is accomplished by removing the galaxy group members from the catalogue of galaxies around the lens, when computing the  weighted count ratio constraints reported in Table \ref{tab:ratios}. We adopt the technique from \citet{Rusu2019} and \citet{Chen2019} to account for spectroscopic incompleteness. That technique consists of two different methods, one which uses the \citet{Andreon2010} relation between group velocity dispersion and richness, and one which assumes Poisson statistics to compute the number of additional galaxy members potentially missed due to sparse spectroscopy. We choose to use only the second method, as the first relies on numerous physical assumptions, and cannot reconcile the small velocity dispersion of the group ($\sim230$ km/s) with the large number of observed members (17). We found a similar mismatch between the two methods in \citet{Chen2019} for the lens PG 1115+080.  

In Figures \ref{fig:kappa} and \ref{fig:kappa2038} we plot the resulting distributions of $P(\kappa_\mathrm{ext})$ for \DESone and \DEStwo, respectively, for a selection of weights, using the \citet{DeLucia2007} galaxy models. The $\zeta_{q}$ constraints are taken from Table~\ref{tab:ratios}, where we marginalize over the DNF and BPZ values. As in \citet{Birrer2019,Rusu2019}, we combine the $\zeta_{q}$ constraints from the 45\arcsec and 120\arcsec apertures. We consider as fiducial distributions, to be used in the cosmological analysis, the ones which use as constraints the most robust $\zeta_{q}$ constraints, i.e. those with $q=1$ and $q=1/r$ in both apertures. In previous work we also used the external shear values corresponding to the best-fit mass models as an additional constraint. At the time our analysis was completed, the final shear values from Yildirim et al. in prep. and Wong et al. in prep, which will complement the cosmographic inference of \citet{Shajib2019_H0}, as well as for \DEStwo, were unknown. Therefore, we choose to report the statistics of $P(\kappa_\mathrm{ext}|\zeta_1^{45\arcsec},\zeta_{1/r}^{45\arcsec},\zeta_1^{120\arcsec},\zeta_{1/r}^{120\arcsec},\gamma)$ for a variety of $\gamma$ values, in Table \ref{tab:kappagamma}.

\begin{figure}
 \includegraphics[width=\columnwidth]{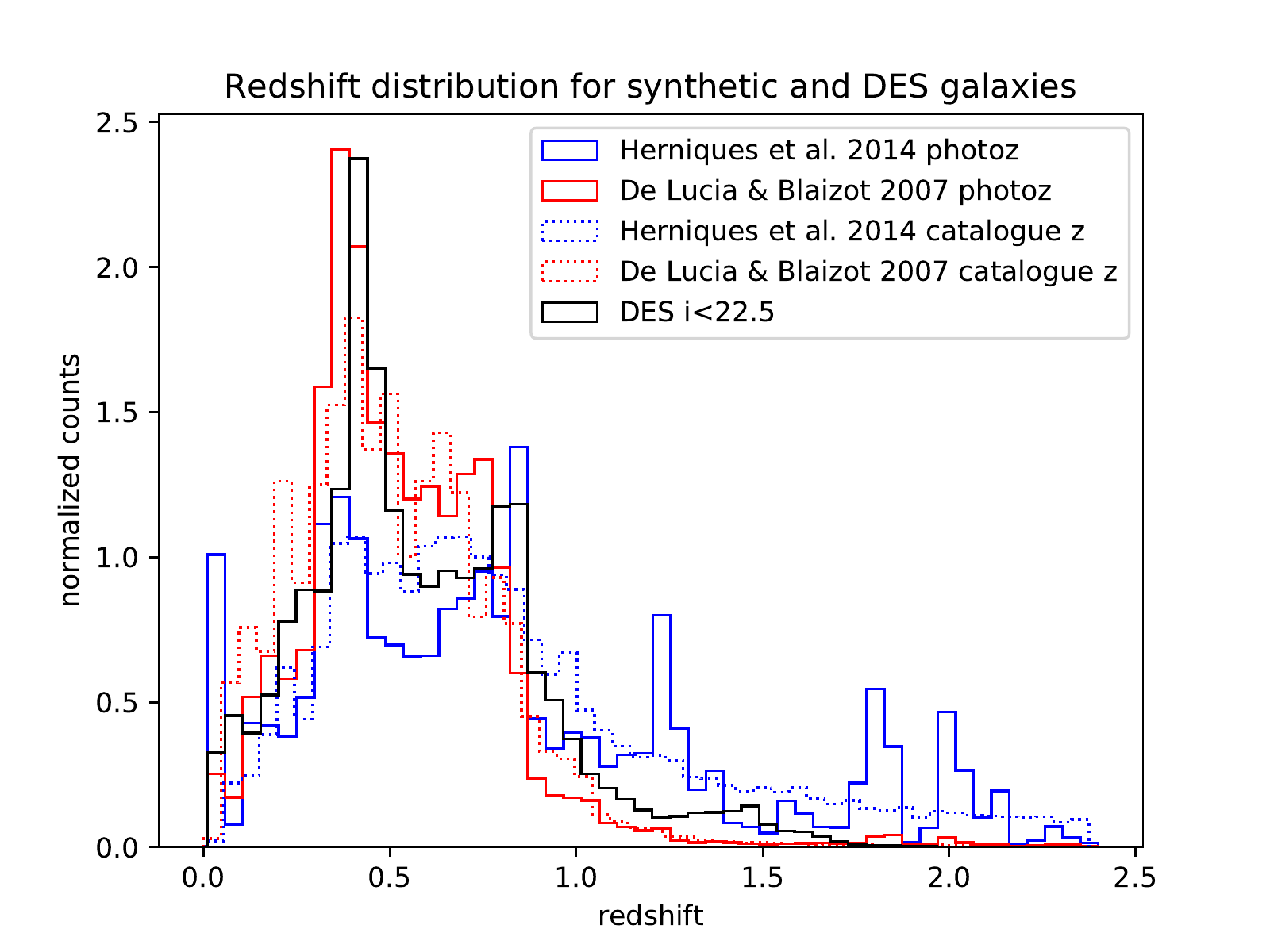}
 \caption{Histograms of the catalogue and BPZ-based photometric redshift distributions for a representative fraction of $\sim500000$ galaxies in the MS, using the semi-analytical models of \citet{DeLucia2007} and \citet{Henriques2015}. The BPZ-based photometric redshift distribution for a similar number of DES galaxies, down to the same $i<22.5$ mag limit, is also shown.}
 \label{fig:SA-Henriques}
\end{figure}

\begin{figure*}
 \includegraphics[width=\textwidth]{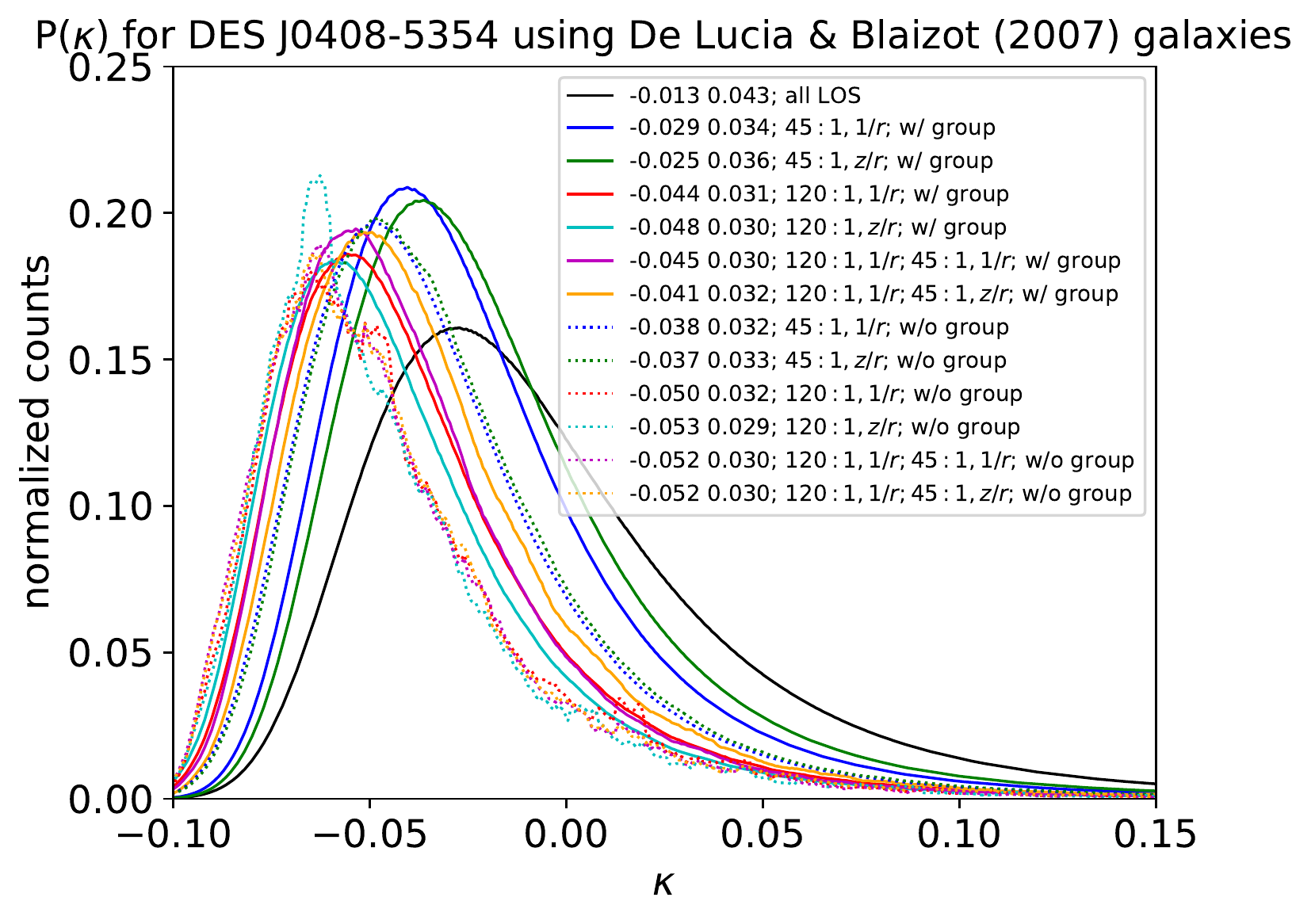}
 \caption{Histograms of smoothed $\kappa_\mathrm{ext}$ distributions for \DESone for a variety of constraints, using the \citet{DeLucia2007} semi-analytical galaxy models. ``w/ group'' and ``w/o group'' refer to the case when the member galaxies of the lens group are kept or not kept as part of the LOS, respectively. The first distribution shown is for the case when all LOS from MS are used, without constraints. The first two numbers in the legend are the median and the semi-difference between the 16th and the 84th percentiles of each distribution, respectively.}
 \label{fig:kappa}
\end{figure*}

\begin{figure*}
 \includegraphics[width=\textwidth]{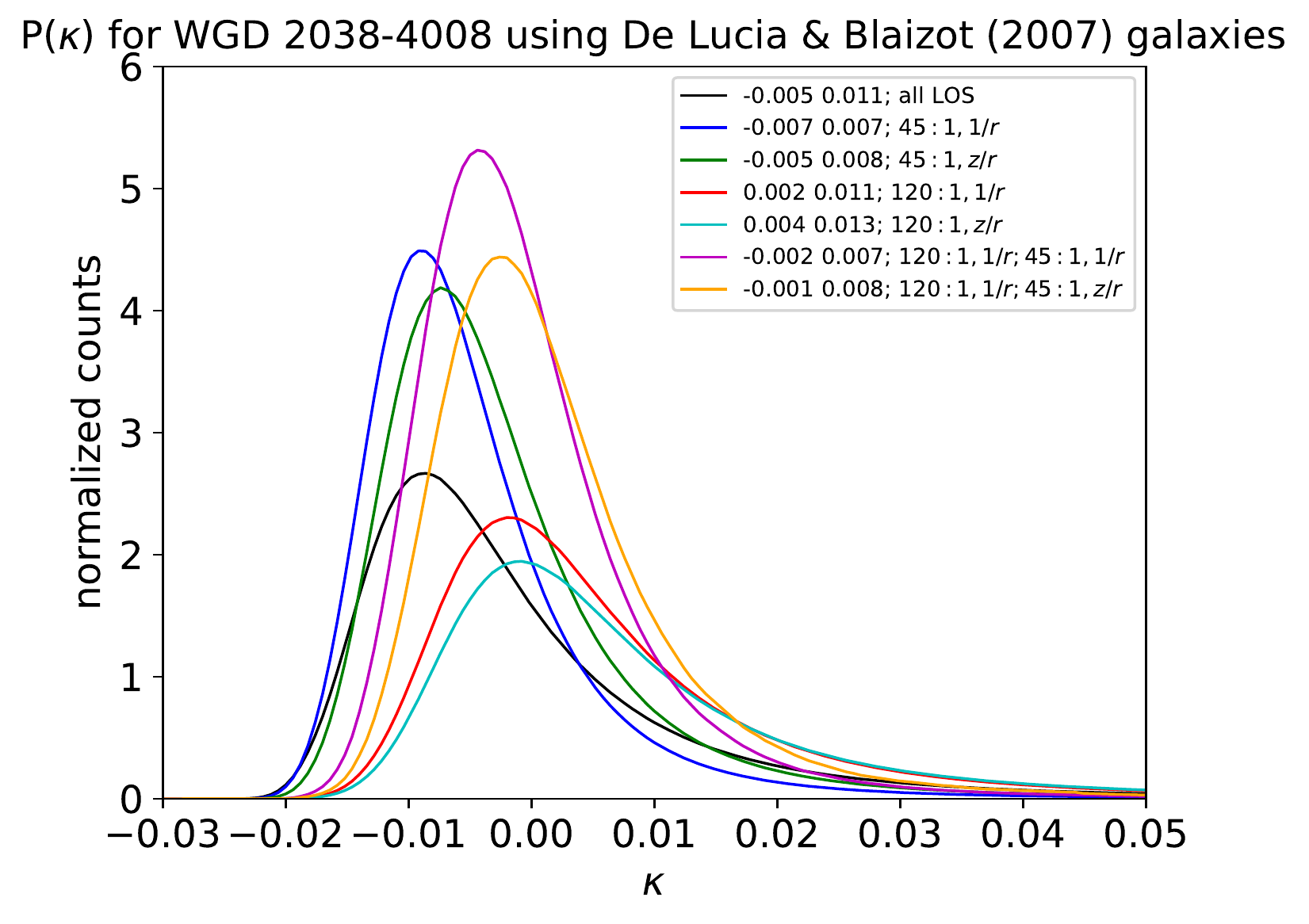}
 \caption{Similar to Figure \ref{fig:kappa}, for the case of \DEStwo.}
 \label{fig:kappa2038}
\end{figure*}

For the choices in Figure \ref{fig:kappa}, we found that the use of the \citet{Henriques2015} models results in $P(\kappa_\mathrm{ext})$ lower by $\kappa_\mathrm{ext}\lesssim 0.01$, therefore at the $\lesssim 1\%$ level. Based on the fact that the photometric redshift distribution of the DES galaxies in Figure \ref{fig:SA-Henriques} is more consistent with that of the \citet{DeLucia2007} models (e.g., the large peak at $z\sim0.5$ and the absence of galaxies above $z\sim1.5$), as well as for consistency with our previous work on H0LiCOW lenses, we adopt the \citet{DeLucia2007} models. We will test in more detail the impact of a particular choice of semi-analytical galaxy models, which appears to be comparatively larger for large source redshifts and over/under-densities, in another work, Mukherjee et al., in prep.

As expected from the significantly underdense field of \DESone, the resulting fiducial $P(\kappa_\mathrm{ext})$ distributions are tight (approximately 0.03 width in $\kappa_\mathrm{ext}$), with medians around $\kappa_\mathrm{ext}\sim-0.04$ - $-0.05$, or $4$ - $5\%$.\footnote{The distributions of $P(\kappa_\mathrm{ext})$ presented in this paper differ slightly from the ones used in the blind cosmographic analysis of \DESone by \citet{Shajib2019_H0}. In addition to that work using the shear constraint derived therein, there is a minor difference owing to a clerical error discovered after unblinding, which corresponds to a change in $H_0$ of 0.13\%, much smaller than the statistical uncertainty of 3.9\%. In order to preserve the blindness of the $H_0$ measurement, this correction has not been propagated through the \citet{Shajib2019_H0} measurement. However, future measurements based on \DESone should use the corrected distribution of $P(\kappa_\mathrm{ext})$ given in this paper.} As the group contribution is removed from the LOS, $P(\kappa_\mathrm{ext})$ decreases by $\lesssim0.01$.

The distributions for \DEStwo are much tighter, with width $\sim 1\%$. This is expected due to the significantly lower source redshift, as there are fewer structures in the MS along the LOS to contribute convergence. The tightening of the distributions as information from multiple apertures is used is evident \citep[see also Figure E2 in][]{Rusu2019}. The medians of the distributions are close to null.

Appendix C of \citet{Rusu2017} shows that our use of the MS to derive 
$P(\kappa_\mathrm{ext})$ can bias the inference because of the different set of assumed cosmological parameters. However, since our $P(\kappa_\mathrm{ext})$ is close to zero, the expected value of the bias, if we assume the cosmological parameters derived from {\it Planck} \citep[e.g.,][]{PlanckCollaboration2016}, is at a level of $\sim0.5\%$, below the 1\% level of accuracy currently aimed at from time delay cosmography \citep{Suyu2017a}.

\begin{table}
\caption{Statistics of $P(\kappa_\mathrm{ext}|\zeta_1^{45\arcsec},\zeta_{1/r}^{45\arcsec},\zeta_1^{120\arcsec},\zeta_{1/r}^{120\arcsec},\gamma)$ as a function of $\gamma$, for the \DESone and \DEStwo systems. We report the variation in the median and semi-difference of the 16th and 84th quantiles. The error bar around each $\gamma$ constraint is 0.005. For \DEStwo  the distributions for large $\gamma$ are too noisy to report.}
\label{tab:kappagamma}
\begin{tabular}{lcccc}
\hline
& \multicolumn{2}{c}{\DESone} & \multicolumn{2}{c}{\DEStwo} \\
$\gamma$ & $\mathrm{med}_\kappa$ & $\sigma_\kappa$ & $\mathrm{med}_\kappa$ & $\sigma_\kappa$\\
\hline
$0.01$ & $-0.053$ & 0.025 & $-0.004$ & 0.006 \\
$0.02$ & $-0.052$ & 0.025 & $-0.002$ & 0.007 \\
$0.03$ & $-0.049$ & 0.026 & 0.002 & 0.008 \\
$0.04$ & $-0.046$ & 0.027 & 0.008 & 0.011 \\
$0.05$ & $-0.041$ & 0.029 & 0.016 & 0.017 \\
$0.06$ & $-0.035$ & 0.032 & 0.032 & 0.028 \\
$0.07$ & $-0.028$ & 0.035 & 0.056 & 0.038 \\
$0.08$ & $-0.020$ & 0.039 & 0.074 & 0.042 \\
$0.09$ & $-0.009$ & 0.043 & 0.101 & 0.046 \\
$0.10$ & $0.002$ & 0.048 & 0.128 & 0.054 \\
$0.11$ & $0.018$ & 0.054 & ... & ... \\
$0.12$ & $0.032$ & 0.060 & ... & ... \\
$0.13$ & $0.046$ & 0.067 & ... & ... \\
$0.14$ & $0.067$ & 0.075 & ... & ... \\
$0.15$ & $0.088$ & 0.083 & ... & ... \\
\hline
\end{tabular}
\end{table}

\section{Conclusions}
\label{sec:conclude}

In this paper we have presented work on three of the ingredients that are necessary to make a high-precision measurement of the Hubble constant $H_0$ using the lensed quasars \DESone and \DEStwo. These are 1) determining the velocity dispersion of the lensing galaxy, 2) identifying galaxies and groups along the line of sight that are close enough to the lens and massive enough that they need to be included in the mass model, and 3) estimating the external convergence $\kappa_\mathrm{ext}$ due to less massive structures that do not need to be included explicitly in the mass model. These ingredients require spectroscopic redshifts for the galaxies in the fields of the two lenses. To obtain these we have carried out spectroscopic observations using Gemini South/GMOS-S, Magellan/LDSS-3 and VLT/MUSE.  As detailed in \S\ref{subsec:specz}-\ref{subsec:completeness}, we obtained a total of 199 high-confidence redshifts from the three instruments for \DESone, corresponding to a redshift completeness of 68\% for galaxies with $18 \leq i < 23$ and $5\arcsec \leq$~radius~$ < 3\arcmin$. For \DEStwo we obtained 54 high-confidence redshifts from the Gemini South/GMOS-S data, with a 16\% redshift completeness for the same $i$-band magnitude and radius ranges. 

As described in \S\ref{sec:velocity_dispersion}, in our redshift survey observations we also set aside slits to measure the stellar velocity dispersions of the main lensing galaxies in our two systems.  The velocity dispersion of the main deflector G1 in \DESone was measured using four independent spectra from the above three instruments, with a consistent result of about 230 \kms (see Table~\ref{tab:velocity_dispersion}). The velocity dispersion for the main lensing galaxy G in \DEStwo was obtained from one spectrum taken using Gemini South/GMOS-S, with a resulting value of about 300 \kms.
The detailed velocity dispersion measurements and uncertainties are given in Table~\ref{tab:velocity_dispersion}.

The galaxy group identification uses the spectroscopic redshifts described above and the same algorithm that was used for the analysis of the \HOLI lenses \HEofor \citep{Sluse2017} and \WFItwenty \citep{Sluse2019}. We find 10 galaxy groups in \DESone for which we then compute the flexion shift $\Delta_3 x$ proposed by \citet{McCully2017}.  \citet{McCully2017} showed that explicitly modeling perturbers with flexion shifts larger than $\Delta_3 x > 10^{-4}\arcsec$ allows one to constrain the bias on $H_0$ due to this uncertainty to the percent level. Out of our 10 groups in \DESone we find one group (labelled 5 in Table \ref{tab:groups}) that has a flexion shift of $\log_{10}\Delta_3 x = -3.86^{+0.97}_{-0.72}$. This group has 17 members, one of which is the lensing galaxy G1 and the centroid of this group is close to the lens. However \citet{Shajib2019_H0} show that the change in $H_0$ of including this group would result in a decrease of approximately 0.4 percent so it is not explicitly included in the mass model.  For \DEStwo, we find two galaxy groups from our spectroscopic redshift sample, but neither group has flexion shift above our cut.

To calculate the flexion shift for individual galaxies we start with the general methodology described in \citet{Sluse2019} and then use two different scaling relations \citep{Zahid2016,Auger2010} to estimate the line-of-sight central velocity dispersion of the galaxy from its stellar mass.  The stellar masses are calculated using galaxy model fitting to DES photometry, as detailed in \S\ref{sec:photozmstar}. 
The \citet{Auger2010} relation between velocity dispersion and stellar mass gives more conservative, larger flexion shift values, resulting in four galaxies in \DESone with flexion shifts larger than our $\log_{10}\Delta_3 x = -4.0$ cut; these galaxies (G3, G4, G5, and G6 in Table~\ref{tab:flexion_galaxy}) are therefore selected for explicit modeling by \citet{Shajib2019_H0}. For \DEStwo, we do not find any individual galaxies with flexion shift greater than our cut.

Our measurement of the external convergence $\kappa_\mathrm{ext}$ starts with determining the line-of-sight under/overdensities for \DESone and \DEStwo using weighted number counts. We use a catalog of galaxy properties for the two fields from the DES Year 3 Gold version 2.2 catalog. As both of these fields are within the DES footprint we are able to select the control fields from the DES catalog as well. This helps us to avoid potential biases due to mismatches in, for example, image resolution between the target and control fields. Where available we use the spectroscopic redshifts for the galaxies in the target fields.  
As detailed in \S\ref{sec:number_counts}, for the galaxy counts we use four different sets of weights (including weighting by redshift and/or radius), two different apertures (one of radius $120\arcsec$ and the other of radius $45\arcsec$), as well as two photometric redshift schemes (DNF and BPZ). For \DESone we remove the four galaxies (G3-G6 in Table~\ref{tab:flexion_galaxy}) which are explicitly incorporated into the lens model. Also for \DESone we calculate the weighted counts for both the case where the galaxy group containing the main deflector G1 is included in the count and the case where we explicitly exclude the group.  We find that \DESone lives in a significantly under-dense environment whereas \DEStwo is in an environment that is closer to mean density. As described in \S\ref{sec:kappa_ext}, we then apply the same observational constraints to the MS, with galaxies from the \citet{DeLucia2007} semi-analytical model, to compute $\zeta_{q}$, the ratio of weighted galaxy counts of target to control fields. We obtain $P(\kappa_\mathrm{ext})$, the probability distribution of the external convergence $\kappa_\mathrm{ext}$, by selecting lines of sight from the MS that match the observed $\zeta_{q}$ constraints. As expected from the significantly underdense field of \DESone, the resulting fiducial $P(\kappa_\mathrm{ext})$ distributions are tight (with width $\approx 0.03$ in $\kappa_\mathrm{ext}$) and medians around $\kappa_\mathrm{ext}\sim-0.04$ - $-0.05$.  
For the case excluding the group containing G1 from the number counts, the median $\kappa_\mathrm{ext}$ decreases by $\lesssim 0.01$.  The distributions for \DEStwo are much tighter, with width $\sim 1\%$. This is expected due to the significantly lower source redshift ($z_{\rm s}=\,$\zstwo\, for \DEStwo vs.\ $z_{\rm s}=\,$\zsone\, for \DESone), as there will be fewer structures along the line of sight to contribute to the convergence.


\section*{Acknowledgments}

J.P. would like to thank Gourav Khullar for their help and insightful discussions that helped improve the analysis in this paper. This work made use of computing resources and support provided by the Research Computing Center at the University of Chicago. J.P. is supported in part by the Kavli Institute for Cosmological Physics at the University of Chicago through grant NSF PHY-1125897 and an endowment from the Kavli Foundation and its founder Fred Kavli.

AJS was supported by the National Aeronautics and Space Administration (NASA) through the Space Telescope Science Institute (STScI) grant HST-GO-15320. AJS was also supported by the Dissertation Year Fellowship from the University of California, Los Angeles (UCLA) graduate division.

TA acknowledges support from Proyecto FONDECYT N: 1190335.

This work was supported by World Premier International Research Center Initiative (WPI Initiative), MEXT, Japan.

CDF acknowledges support for this work from the National Science Foundation under Grant No. AST-1715611.

SM acknowledges the funding from the European Research Council (ERC) under the EUs Horizon 2020 research and innovation program (COSMICLENS; grant agreement No. 787886).

Funding for the DES Projects has been provided by the U.S. Department of Energy, the U.S. National Science Foundation, the Ministry of Science and Education of Spain,
the Science and Technology Facilities Council of the United Kingdom, the Higher Education Funding Council for England, the National Center for Supercomputing
Applications at the University of Illinois at Urbana-Champaign, the Kavli Institute of Cosmological Physics at the University of Chicago,
the Center for Cosmology and Astro-Particle Physics at the Ohio State University,
the Mitchell Institute for Fundamental Physics and Astronomy at Texas A\&M University, Financiadora de Estudos e Projetos,
Funda{\c c}{\~a}o Carlos Chagas Filho de Amparo {\`a} Pesquisa do Estado do Rio de Janeiro, Conselho Nacional de Desenvolvimento Cient{\'i}fico e Tecnol{\'o}gico and
the Minist{\'e}rio da Ci{\^e}ncia, Tecnologia e Inova{\c c}{\~a}o, the Deutsche Forschungsgemeinschaft and the Collaborating Institutions in the Dark Energy Survey.

The Collaborating Institutions are Argonne National Laboratory, the University of California at Santa Cruz, the University of Cambridge, Centro de Investigaciones Energ{\'e}ticas,
Medioambientales y Tecnol{\'o}gicas-Madrid, the University of Chicago, University College London, the DES-Brazil Consortium, the University of Edinburgh,
the Eidgen{\"o}ssische Technische Hochschule (ETH) Z{\"u}rich,
Fermi National Accelerator Laboratory, the University of Illinois at Urbana-Champaign, the Institut de Ci{\`e}ncies de l'Espai (IEEC/CSIC),
the Institut de F{\'i}sica d'Altes Energies, Lawrence Berkeley National Laboratory, the Ludwig-Maximilians Universit{\"a}t M{\"u}nchen and the associated Excellence Cluster Universe,
the University of Michigan, the National Optical Astronomy Observatory, the University of Nottingham, The Ohio State University, the University of Pennsylvania, the University of Portsmouth,
SLAC National Accelerator Laboratory, Stanford University, the University of Sussex, Texas A\&M University, and the OzDES Membership Consortium.

Based in part on observations at Cerro Tololo Inter-American Observatory, National Optical Astronomy Observatory, which is operated by the Association of
Universities for Research in Astronomy (AURA) under a cooperative agreement with the National Science Foundation.

The DES data management system is supported by the National Science Foundation under Grant Numbers AST-1138766 and AST-1536171.
The DES participants from Spanish institutions are partially supported by MINECO under grants AYA2015-71825, ESP2015-88861, FPA2015-68048, SEV-2012-0234, SEV-2016-0597, and MDM-2015-0509,
some of which include ERDF funds from the European Union. IFAE is partially funded by the CERCA program of the Generalitat de Catalunya.
Research leading to these results has received funding from the European Research
Council under the European Union's Seventh Framework Program (FP7/2007-2013) including ERC grant agreements 240672, 291329, and 306478.
We  acknowledge support from the Australian Research Council Centre of Excellence for All-sky Astrophysics (CAASTRO), through project number CE110001020.

This manuscript has been authored by Fermi Research Alliance, LLC under Contract No. DE-AC02-07CH11359 with the U.S. Department of Energy, Office of Science, Office of High Energy Physics. The United States Government retains and the publisher, by accepting the article for publication, acknowledges that the United States Government retains a non-exclusive, paid-up, irrevocable, world-wide license to publish or reproduce the published form of this manuscript, or allow others to do so, for United States Government purposes.

Based on observations obtained at the Gemini Observatory (acquired through the Gemini Observatory Archive and processed using the Gemini IRAF package), which is operated by the Association of Universities for Research in Astronomy, Inc., under a cooperative agreement with the NSF on behalf of the Gemini partnership: the National Science Foundation (United States), National Research Council (Canada), CONICYT (Chile), Ministerio de Ciencia, Tecnolog\'{i}a e Innovaci\'{o}n Productiva (Argentina), Minist\'{e}rio da Ci\^{e}ncia, Tecnologia e Inova\c{c}\~{a}o (Brazil), and Korea Astronomy and Space Science Institute (Republic of Korea).

This paper includes data gathered with the 6.5 meter Magellan Telescopes located at Las Campanas Observatory, Chile.

Based on observations collected at the European Southern Observatory under ESO programme 0102.A-0600(E).

This work made extensive use of the Astropy library, a community-developed core Python package for Astronomy \citep{Astropy2013}. 



\bibliographystyle{mnras}
\bibliography{main}

\section*{Affiliations}
  $^1$Fermi National Accelerator Laboratory, P. O. Box 500, Batavia, IL 60510, USA \\
  $^{2}$National Astronomical Observatory of Japan, 2-21-1 Osawa, Mitaka, Tokyo 181-8588, Japan \\
  $^{3}$Department of Astronomy \& Astrophysics, University of Chicago, Chicago, IL 60637 \\
  $^{4}$Kavli Institute for Cosmological Physics, University of Chicago, Chicago, IL 60637\\
  $^{5}$DARK, Niels Bohr Institute, University of Copenhagen, Lyngbyvej 2, DK-2100 Copenhagen, Denmark \\
  $^6$\ucla\\
  $^7$\ports\\
  $^8$\kipac\\
  $^9$Departamento de Ciencias Fisicas, Universidad Andres Bello Fernandez Concha 700, Las Condes, Santiago, Chile\\
$^{10}$Millennium Institute of Astrophysics, Chile\\
$^{11}$\ucd\\
 $^{12}$\epfl\\
 $^{13}$STAR Institute, Quartier Agora - All\'{e}e du six Aout, 19c B-4000 Li\`{e}ge, Belgium
 $^{14}$Kavli IPMU (WPI), UTIAS, The University of Tokyo, Kashiwa, Chiba 277-8583, Japan\\ 
  $^{15}$ Departamento de F\'isica Matem\'atica, Instituto de F\'isica, Universidade de S\~ao Paulo, CP 66318, S\~ao Paulo, SP, 05314-970, Brazil\\
$^{16}$ Laborat\'orio Interinstitucional de e-Astronomia - LIneA, Rua Gal. Jos\'e Cristino 77, Rio de Janeiro, RJ - 20921-400, Brazil\\
$^{17}$ Instituto de Fisica Teorica UAM/CSIC, Universidad Autonoma de Madrid, 28049 Madrid, Spain\\
$^{18}$ CNRS, UMR 7095, Institut d'Astrophysique de Paris, F-75014, Paris, France\\
$^{19}$ Sorbonne Universit\'es, UPMC Univ Paris 06, UMR 7095, Institut d'Astrophysique de Paris, F-75014, Paris, France\\
$^{20}$ Department of Physics and Astronomy, Pevensey Building, University of Sussex, Brighton, BN1 9QH, UK\\
$^{21}$ Department of Physics \& Astronomy, University College London, Gower Street, London, WC1E 6BT, UK\\
$^{22}$ Centro de Investigaciones Energ\'eticas, Medioambientales y Tecnol\'ogicas (CIEMAT), Madrid, Spain\\
$^{23}$ Department of Astronomy, University of Illinois at Urbana-Champaign, 1002 W. Green Street, Urbana, IL 61801, USA\\
$^{24}$ National Center for Supercomputing Applications, 1205 West Clark St., Urbana, IL 61801, USA\\
$^{25}$ Institut de F\'{\i}sica d'Altes Energies (IFAE), The Barcelona Institute of Science and Technology, Campus UAB, 08193 Bellaterra (Barcelona) Spain\\
$^{26}$ Institut d'Estudis Espacials de Catalunya (IEEC), 08034 Barcelona, Spain\\
$^{27}$ Institute of Space Sciences (ICE, CSIC),  Campus UAB, Carrer de Can Magrans, s/n,  08193 Barcelona, Spain\\
$^{28}$ INAF-Osservatorio Astronomico di Trieste, via G. B. Tiepolo 11, I-34143 Trieste, Italy\\
$^{29}$ Institute for Fundamental Physics of the Universe, Via Beirut 2, 34014 Trieste, Italy\\
$^{30}$ Observat\'orio Nacional, Rua Gal. Jos\'e Cristino 77, Rio de Janeiro, RJ - 20921-400, Brazil\\
$^{31}$ Department of Physics, IIT Hyderabad, Kandi, Telangana 502285, India\\
$^{32}$ Department of Astronomy/Steward Observatory, University of Arizona, 933 North Cherry Avenue, Tucson, AZ 85721-0065, USA\\
$^{33}$ Jet Propulsion Laboratory, California Institute of Technology, 4800 Oak Grove Dr., Pasadena, CA 91109, USA\\
$^{34}$ Santa Cruz Institute for Particle Physics, Santa Cruz, CA 95064, USA\\
$^{35}$ Department of Physics, Stanford University, 382 Via Pueblo Mall, Stanford, CA 94305, USA\\
$^{36}$ Kavli Institute for Particle Astrophysics \& Cosmology, P. O. Box 2450, Stanford University, Stanford, CA 94305, USA\\
$^{37}$ SLAC National Accelerator Laboratory, Menlo Park, CA 94025, USA\\
$^{38}$ School of Mathematics and Physics, University of Queensland,  Brisbane, QLD 4072, Australia\\
$^{39}$ Center for Cosmology and Astro-Particle Physics, The Ohio State University, Columbus, OH 43210, USA\\
$^{40}$ Department of Physics, The Ohio State University, Columbus, OH 43210, USA\\
$^{41}$ Center for Astrophysics $\vert$ Harvard \& Smithsonian, 60 Garden Street, Cambridge, MA 02138, USA\\
$^{42}$ Australian Astronomical Optics, Macquarie University, North Ryde, NSW 2113, Australia\\
$^{43}$ Lowell Observatory, 1400 Mars Hill Rd, Flagstaff, AZ 86001, USA\\
$^{44}$ George P. and Cynthia Woods Mitchell Institute for Fundamental Physics and Astronomy, and Department of Physics and Astronomy, Texas A\&M University, College Station, TX 77843,  USA\\
$^{45}$ Department of Astrophysical Sciences, Princeton University, Peyton Hall, Princeton, NJ 08544, USA\\
$^{46}$ Instituci\'o Catalana de Recerca i Estudis Avan\c{c}ats, E-08010 Barcelona, Spain\\
$^{47}$ Department of Physics, University of Michigan, Ann Arbor, MI 48109, USA\\
$^{48}$ School of Physics and Astronomy, University of Southampton,  Southampton, SO17 1BJ, UK\\
$^{49}$ Brandeis University, Physics Department, 415 South Street, Waltham MA 02453\\
$^{50}$ Computer Science and Mathematics Division, Oak Ridge National Laboratory, Oak Ridge, TN 37831\\
$^{51}$ Max Planck Institute for Extraterrestrial Physics, Giessenbachstrasse, 85748 Garching, Germany\\
$^{52}$ Universit\"ats-Sternwarte, Fakult\"at f\"ur Physik, Ludwig-Maximilians Universit\"at M\"unchen, Scheinerstr. 1, 81679 M\"unchen, Germany


\appendix


\section{Properties for all Galaxies and Galaxy Groups}

We present the full sample of galaxies in the spectroscopic sample as well as the group membership of each identified group. 
\onecolumn
\tabdesgalflexall

\tabgroupmem

\twocolumn
\tabphotozgroup
\bsp	
\label{lastpage}
\end{document}